\documentclass{elsart}
\usepackage{natbib}
\usepackage{graphicx}
\usepackage{graphics}
\usepackage{latexsym}
\usepackage{amsbsy}
\usepackage{amsmath}
\usepackage{amsfonts}
\usepackage{amssymb}
\usepackage{color}
\usepackage{subeqnarray}
\usepackage{bm}
\usepackage[titles]{tocloft}

\begin{document}
\begin{frontmatter}
\title{Link-wise Artificial Compressibility Method}
\author[POLITO]{Pietro Asinari}\footnote{Corresponding author: pietro.asinari@polito.it},
\author[KYOTO]{Taku Ohwada},
\author[POLITO]{Eliodoro Chiavazzo}
and
\author[POLITO]{Antonio Fabio Di Rienzo} 
\address[POLITO]{Dipartimento di Energetica, Politecnico di Torino, Torino 10129, Italy}
\address[KYOTO]{Department of Aeronautics and Astronautics, Graduate School of Engineering,
Kyoto University, Kyoto 606-8501, Japan}
\begin{abstract}
%\blue{[PIETRO,ELIO: To be re-formulated and integrated!!!]}

The Artificial Compressibility Method (ACM) for the incompressible Navier-Stokes equations is (link-wise) reformulated (referred to as LW-ACM) by a finite set of discrete directions (links) on a regular Cartesian mesh, in analogy with the Lattice Boltzmann Method (LBM). 
%The proposed numerical scheme deals with complex boundaries by tracing the intersections of each link with the wall and updating the corresponding variable by a local rule.
%%%%%%
The main advantage is the possibility of exploiting well established technologies originally developed for LBM and classical computational fluid dynamics, with special emphasis on finite differences (at least in the present paper), at the cost of minor changes. {For instance, wall boundaries not aligned with the background Cartesian mesh can be taken into account by tracing the intersections of each link with the wall (analogously to LBM technology).} 
LW-ACM requires no high-order moments beyond hydrodynamics (often referred to as {\em ghost} moments) and no kinetic expansion. Like finite difference schemes, only standard Taylor expansion is needed for analyzing consistency.
Preliminary efforts towards optimal implementations have shown that LW-ACM is capable of similar computational speed as optimized (BGK-) LBM. In addition, the memory demand is significantly smaller than (BGK-) LBM. Importantly, with an efficient implementation, this algorithm may be one of the few which is compute-bound and not memory-bound.
%%%%%%
Two- and three-dimensional benchmarks are investigated, and an extensive comparative study between the present approach and state of the art methods from the literature is carried out. Numerical evidences suggest that LW-ACM represents an excellent alternative in terms of simplicity, stability and accuracy. %Preliminary results on possible extensions to incompressible thermal fluid dynamics are reported as well.
\end{abstract}
\end{frontmatter}

%\red{[Comments for PIETRO]}, \blue{[Comments for ELIO]}
%\\
\newpage
\addtocontents{toc}{\protect\setcounter{tocdepth}{7}}
\tableofcontents

\newpage
\section{Introduction}

Despite a large variety of mesh generation techniques for numerical solvers of the fluid dynamics governing equations \cite{Liseikin2010}, addressing complex geometries remains a difficult duty. To this aim, several approaches were proposed for adapting computational grids to complex geometries by unstructured meshes. Generating unstructured meshes of high quality though, is a challenging computational task {\em per se}, which involves quite advanced algorithms (Ruppert's algorithm, Chew's second algorithm, Delaunay triangulation, etc.) \cite{Liseikin2010}. While those approaches simplify the treatment of boundaries, in turn, each of them introduces new difficulties such as extra terms in the equations, extra interpolations, larger computational molecules, and problems associated with the transfer of information across grid interfaces. The added complexity makes code development even more difficult and increases computation time \cite{Kirkpatrick2003}, with an additional risk that those algorithms may not lead to an acceptable solution.

An alternative approach which has attracted an increasing interest in recent years makes use of Cartesian grids for all cells with the exception of those that present intersections with boundaries, which are thus truncated according to the shape of the boundary surface. The advantages of Cartesian grids can be retained for all cells in the bulk fluid, and a special treatment is only reserved to boundary cells. On the contrary, cells fully outside the flow can be simply ignored during computations \cite{Kirkpatrick2003}. In the literature, this approach is typically referred to as the ``embedded boundary method'', the ``Cartesian grid method'' or the ``cut-cell method'' \cite{Young1991,Zeeuw1993,Ye1999,Ingram2003,Luo2011}. Clearly, the challenging point is to make the method accurate in dealing with curved and planar boundaries transversal to the grid, even though such boundaries are conveniently approximated in a staircase fashion. More specifically, after determining the intersection between the Cartesian grid and a boundary, cells whose center lies in the fluid are reshaped by discarding their part belonging to the solid wall, while pieces of cut cells with the center in the solid are absorbed by neighboring cells \cite{Ye1999}. This results in the formation of control-volumes which are trapezoidal in shape. 

Classical approaches to the incompressible limit of Navier-Stokes equations require (a) dedicated techniques for solving a pressure Poisson equation in order to take advantage of the underlying structured nature of the mesh and {thus speed-up convergence} \cite{Ye1999}. {Moreover}, (b) compact multi-dimensional polynomial interpolating functions are used for obtaining a second-order accurate approximation of the fluxes and gradients on the faces of the trapezoidal boundary cells from available neighboring cell-center values \cite{Ye1999}. Recent developments to this also follows a similar approach \cite{Ingram2003,Luo2011}.

Both (a) the need of a dedicated solver for the pressure Poisson equation and (b) the use of compact multi-dimensional interpolations, can be overcome by the lattice Boltzmann method (LBM) \cite{Succi2001}, while preserving the main features of the Cartesian cut-cell method for mesh generation and boundary treatment. However, this comes at a price of dealing with specific features inherited from the kinetic theory of gases, which are unessential as far as the continuum description of incompressible Navier-Stokes equations is the only concern. 

For this reason, we propose a novel formulation of the artificial compressibility method (ACM), which retains the convenient features of LBM, namely (a) the artificial compressibility and (b) the link-wise formulation based on the theory of characteristics, but concurrently gets rid of unessential heritages of the kinetic theory of gases.

%%%%%%%%%%%%%%%%%%%%%%%%%
Similarities between LBM and ACM \cite{Chorin76} are sometimes reminded in the literature. It is well known indeed that the Chapman-Enskog expansion of the LBM updating rule delivers the governing equation of ACM: the artificial compressibility equations (ACE). The latter consist of the same momentum equations as the incompressible Navier-Stokes equations (INSE), in addition to an artificial continuity equation including pressure time derivative. ACE can be also recovered by the more systematic expansion such as the Hilbert method under diffusive scaling \cite{junk05}.

The lattice kinetic scheme (LKS) \cite{inamuro02} (a variant of LBM) also shows similarities with ACM at the level of computer programming, despite the fact that the former deals with distribution functions of gas molecules, while the latter only with hydrodynamic (macroscopic) variables. 

For a special value of the relaxation parameter in the LBM updating rule, an updated value of the distribution function depends only on the previous equilibrium function at an arbitrary mesh point in the stencil. Since equilibria are in turn function of macroscopic variables only, the LKS updating rule can be immediately recognized as a kind of finite difference scheme, acting on hydrodynamic variables. As a result, the moment system of LKS delivers a variant of ACM. Recently, taking advantage of the similarities between LBM and ACM, the latter was reformulated as a high order accurate numerical method (fourth order in space and second order in time) \cite{ohwada10}. 

Motivated by the common belief that an important reason of success of the LBM (in particular MRT-LBM \cite{dHumieres92}) is its remarkable robustness for simulating the various complex flows, the stability of the revived ACM has been further enhanced \cite{ohwada11}.

In this paper, in an attempt of making ACM even more similar to LBM, we propose yet a new formulation of ACM referred to as link-wise ACM (LW-ACM) in the following text. For the sake of completeness, we summarize both the main features of the revived ACM \cite{ohwada10,ohwada11}, still valid for the present LW-ACM, and the additional advantages due to a link-wise formulation.
\begin{enumerate}
\item ACM deals with macroscopic variables only, thus offering the opportunity of exploiting all the pre-existing finite-difference (FD) technologies: This is, for instance, a clear advantage when imposing inlet and outlet boundary conditions. On the contrary, LBM needs to account for nonhydrodynamic quantities (sometimes called {\em ghost} quantities) which, though may not have direct impact on the hydrodynamic behavior, they can still be responsible of numerical instabilities \cite{Dellar02}. Unfortunately, owing to nonlinearities, there are no clear and general recipes yet, on how to optimally design ghost quantities with desired stability properties. As far as the popular compact stencils are concerned, such as D2Q9, D3Q15 and D3Q19 \cite{qian92} with no special corrections, LBM ghost quantities remain numerical artifacts: Positive effects of such quantities for enhancing stability of usual FD schemes are still far from being clearly demonstrated. ACM fully overcomes this issue, focusing instead on the minimum set of information for incompressible fluid dynamics. 

\item Similarly to LBM, ACM posses the ability of computing transient solutions of incompressible Navier-Stokes equations (INSE), without resorting to a Poisson equation for pressure. The underlying idea, directly inspired by the asymptotic analysis of LBM schemes, is to multiply the pressure time derivative of artificial continuity equation by a mesh-dependent parameter. In this way, the numerical Mach number, which is a mere numerical artifact for INSE (rigorously valid in the limit of vanishing Mach number) is linked to the mesh spacing. Higher accuracy than LBM schemes can be also achieved by exploiting the asymptotic behavior of the solution of the artificial compressibility equations for small Mach numbers \cite{ohwada10}.

\item ACM can use different meshing techniques. For example, it is possible to use simple lattice structures, namely Cartesian structured meshes, eventually recursively refined like those also used by LBM, or it can be even formulated in a finite-volume fashion including unstructured body-fitted meshes. In the latter case, the same comments discussed at the beginning of this section about the computational overhead for generating unstructured meshes hold as well. On the other hand, adopting simple lattice structures is not so simple as in LBM: (a) the boundary conditions must discriminate all possible orientations of the wall with regards to the lattice structure by a look-up table made of discrete cases, and (b) the complex wall treatment depends on the dimensionality of the problem (namely the look-up table for 2D is different from the one for 3D). Both previous problems can be overcome by LBM thanks to the link-wise formulation. A ``link'' is a generic direction identified by a discrete velocity of the lattice and coincides with one of the characteristics along which advection is performed (consistently with the method of characteristics -- MOC). Such a numerical scheme based on a finite set of links can cope with a complex boundaries by (a) identifying the intersections of each link with the wall and (b) updating the variable corresponding to such a link by a local rule. The local rule is always the same (no need for look-up tables) and the intersections can be computed once for all during pre-processing. The previous procedure easily applies to any orientations of the wall with respect to the lattice, as well as to any dimensions. In this paper, the above advantages of LBM are made available to ACM.

\item ACM deals with the minimum number of fields describing incompressible fluid dynamics: D+1, where D is the physical dimension of the problem. On the other hand, LBM deals with discrete distribution functions $f_i$, which are as many as the lattice velocities Q. LBM has thus a memory overhead {due to:} D+1$<$Q. Between these two sets of variables, there is a simple connection: Local equilibria $f_i^{(e)}$ (Q variables) can be computed by means of macroscopic quantities only (D+1 variables). {Introducing a larger set of variables} $f_i^{(e)}$ may seem a redundant and useless artifact. However, this work aims at demonstrating that formulating ACM in terms of $f_i^{(e)}$ offers advantages as well. In particular, as far as the updating rule of the algorithm is similar to LBM, it is possible (eventually with minor changes) to take advantage of most of LBM technology. For example, link-wise ACM can also be formulated in terms of local equilibrium $f_i^{(e)}$ and this enables a convenient treatment of complex moving boundaries typical of the LBM (see next). In conclusion, link-wise ACM has two possible (and fully equivalent) formulations: (a) in terms of macroscopic variables like standard ACM (capable of exploiting pre-existing FD technology) and (b) in terms of local equilibrium (capable of exploiting pre-existing LBM technology).
\end{enumerate}

This paper is organized in sections as follow. The link-wise artificial compressibility algorithm for incompressible isothermal fluid dynamics is introduced In Section \ref{INSE}, where some {classical benchmarks} are presented (isothermal Couette flow, generalized Green-Taylor vortex flow and Minion \& Brown flow) as well. In Section \ref{WBC} the link-wise wall boundary conditions are discussed, including moving and complex walls, and some numerical tests are presented (2D lid driven cavity flow, 3D diagonally driven cavity flow and Circular Couette flow). Finally, Section \ref{conclusions} reports some concluding remarks.

\section{\label{LWACM}Link-wise Artificial Compressibility Method}

\subsection{\label{INSE}The main algorithm: Link-wise formulation}

Link-wise re-formulation of the Artificial Compressibility Method (ACM) for the incompressible isothermal fluid dynamics yields the following system of algebraic equations
{
\begin{equation}\label{ACM2.0}
\begin{split}
f_i(\hat{\mathbf{x}},\hat{t}+1)=
f_i^{(e)}(\hat{\mathbf{x}}-\hat{\mathbf{v}}_i,\hat{t})
+2\left(\frac{\omega-1}{\omega}\right)\left(
f_i^{(e,o)}(\hat{\mathbf{x}},\hat{t})-
f_i^{(e,o)}(\hat{\mathbf{x}}-\hat{\mathbf{v}}_i,\hat{t})\right), \\ i=0,\dots,{Q}-1
\end{split}
\end{equation}}
where ${Q}$ is the number of lattice velocities. {Full derivation of the equations (\ref{ACM2.0}) is provided in the Appendix \ref{appPHYS}}. The {fictitious} variables $f_i$ are defined at a discrete set\footnote{If not evident otherwise, we use ``hat'' notation for lattice quantities expressed by means of integer values and ``prime'' notation for quantities expressed in physical units.} of spatial points $\hat{\mathbf{x}}=\mathbf{x}/\Delta x$, where $\Delta x$ is the dimensionless mesh spacing, $\Delta x=\Delta x'/L$, {with} $\Delta x'$ the mesh spacing in physical units and $L$ the characteristic length scale of the flow field. Similarly, time levels are defined as $\hat{t}=t/\Delta t$, where $\Delta t$ is the dimensionless time step, $\Delta t=\Delta t'/(L/U)$, {with} $\Delta t'$ the time step in physical units and $U$ a characteristic flow speed. 

The ${Q}$ lattice velocities $\hat{\mathbf{v}}_i$ are defined according to the considered scheme \cite{qian92}. {All} points $\hat{\mathbf{x}}$ form a regular lattice such that $\hat{\mathbf{x}}-\hat{\mathbf{v}}_i$ belongs to the lattice, {regardless of} $\hat{\mathbf{x}}$ and $\hat{\mathbf{v}}_i$. The quantities $f_i^{(e)}$ are local functions of density $\rho=\sum_i f_i$ and momentum $\rho\mathbf{u}=\sum_i \hat{\mathbf{v}}_i f_i$ {computed} at $\hat{\mathbf{x}}$ and $\hat{t}$, namely $f_i^{(e)}=f_i^{(e)}(\rho,\mathbf{u})$ (see Appendix \ref{appEQ} for some examples). 

The quantities $f_i^{(e)}$ are designed in order to recover the incompressible isothermal fluid dynamics \cite{qian92}. In particular, recovering incompressible Euler equations requires that $\sum_i f_i^{(e)}=\rho$ and $\sum_i\hat{\mathbf{v}}_i f_i^{(e)}=\rho\mathbf{u}$, i.e. {conservation of hydrodynamic moments, and $\sum_i\hat{\mathbf{v}}_i\hat{\mathbf{v}}_i f_i^{(e)}=\mathbf{\Pi}^{(e)}=\rho\mathbf{u}\mathbf{u}+p\,\mathbf{I}$, with $p$ function of density only (isothermal case): $p=p(\rho)$.} Further constraints can be found by asymptotic analysis (see Appendix \ref{appA} for details). However consistency leaves some degrees of freedom in designing these functions, which can be used for improving stability (one possible strategy is discussed in Appendix \ref{appD}). 

On the other hand, the quantities $f_i^{(e,o)}$, which are the odd parts of equilibria, are defined as
\begin{equation}\label{feqo}
f_i^{(e,o)}(\rho,\mathbf{u})=\frac{1}{2}\left(f_i^{(e)}(\rho,\mathbf{u})-f_i^{(e)}(\rho,-\mathbf{u})\right).
\end{equation}
{A clear advantage of the {above} scheme is that all quantities appearing in Eqs. (\ref{ACM2.0}) and (\ref{feqo}) only depends on known (equilibrium) functions at a mesh node and its close neighbors.} This {introduces a significant simplification in} the treatment of boundary conditions, which can be directly borrowed from finite-difference technology (see e.g. the isothermal Couette flow test case reported in Section \ref{isocouette}).

{Similarly to LBM, the algebraic equations (\ref{ACM2.0}) can be implemented in three subsequent steps (``pull'' formulation), namely pre-combining, streaming and post-combining,}
\begin{subeqnarray}\label{ACM2.0b}
f_i^*(\hat{\mathbf{x}}-\hat{\mathbf{v}}_i,\hat{t})&=&
f_i^{(e)}(\hat{\mathbf{x}}-\hat{\mathbf{v}}_i,\hat{t})
-2\left(\frac{\omega-1}{\omega}\right)
f_i^{(e,o)}(\hat{\mathbf{x}}-\hat{\mathbf{v}}_i,\hat{t}),\\
f_i^{**}(\hat{\mathbf{x}},\hat{t}+1)&=&
f_i^*(\hat{\mathbf{x}}-\hat{\mathbf{v}}_i,\hat{t}),\\
f_i(\hat{\mathbf{x}},\hat{t}+1)&=&
f_i^{**}(\hat{\mathbf{x}},\hat{t}+1)
+2\left(\frac{\omega-1}{\omega}\right)f_i^{(e,o)}(\hat{\mathbf{x}},\hat{t}).
\end{subeqnarray}
{Pre- and post-combining are local processes involving arithmetic operators, whereas streaming alone takes care of data exchange among the nearest neighbors of an arbitrary cell}.

The implementation strategy given by Eqs. (\ref{ACM2.0b}) {admits a straightforward inclusion of external forcing, by considering the additional step}
\begin{equation}\label{force}
f_i^{\text{force}}(\hat{\mathbf{x}},\hat{t}+1)=f_i(\hat{\mathbf{x}},\hat{t}+1)+
f_i^{(e,o)}\left(\rho(\hat{\mathbf{x}},\hat{t}),\mathbf{g}(\hat{\mathbf{x}},\hat{t})\right),
\end{equation}
where $\mathbf{g}=(g_x,g_y)^T$ is the external acceleration. The previous correction is local: Similarly to finite-difference schemes, the external forcing is applied to the point where it is supposed to act. The functions $f_i^{(e,o)}$ are used for convenience (they are already known), for ensuring that the force only applies to the momentum equations. 

{We notice that,} the same simple procedure cannot be applied to the lattice Boltzmann method, because a correction to the distribution function may affect the dynamics of the higher order moments as well. Consistent treatment of the forcing {typically involves} some special (non-trivial) techniques \cite{Guo08}. Details on the effects due to the correction (\ref{force}), by asymptotic analysis, are reported in the Appendix \ref{appA}. {Imposing} a given physical acceleration $\bar{\mathbf{g}}$ {within a} flow, requires the tuning of numerical acceleration $\mathbf{g}$, namely $\mathbf{g}=\epsilon^3\,\bar{\mathbf{g}}$ (in case of diffusive scaling). The same approach can be {adopted} to implement mass sources in the numerical scheme.

\subsection{\label{INSE.fd}The main algorithm: Finite difference formulation}
Since the right hand side of Eq. (\ref{ACM2.0}) only depends on the equilibrium condition, which is {in turn} a function of the macroscopic quantities, it is possible to provide a finite difference formula, expressed in terms of macroscopic quantities, which is fully equivalent to (\ref{ACM2.0}). {As commonly done} in the finite-difference literature, {we denote} by $\{P\}$ the set of computational points surrounding a generic point $P$ {(otherwise stated, the generic computational stencil)}. All the quantities are intended computed at the generic time level $\hat{t}$, while the superscript ``$+$'' {denotes} a quantity at the next time level $\hat{t}+1$. The unknown quantities are given by the velocity components $\mathbf{u}=(u,v)^T$ and the pressure $p$. Hence the equivalent finite-difference formulas must provide a way to compute $u_P^+$, $v_P^+$ and $p_P^+$ namely
\begin{subeqnarray}\label{FDcompact}
u_P^+ &=& f_u\left(u_{\{P\}},v_{\{P\}},p_{\{P\}};\omega\right), \\
v_P^+ &=& f_v\left(u_{\{P\}},v_{\{P\}},p_{\{P\}};\omega\right), \\
p_P^+ &=& f_p\left(u_{\{P\}},v_{\{P\}},p_{\{P\}};\omega\right).
\end{subeqnarray}
See Appendix \ref{appC} for a complete example based on the D2Q9 lattice \cite{qian92}. The same finite-difference counterpart can be found for the Lattice Kinetic Scheme (LKS) \cite{inamuro02}, recovered in case $\omega=1$, but Eq. (\ref{ACM2.0}) is also valid for tunable $\omega$ and consequently tunable viscosity $\nu$ (in particular, for high Reynolds number flows). Moreover, the same derivation can be done for the FD-LKS$\nu$ proposed in \cite{asinari09}, but Eq. (\ref{ACM2.0}) is formulated only along a particular lattice link and hence it can also take advantage of most of LBM technology (which is link-wise). 
\begin{enumerate}
\item {Availability of} two alternative formulations of the same numerical scheme is very convenient. For example, it is possible to commute (even dynamically) between the formulation based on Eq. (\ref{ACM2.0}) and that based on Eqs. (\ref{FDcompact}), depending on the best {option} in dealing with the local boundary conditions. 
\item Similarly to conventional ACM \cite{asinari09,ohwada10,ohwada11}, the formulation based on Eqs. (\ref{FDcompact}) can be improved by introducing a semi-implicit step for updating the pressure field. Essentially the step in Eq. (\ref{FDcompact}c) can be substituted by 
\begin{equation}\label{FDcompact_si}
p_P^+ = f_p\left(u^+_{\{P\}},v^+_{\{P\}},p_{\{P\}};\omega\right),
\end{equation}
using the already updated velocity field.
\item The finite-difference formulation allows to {choose different} relaxation parameters $\omega$ for different (macroscopic) equations. Let us define $\omega_u$ the relaxation frequency used in Eq. (\ref{FDcompact}a) and (\ref{FDcompact}b), while let us define $\omega_p\neq\omega_u$ the relaxation frequency used in Eq. (\ref{FDcompact}c). Consequently two kinematic viscosities follow, namely $\nu = \nu(\omega)$ and $\nu_p = \nu(\omega_p)$, where the function $\nu=\nu(\omega)$ is given by Eq. (\ref{viscosity2}). By introducing these relaxation frequencies in Eqs. (\ref{FDcompact}) and applying Taylor expansion to such novel expressions (see Appendix \ref{appA} for details), the equivalent system of macroscopic equations can be recovered in the continuum limit ($\epsilon\rightarrow 0$). The momentum equation involves the kinematic viscosity $\nu$ (as previously), while the pseudo-compressibility term in the artificially compressible continuity equation, namely the first term in Eq. (\ref{AArho4}), becomes proportional to $\epsilon^2/\nu_p$. Hence, for high Reynold number flows, it is possible to realize $\nu\ll 1$, while $\nu_p\sim 1$ {with a more accurate fulfillment of} the diverge-free condition for the velocity field.  
\end{enumerate}
%
%\item From the computational point of view, it may appear that updating formulas given by Eqs. (\ref{FDcompact}) are not suitable for efficient implementation, because they involve many floating point operations. However, because they are derived from Eq. (\ref{ACM2.0}), it is possible to simplify them using (by-hand) common subexpression elimination (CSE). See Appendix \ref{appC} for a complete example based on the D2Q9 lattice \cite{qian92}. The link-wise ACM on D2Q9 lattice requires about 140 floating point operations (80 additions/subtractions and 60 multiplications) per cell and time step (instead of 90-100 floating point operations required by BGK-LBM). However the required memory demand is exactly one third of that of BGK-LBM and the single-loop implementation is straightforward. In order to check if the advantages balance the drawbacks, a simple test has been performed on a workstation with Intel\textregistered\;Core i7-820QM processor (1.73GHz Quad Core, 8M cache) with 8GB SDRAM memory (1066MHz DDR3). On a mesh with $512\times512$ nodes/sites, performing 12800 iterations required 13 minutes, leading to 4.3 million fluid lattice cell updates per second (MLUPS), which is the standard way to measure the performance of LBM implementations \cite{Wellein2006}. This values is roughly half of the best-in-class LBM implementations, but there is room for further improvements. The important point of this test is that it confirms that updating formulas given by Eqs. (\ref{FDcompact}) can be very efficiently implemented.

\subsection{\label{laphUNITS}Physical and lattice units}
%%%%%%%%%%%%%%%%%%%%%%%%%%%%%%%%%%%%%%%%%%%%%%%%%%%%%%%%%%%%%%
%%%%%%%%%%%%%%%%%%%%%%%%%%%%%%%%%UNITS%%%%%%%%%%%%%%%%%%%%%%%%
%%%%%%%%%%%%%%%%%%%%%%%%%%%%%%%%%%%%%%%%%%%%%%%%%%%%%%%%%%%%%%
\vspace{0.5cm}
\begin{table}[ht]
\caption{{Conversion table from lattice to physical units in case of diffusive scaling, i.e. $\Delta x=\epsilon$ and $\Delta t=\epsilon^2$ with $\epsilon=\Delta x/L=1/N$ and $N$ the number of mesh points along one axis. Quantities in lattice units are readily computed in the code, but they are mesh-dependent. Corresponding quantities in physical units are mesh-independent and can be computed during post-processing. In the text, in case of ambiguity, quantities in physical units are denoted by over-bar.}}
\label{tab:scaling}
\vspace{0.5cm}
\begin{center}
\begin{tabular}{lcc}
	\hline
 Quantity	& Lattice units & Physical units \\
	\hline
Pressure & $p = 1/3\,\sum_i f_i$ & $\bar{p}=(p-p_0)/\epsilon^2$ \\
Velocity & $\mathbf{u} = \sum_i \hat{\mathbf{v}}_i f_i/\sum_i f_i$ & $\bar{\mathbf{u}}=\mathbf{u}/\epsilon$ \\
Force by Eqs. (\ref{MEA4}, \ref{MEA6}) & $\mathbf{\mathcal{F}}=\sum_{i\in \textsf{S}}\hat{\mathbf{v}}_i\mathbf{p}_i$ & 
$\Delta\bar{\mathbf{\mathcal{F}}} = (\mathbf{\mathcal{F}}-\mathbf{\mathcal{F}}_0)/\epsilon^2$ \\
Torque by Eqs. (\ref{MEA5}, \ref{MEA7}) & $\mathbf{\mathcal{T}}=\sum_{i\in \textsf{S}}\left(\hat{\mathbf{x}}
-\hat{\mathbf{x}}_c\right)\times\mathbf{p}_i$ & $\bar{\mathbf{\mathcal{T}}}=\mathcal{T}$ \\
Temperature & $T=\sum_i g_i$ & $\bar{T}=T$
\end{tabular}
\end{center}
\end{table}

%\red{
Let us assume $\Delta x=\epsilon$ and $\Delta t=\epsilon^2$ (diffusive scaling), with $\epsilon=\Delta x/L=1/N$ and $N$ the number of mesh points. {As reported in the Appendix \ref{appA}, asymptotic analysis \cite{junk05} of (\ref{ACM2.0}) and (\ref{feqo}) shows that, in the limit of vanishing grid spacing, $\epsilon\ll 1$, the quantities $\bar{p}=(p-p_0)/\epsilon^2=(p'-p'_0)/U^2$ and $\bar{\mathbf{u}}=\mathbf{u}/\epsilon=\mathbf{u}'/U$ satisfy the incompressible isothermal Navier-Stokes equations, with viscosity}
\begin{equation}\label{viscosity2}
\nu=\frac{1}{3}\left(\frac{1}{\omega}-\frac{1}{2}\right),
\end{equation}
{where the subscript $0$ denotes mean value over the whole computational domain. Here, we stress that for a correct use of the proposed algorithm}, it is important to consider a proper post-processing of the numerical results. {To this respect, we notice that, for instance, the velocity field $\mathbf{u} = \sum_i \hat{\mathbf{v}}_i f_i/\sum_i f_i$ is mesh-dependent: $\mathbf{u} = \mathbf{u}\left(\epsilon\right)$, with $\mathbf{u}$ going to zero as the mesh spacing $\epsilon$ vanishes. As a result, $\mathbf{u}$ is not the proper choice for describing the velocity field of incompressible flows: To this aim, instead, the quantity $\bar{\mathbf{u}}=\mathbf{u}/\epsilon$ is adopted due to mesh-independence. Similar considerations apply also to other fields. For consistency with the LBM literature, in this work, the units of quantities directly based on $f_i$ are referred to {\em lattice units}, while units of the corresponding mesh-independent quantities are termed {\em physical units}.}

{For the sake of clarity,} the complete set {of formulas for converting units of all relevant variables} are reported in Table \ref{tab:scaling}. {We stress that finite-difference formulation of the proposed method can be carried out directly in physical units, thus avoiding the above post-processing. Nevertheless, here we prefer to keep the above post-processing for consistency with the Lattice Boltzmann community.} 

{Finally, similarly to LBM, accurate solution of INSE requires $\epsilon^2/\nu\ll 1$ (see Appendix \ref{appA} for details and the following discussion about the Minion \& Brown flow in Section \ref{minion}). Numerical evidences suggest that LW-ACM is stable for $1\leq \omega<2$, which corresponds only to half of the stability range for the relaxation frequency of the LBM. However, this is the most interesting range from the practical point of view, since it corresponds to the limit of vanishing viscosity (high Reynolds numbers).}
%%%%%%%%%%%%%%%%%%%%%%%%%%%%%%%%%%%%%%%%%%%%%%%%%%%%%%%%%%%%%%
%%%%%%%%%%%%%%%%%%%%%%%%%%%%%%%%%UNITS%%%%%%%%%%%%%%%%%%%%%%%%
%%%%%%%%%%%%%%%%%%%%%%%%%%%%%%%%%%%%%%%%%%%%%%%%%%%%%%%%%%%%%%

\subsection{\label{optimization}Optimized computer implementation}
%%%%%%%%%%%%%%%%%%%%%%%%%%%%%%%%%%%%%%%%%%%%%%%
%%%%%%%%%%COMPUTER SCIENCE PART%%%%%%%%%%%%%%%%
%%%%%%%%%%%%%%%%%%%%%%%%%%%%%%%%%%%%%%%%%%%%%%%
{In the following, we discuss a few strategies useful for reducing the computational time, thanks to an optimized implementation of the algorithm (\ref{ACM2.0b}) and (\ref{FDcompact}).} Similarly to LBM, the performance characteristics of single-processor implementations depends on the effect of different data layouts \cite{Wellein2006} (multi-processor optimization strategies are not considered in the present work, see Ref. \cite{Hager2010}). 

First of all, the streaming step should not overwrite data required for updating neighboring sites. A usual way to work around the resulting data dependencies owing to the propagation step is the use of two arrays (one for the current and the other for the next time step), and toggling between them \cite{Wellein2006}. It is also possible (and even more efficient) using a single array, with proper ordering of the sequence of streamed lattice directions, though this may become cumbersome when dealing with wall boundary conditions. The best data layout requires that the distribution functions of the current cell are contiguously located in memory (e.g. by using the first index in Fortran, which addresses consecutive memory locations due to column major order \cite{Wellein2006}). 

Secondly, to reduce the memory traffic, it is important that pre-combining, streaming and post-combining are executed in a single loop and not independently of each other in separate loops or routines, similarly to LBM \cite{Wellein2006}. This goal can be easily accomplished by reformulating Eqs. (\ref{ACM2.0b}) in term of $f_i^*$: In fact, the hydrodynamic moments of $f_i^*$ are not exactly equal to the hydrodynamic quantities, but the former are known functions of the latter. Hence it is convenient to compute directly $f_i^*$, which are ready to be streamed, and to extract the hydrodynamic quantities from $f_i^*$. This simplifies the implementation of a single updating loop through all computational sites at each time step.

Finally, it is important to reduce as much as possible the number of floating point operations and memory accesses per updated site. In LBM, the D3Q19 lattice (see Appendix \ref{appEQ} for details) with BGK collision operator requires about 180-200 floating point operations per cell and time step as well as reading 19 floating point values and writing to 19 different memory locations \cite{Wellein2006}. Roughly half of the floating point operations and half of the memory accesses are required by the D2Q9 lattice. 

Let us considering the D2Q9 lattice and the LBM-style formulation of LW-ACM, as dictated by Eqs. (\ref{ACM2.0b}), with the optimization tricks reviewed above. Such an implementation of LW-ACM requires 115 floating point operations (+28\% compared to BGK-LBM) and 26 memory accesses (+44\%) per cell. 
%However external size of the computational stencil for link-wise ACM by LBM-style formulation is the same as BGK-LBM (namely 3$\times$3$\times$9). In spite of more floating point operations and more memory accesses, link-wise ACM by LBM-style formulation has roughly the same performance in terms of computational time as BGK-LBM (results not reported). This means that the external size of the computational stencil is the real bottleneck for performance. In fact, link-wise ACM by FD-style formulation will achieve better performance by a smaller computational stencil, as shown in the following.
%{Therefore, due to superior features of LW-ACM (compared to BGK-LBM), only the performances of the FD-style LW-ACM will be further investigated in the sequel.}
{However, in the following, only the computational performances of LW-ACM in the FD formulation are further investigated, and its superior capabilities are demonstrated (compared to BGK-LBM).}
%%%%%%%%%%%%%%%%%%%%%%%%%%%%%%%%%%%%%%%%%%%%%%%
%%%%%%%%%%COMPUTER SCIENCE PART%%%%%%%%%%%%%%%%
%%%%%%%%%%%%%%%%%%%%%%%%%%%%%%%%%%%%%%%%%%%%%%%
%%%%%%%%%%%%%%%%%%%%%%%%%%%%%%%%%%%%%%%%%%%%%%%
%%%%%%%%%%COMPUTER SCIENCE PART%%%%%%%%%%%%%%%%
%%%%%%%%%%%%%%%%%%%%%%%%%%%%%%%%%%%%%%%%%%%%%%%
%%%%%%%%%%%%%%%%%
\vspace{0.5cm}
\begin{table}[ht]
\caption{Performance test of the link-wise ACM (by FD formulation) vs. BGK-LBM, based on the Minion \& Brown flow \cite{Minion97} with $\mbox{Re} = 10,000$ in the time {range $t \in [0,1]$}, solved by a mesh with $512\times512$ nodes/sites and performing 12,800 iterations ($\mbox{Ma} = 0.04$). Both codes are serial and use double precision. The considered workstation has Intel\textregistered\; Core\texttrademark\; i7-920 (Bloomfield, 4 physical cores, 8MB L3) with clock rate 2.67GHz (due to TurboMode\texttrademark\; actually running at 2.80 GHz) and 12 GB of DDR3 memory (1333 MHz). The used Fortran compiler is Intel\textregistered version 11.1up8 (optimization level option ``-O3'') and the operative system is Ubuntu Linux i10.04 LTS (64 bit). The million fluid lattice cell updates per second (MLUPS) are reported for both methods.}
\label{performance}
\vspace{0.5cm}
\begin{center}
\begin{tabular}{lcc}
	\hline
 Elementary stencil	& Link-wise ACM by FD formulation & BGK-LBM \\
	\hline
	\# of additions/subtractions & 80 & 70 \\
	\# of multiplications & 60 & 40 \\
	\# of floating point operations & 140 & 110 \\
	\hline
	\# of actual data ($t$) & 27 & 9 \\
	\# of updated data ($t+1$) & 3 & 9 \\
%	\# of data (input$+$output) & 30 & 18 \\
	external size of the stencil & 3$\times$3$\times$3 & 3$\times$3$\times$9 \\
	\hline
	MLUPS & 29.43 & 27.28 \\
\end{tabular}
\end{center}
\end{table}
From a computational point of view, it may appear that formulas (\ref{FDcompact}) are not suitable for an efficient implementation, since they involve many floating point operations. However, because they are derived from Eq. (\ref{ACM2.0}), it is possible to simplify them using (by-hand) common subexpression elimination (CSE) \cite{Hager2010}. See Appendix \ref{appC} for a complete example based on the D2Q9 lattice \cite{qian92}. Moreover, the same implementation tricks discussed above can be {properly} applied here. 

First of all, the memory storage required by link-wise ACM is exactly one third of that of BGK-LBM (only hydrodynamic variables are needed). At each time step, it is enough to go through all computational cells/sites once and this can be done straightforwardly, because updating formulas are already expressed in terms of hydrodynamic variables. Finally, for locating the macroscopic quantities $(p,u,v)$ contiguously in memory, it is possible to collect them in a single array and to use the first index for addressing them. This leads to an optimized FD-style implementation of Eqs. (\ref{FDcompact}). On the D2Q9 lattice, a comparison between the FD-style implementation of link-wise ACM and BGK-LBM is reported in Table \ref{performance}. Link-wise ACM requires more floating point operations but less memory accesses than BGK-LBM. 

{For clarity, let us analyze the updating process at each time.} The external size of the stencil of LW-ACM is smaller than the one of LBM (3$\times$3$\times$3 instead of 3$\times$3$\times$9 respectively, where 3$\times$3 is due to the D2Q9 lattice, 3 is the number of hydrodynamic quantities and 9 the number of discrete distribution functions). During the generic updating process, if the cache is large enough to hold 3 ``lines'' (or 3 planes in 3D) of the computational domain, then updating the hydrodynamic quantities in a cell requires {the loading of} only the actual values of a further cell in the cache (3 loads). In the worst case, updating the hydrodynamic quantities in a cell requires {the loading of} the actual values of three {additional} cells in the cache:  This {amounts to} 9 loads from actual array (3 physical quantities from the current, previous and next ``line''). In any case, 3 write-allocate transfers from main memory to cache and 3 stores to get the updated array from cache back to main memory are always required. This leads to 9--15 in total per nodal updates. 

On the other hand, BGK-LBM requires 9 loads from the actual array (discrete distribution function), 9 write-allocate transfers and 9 stores to the updated array, leading to 27 memory transfers. For BGK-LBM, there is no reuse of data from cache because every discrete distribution function is only used once. As the number of memory transfers usually affects the performance more than the number of floating point operations, the performance of link-wise ACM is superior than that of BGK-LBM. Some performance data are reported in Table \ref{performance}. FD-style implementation of link-wise ACM was able to achieve 29.43 million fluid lattice cell updates per second (MLUPS), which is the standard way to measure the performance of LBM implementations \cite{Wellein2006}. For the previous test, this value is roughly 8\% faster than BGK-LBM. 

{\em Remark:} In our opinion, there is still room for improvement according to the performance model (based on assuming either infinitely fast memory or infinitely fast compute units). For example, the numerical code for solving the 2D lid driven cavity test case achieved 32.3 MLUPS and this was essential for simulating very high Reynolds number flows. Importantly, with an efficient implementation, this algorithm may be one of the few which is compute-bound and not memory-bound. The latter {observation} is of particular interest for General-Purpose computing on Graphics Processing Units (GPGPU). %In conclusions, we stress that this test confirms that updating formulas given by Eqs. (\ref{FDcompact}) can be very efficiently implemented.

%\blue{[PIETRO: Do you also want to mention that with boundaries, our best performance was: 32.3 MLUPS???]}
%%%%%%%%%%%%%%%%%%%%%%%%%%%%%%%%%%%%%%%%%%%%%%%
%%%%%%%%%%COMPUTER SCIENCE PART%%%%%%%%%%%%%%%%
%%%%%%%%%%%%%%%%%%%%%%%%%%%%%%%%%%%%%%%%%%%%%%%

\subsection{\label{simple}Simple boundary conditions}

For sake of simplicity, numerical tests with simple boundary conditions are discussed first. Here, by simple boundary conditions, we mean either finite-difference boundary conditions (isothermal Couette flow) or periodic (generalized Green-Taylor vortex flow and Minion \& Brown flow). More general boundary conditions taking advantage of the LBM technology will be discussed in Section \ref{WBC}.

\subsubsection{\label{isocouette}Isothermal Couette flow}

In this section, we consider the plane Couette flow where a viscous fluid is confined in a gap between two parallel plates, with the one moving in its own plane with respect to the other. Here, two configurations are simulated by the present LW-ACM method on several meshes: Couette flow without wall injection (referred to as Test 1), and Couette flow with wall injection (referred to as Test 2). In the latter configuration (Test 2) fluid is injected from the bottom wall into the gap and extracted from the top wall with a constant orthogonal velocity $\bar{v}_0$. At the stationary condition, the above configurations admit the following exact solutions:
\begin{eqnarray}\label{Couette.exact}
 \bar{u}(y) = &\frac{1}{2}\hat{u}_L \left( 1 + \frac{y}{L} \right) + \bar{u}_0 \left( {1 - \frac{{y^2 }}{{L^2 }}} \right), \qquad &\text{(Test 1)}, \\ 
 \bar{u}(y) = &\bar{u}_L\left( {\frac{{\exp(y\,\mbox{Re} /L)  - 1}}{{\exp(\mbox{Re})  - 1}}} \right), \qquad  &\text{(Test 2)},
\end{eqnarray}
where the Reynolds number $\mbox{Re}$ is the main control parameter, $L$ is a characteristic length depending on the considered test and $\nu$ is the kinematic viscosity. 

For Test 1, $L$ is half the gap height, while $\bar{u}_0$ represents a velocity based on the imposed pressure gradient $\nabla \bar{p}$:
\begin{equation}\nonumber
\bar{u}_0=\frac{L^2 \nabla \bar{p}}{2 \rho \nu},
\end{equation}
$\bar{u}_L$ is the velocity of the top wall $\bar{u}_L=\bar{u}(y=L)$ and $\rho$ is the density. The bottom wall is assumed stationary: $\bar{u}(y=-L)=0$. 

In case of wall injection (Test 2), $L$ is the gap height, the Reynolds number $\mbox{Re}$ in (\ref{Couette.exact}) is defined on the basis of the injection velocity $\bar{v}_0$, namely $\mbox{Re}=\bar{v}_0 L/\nu$. Velocities at the top and bottom walls are: $\bar{u}(y=L)=\bar{u}_L$ and $\bar{u}(y=0)=0$, respectively. In all simulations, no-slip boundary conditions are applied along the wall by simply imposing the local equilibrium with the desired velocity, while periodic boundary conditions are adopted at the inlet and outlet.

%%%%%%%%%%%%%%%%
%% BOUNDARIES??????????
%%%%%%%%%%%%%%%%
%In this section, some numerical results are reported for Couette flow without (Test 1) and with (Test 2) wall injection in case of different scalings. {[ELIO: add a case test description; mention that boundary conditions are based on equilibrium only, which is exact for link-wise because the latter depends on equilibrium only] Details about this test case can be found in standard textbooks}. Some screen shots are reported in Figure \ref{couette-overview}.
%
%\begin{figure}[ht]
%\begin{center}
%\includegraphics[width=14cm]{couette-overview}
%\end{center}
%\caption{{[ELIO: eliminate this figure]} (Top) Couette flow with positive (left) and negative (right) pressure gradient (Test 1). (Bottom) Streamlines of the Couette flow with wall injection for Reynolds number equal to 1 (left) and 5 (right) respectively (Test 2).}
%\label{couette-overview}
%\end{figure}

\vspace{0.5cm}
\begin{table}[ht]
\caption{Convergence analysis for Couette flow without (Test 1) and with (Test 2) wall injection in case of both diffusive and acoustic scaling.}
\label{Couette-scaling}
\vspace{0.5cm}
\begin{center}
\begin{tabular}{ccccc}
\hline
\multicolumn{5}{c}{$\Delta t \propto \Delta x^2$ (diffusive scaling)} \\
  \hline
  & & & \multicolumn{2}{c}{Error $L^2[\bar{u}]$} \\
$\epsilon\equiv\Delta x$ & $\mbox{Ma}\propto \Delta t/\Delta x$ & $\nu\propto\mbox{Re}^{-1}$ & Test 1 & Test 2\\
  \hline
$1/10$ & $3.0\times 10^{-2}$ & $3.0\times 10^{-2}$ & $1.74\times10^{-3}$ & $4.59\times10^{-4}$\\
$1/20$ & $1.5\times 10^{-2}$ & $3.0\times 10^{-2}$ & $4.49\times10^{-4}$ & $1.21\times10^{-4}$\\
$1/40$ & $7.5\times 10^{-3}$ & $3.0\times 10^{-2}$ & $1.20\times10^{-4}$ & $3.11\times10^{-5}$\\
%%%%%%
\hline
\multicolumn{5}{c}{$\Delta t \propto \Delta x$ (acoustic scaling)} \\
 \hline
  & & & \multicolumn{2}{c}{Error $L^2[\bar{u}]$} \\
$\epsilon\equiv\Delta x$ & $\mbox{Ma}\propto \Delta t/\Delta x=\mbox{const.}$ & $\nu\propto\mbox{Re}^{-1}$ & Test 1 & Test 2\\
  \hline
$1/10$ & $3.0\times 10^{-1}$ & $3.0\times 10^{-3}$ & $4.27\times10^{-2}$ & $1.05\times10^{-2}$\\
$1/20$ & $3.0\times 10^{-1}$ & $1.5\times 10^{-3}$ & $2.76\times10^{-2}$ & $5.19\times10^{-3}$\\
$1/40$ & $3.0\times 10^{-1}$ & $7.5\times 10^{-4}$ & $1.54\times10^{-2}$ & $2.66\times10^{-3}$\\
%%%%%%
\end{tabular}
\end{center}
\end{table}

First of all, diffusive scaling is considered: This strategy consists in scaling the velocity field on different meshes, keeping fixed the relaxation frequency (see Appendix \ref{appA} for details). This scaling ensures second order convergence in the accuracy, as reported in upper part of Table \ref{Couette-scaling}, where the $L^2$ norm of deviation of numerical results from exact solution (\ref{Couette.exact}) is shown.
%
%{\begin{equation}\nonumber
%{\rm Error}=\frac{{\sqrt {\left( {U_{num}  - U_{exact} } \right)^2 } }}{{U_{top} }}
%\end{equation}}
%
%{is shown.} {[FABIO, PIETRO, please check if this is what you have done!!!]}
%
In addition, acoustic scaling is considered. This strategy consists in tuning the relaxation frequency on different meshes in order to keep constant the computed velocity field (see Appendix \ref{appA} for details). This scaling ensures first order convergence in the accuracy, as reported in the lower part of Table \ref{Couette-scaling}.
%
%\vspace{0.5cm}
%\begin{table}[ht]
%\caption{ Convergence analysis for Couette flow without (Test 1) and with (Test 2) wall injection in case of acoustic scaling.}
%\label{Couette-acoustic}
%\vspace{0.5cm}
%\begin{center}
%\begin{tabular}{ccccc}
%\hline
%\multicolumn{5}{c}{$\Delta t \propto \Delta x$ (acoustic scaling)} \\
% \hline
% & & & \multicolumn{2}{c}{{Error $L^2[\bar{u}]$}} \\
%$\epsilon\equiv\Delta x$ & $\mbox{Ma}\propto \Delta t/\Delta x=\mbox{const.}$ & $\nu\propto\mbox{Re}^{-1}$ & Test 1 & Test 2\\
%  \hline
%$1/10$ & $3.0\times 10^{-1}$ & $3.0\times 10^{-3}$ & \red{$4.27\times10^{-2}$} & \red{$1.05\times10^{-2}$}\\
%$1/20$ & $3.0\times 10^{-1}$ & $1.5\times 10^{-3}$ & \red{$2.76\times10^{-2}$} & \red{$5.19\times10^{-3}$}\\
%$1/40$ & $3.0\times 10^{-1}$ & $7.5\times 10^{-4}$ & \red{$1.54\times10^{-2}$} & \red{$2.66\times10^{-3}$}\\
%\end{tabular}
%\end{center}
%\end{table} 
%%%%%%%%
%%%%%%%%
%The numerical results reported in Table \ref{Couette-diffusive} and \ref{Couette-acoustic} are plotted in Figure \ref{couette} for making more evident the order of convergence. Moreover the numerical results obtained by standard Lattice Boltzmann method are reported as well for comparison.
%%%%%%%%
%%%%%%%%
%%%%%%%%
%\begin{figure}[ht]
%\begin{center}
%\includegraphics[width=14cm]{couette}
%\end{center}
%\caption{{[ELIO: improve this figure]} Normalized error analysis for Couette flow (left) and Couette flow with wall injection (right).}
%\label{couette}
%\end{figure}

\subsubsection{Generalized Green-Taylor vortex flow}

{In this section, some numerical results of the Taylor-Green vortex flow are reported.} The latter problem is widely employed as a {benchmark} for various incompressible Navier-Stokes solvers, {owing to the existence of a} simple analytical solution. {The original problem is characterized by the exponential decay in time due to} viscous dissipation. {However, here} the original problem is modified such that it becomes periodic in time by {the introduction of} a proper external acceleration $\bar{\mathbf{g}}=(\bar{g}_x,\bar{g}_y)^T$ to Eq. (\ref{AAu3}), where 
\begin{equation}\label{external-force}
\begin{split}
\bar{g}_x(t,x,y)&=\sin(x-\bar{u}_0 t)\cos(y-\bar{v}_0 t)(2\nu\cos t-\sin t),\\
\bar{g}_y(t,x,y)&=-\cos(x-\bar{u}_0 t)\sin(y-\bar{v}_0 t)(2\nu\cos t-\sin t),
\end{split}
\end{equation}
{with} $\bar{u}_0$ and $\bar{v}_0$ {being} constants {aiming at preventing that} the advection term balances with the pressure gradient. The modified problem admits the following analytical solution
\begin{equation}\label{exactsolution}
\begin{split}
\bar{u}(t,x,y)& = \bar{u}_0+\sin(x-\bar{u}_0 t)\cos(y-\bar{v}_0 t)\cos t,\\
\bar{v}(t,x,y)& = \bar{v}_0-\cos(x-\bar{u}_0 t)\sin(y-\bar{v}_0 t)\cos t,\\
\bar{p}(t,x,y)& = \frac{1}{4}[\cos 2(x-\bar{u}_0 t)+\cos 2(y-\bar{v}_0 t)]\cos^2 t.
\end{split}
\end{equation}
We solve the above modified problem numerically in the domain $\Omega=[0\leq x\leq 2\pi]\times[0\leq y\leq 2\pi]$ {using} periodic boundary condition, and $(\bar{u}_0,\bar{v}_0)=(0.3,0.6)$. The kinematic viscosity is $\nu=0.1$. Diffusive scaling is adopted for all simulations, namely $\mbox{Ma}=2\pi\,\mbox{Kn}$ or equivalently $\Delta t = \Delta x^2$ (see Appendix \ref{appA} for details).

\vspace{0.5cm}
\begin{table}[ht]
\caption{The $L^1$ norm of the error versus $\epsilon\equiv\Delta x\equiv\mbox{Ma}/2\pi$ at $t=60$ in the problem of the generalized Taylor-Green vortex problem for $\nu=0.1$.}
\label{gtg}
\vspace{0.5cm}
\begin{center}
\begin{tabular}{cccc}
  \hline
\multicolumn{4}{c}{Link-wise ACM} \\ 
  \hline
 $\epsilon\equiv\Delta x$ & Error $L^1[\bar{u}]$ & Error $L^1[\bar{v}]$ & Error $L^1[\bar{p}]$  \\
  \hline
$\pi/16$ & $1.69262\times 10^{-2}$ & $1.96356\times 10^{-2}$ & $3.45590\times10^{-2}$\\
$\pi/32$ & $4.38763\times 10^{-3}$ & $4.70793\times 10^{-3}$ & $7.91104\times10^{-3}$\\
$\pi/64$ & $1.41952\times 10^{-3}$ & $1.50330\times 10^{-3}$ & $2.21013\times10^{-3}$\\
  \hline\hline
\multicolumn{4}{c}{Link-wise ACM (semi-implicit, see Section \ref{INSE})} \\ 
  \hline
 $\epsilon\equiv\Delta x$ & Error $L^1[\bar{u}]$ & Error $L^1[\bar{v}]$ & Error $L^1[\bar{p}]$  \\
  \hline
$\pi/16$ & $1.70353\times 10^{-2}$ & $2.02203\times 10^{-2}$ & $3.09165\times10^{-2}$\\
$\pi/32$ & $4.41348\times 10^{-3}$ & $4.70412\times 10^{-3}$ & $6.11065\times10^{-3}$\\
$\pi/64$ & $1.41906\times 10^{-3}$ & $1.50227\times 10^{-3}$ & $1.73786\times10^{-3}$\\
  \hline\hline
\multicolumn{4}{c}{ACM} \\ 
  \hline
 $\epsilon\equiv\Delta x$ & Error $L^1[\bar{u}]$ & Error $L^1[\bar{v}]$ & Error $L^1[\bar{p}]$  \\
  \hline
$\pi/16$ & $8.03750\times 10^{-3}$ & $1.00313\times 10^{-2}$ & $9.70682\times10^{-3}$\\
$\pi/32$ & $1.92844\times 10^{-3}$ & $2.47186\times 10^{-3}$ & $1.93893\times10^{-3}$\\
$\pi/64$ & $5.64682\times 10^{-4}$ & $6.61285\times 10^{-4}$ & $4.34911\times10^{-4}$\\
  \hline\hline
\multicolumn{4}{c}{MRT-LBM} \\ 
  \hline
 $\epsilon\equiv\Delta x$ & Error $L^1[\bar{u}]$ & Error $L^1[\bar{v}]$ & Error $L^1[\bar{p}]$  \\
  \hline
$\pi/16$ & $7.67964\times 10^{-3}$ & $8.90307\times 10^{-3}$ & $1.67526\times10^{-2}$\\
$\pi/32$ & $1.84928\times 10^{-3}$ & $2.17164\times 10^{-3}$ & $3.63345\times10^{-3}$\\
$\pi/64$ & $6.12810\times 10^{-4}$ & $6.81060\times 10^{-4}$ & $1.11633\times10^{-3}$\\
\end{tabular}
\end{center}
\end{table}

The $L^1$ error data are {reported} in Table \ref{gtg} at $t=60$ for link-wise ACM, link-wise ACM with semi-implicit formulation (see the end of Section \ref{INSE} where this formulation is presented), standard (second-order) ACM and multiple-relaxation-time (MRT) LBM. The standard ACM is described in Ref. \cite{ohwada10}. In the MRT-LBM \cite{dHumieres92,lallemand00}, the consistent treatment of forcing is based on Ref.~\cite{Guo08}. The time step employed in the LBM computation is the same as {the one of} ACM, i.e. $\mbox{Ma}=2\pi\,\mbox{Kn}$ or equivalently $\Delta t=\epsilon^2$, and the tuning parameters of MRT are $s_1=s_4=s_6(={\tau_\rho}^{-1}={\tau_j}^{-1})=0$, $s_2(={\tau_e}^{-1})=1.63$, $s_3(={\tau_\epsilon}^{-1})=1.14$, $s_5=s_7(={\tau_q}^{-1})=1.92$ (see Refs.~\cite{lallemand00,Guo08}). All methods show nearly second-order convergence. Link-wise ACM shows larger numerical errors than ACM and MRT-LBM. However it must be stressed that the implementation of forcing in link-wise ACM is straightforward, while the consistent treatment of forcing in LBM is much more complicated \cite{Guo08}. Moreover in link-wise ACM, spatial operators (gradient and Laplace operator) do not need to be discretized individually {(unlike ACM)} but it is enough to discretize along the generic lattice direction: This makes much easier {the treatment of} three-dimensional cases.

\subsubsection{\label{minion}Minion \& Brown flow}
%Even though both link-wise ACM and LBM are methods based on artificially compressibility, {this} quantity is (quantitatively) different in the two cases, 
{LW-ACM and LBM are characterized by different values of artificial compressibility, leading to a different robustness with {respect} to under-resolved simulations. {By referring to the pseudo-continuity equation} for LW-ACM (see Eq. (\ref{AArho4}) in the Appendix \ref{appA}), it follows that accurate solution of divergence-free condition for the velocity field requires: $\epsilon^2/\nu\ll 1$. This is even a more severe condition than the one of LBM, in case of high Reynolds number flows. In the following, we further explore this issue, by means of} the Minion \& Brown flow.

Minion \& Brown \cite{Minion97} studied the performance of various numerical schemes {for} under-resolved simulations of the 2D incompressible Navier–Stokes equations. The relevance of this flow for testing robustness and accuracy of LBM schemes was first pointed out by Dellar \cite{Dellar01}. Minion \& Brown considered initial conditions {corresponding} to the perturbed shear layer
\begin{eqnarray}\label{initial}
\bar{u}(t,x,y) &=& \left\{
\begin{array}{lll}
&\tanh (k(y-1/4)),&\;\;\;y\leq 1/2,\\
&\tanh (k(3/4-y)),&\;\;\;y>1/2,\\
\end{array}\right.\\
\bar{v}(t,x,y) &=& \delta\sin(2\pi(x+1/4)),
\end{eqnarray}
in the periodic domain $\Omega=[0\leq x\leq 1]\times[0\leq y\leq 1]$. The parameter $k$ controls the shear layer width, while $\delta$ the magnitude of the initial perturbation. The shear layer is expected to roll up due to Kelvin-Helmholtz instability. With $k=80$, $\delta=0.05$, and Reynolds number $\mbox{Re} = 1/\nu = 10000$, the thinning shear layer between the two rolling up vortices becomes under-resolved on a $128 \times 128$ grid. Minion \& Brown \cite{Minion97} found that conventional numerical schemes using centered differences became unstable for this under-resolved flow, whereas ``robust'' or ``upwind'' schemes that actively suppress  grid-scale oscillations all produce two spurious secondary vortices at the thinnest points of the two shear layers at $t=1$. Dellar \cite{Dellar01} found that, even though it is stable on the previous mesh, two spurious vortices are generated by the BGK LBM scheme based on the D2Q9 lattice and with unmodified bulk viscosity. The same author proposed a way to increase the bulk viscosity for {overcoming} this problem, and {verified} that the same result can be achieved by using MRT-LBM.

Link-wise ACM is stable for the previous test, but the velocity field is not {accurate enough due to} numerical viscosity. {For $\epsilon=1/128$ and $\nu=1/10000$, indeed, the term that multiply} the pressure time derivative in the pseudo-continuity equation (Eq. (\ref{AArho4}) in the Appendix \ref{appA}), namely $\epsilon^2/\nu$, {reaches the value of} $0.61$, {thus preventing an accurate fulfillment of the diverge-free condition for the velocity field}. {In the following, two improvements are worked out for circumventing the above problem.}

\begin{figure}[ht]
\begin{center}
\includegraphics[width=10cm]{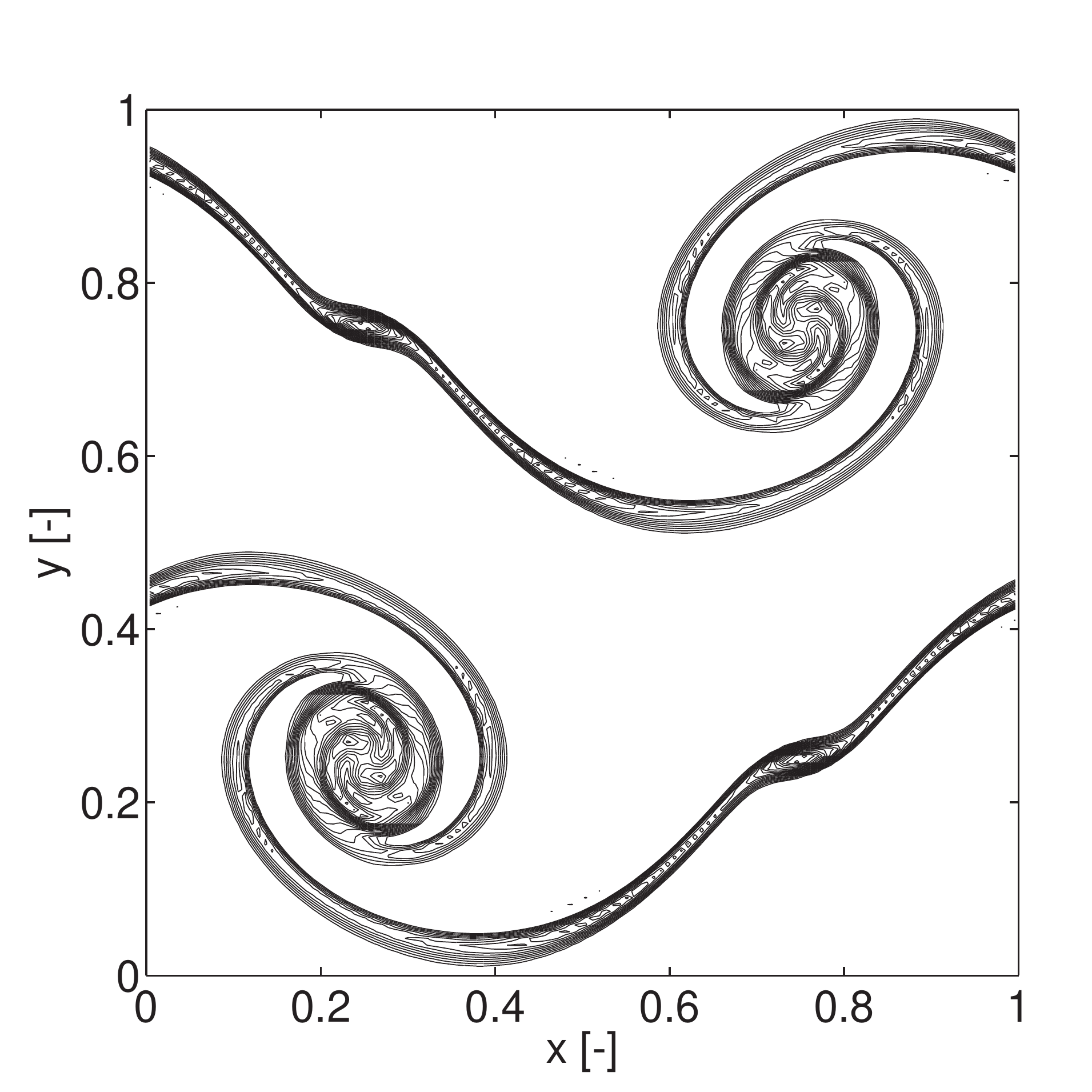}
\end{center}
  \caption{Contours of vorticity at $t = 1$ from the link-wise ACM with the first improvement ($\nu_p = 170\,\nu$, see Section \ref{INSE}) on a $128\times128$ grid with $\mbox{Ma} = 0.04$ and $\mbox{Re} = 10000$. {See also} Fig. 8 in \cite{Minion97}.}
\label{fig:minion1}
\end{figure}

{First, we apply the strategy discussed at the end of Section \ref{INSE.fd}, introducing a numerical fictitious viscosity $\nu_p = 170\,\nu$, with} the corresponding relaxation frequency $\omega_p$ computed by means of (\ref{viscosity2}). The relaxation frequency $\omega_p$ is used in the macroscopic updating rule (\ref{FDcompact}c). {Hence}, the pseudo-compressibility term in the continuity equation, namely the first term in Eq. (\ref{AArho4}), becomes proportional to $\epsilon^2/\nu_p=0.0036\ll\epsilon^2/\nu$ and this improves the quality of the solution. The vorticity at $t=1$ is reported in Figure \ref{fig:minion1}, where the velocity field looks much better, but the two spurious secondary vortices at the thinnest points of the two shear layers are still {present} (similarly to BGK-LBM).

The second improvement {follows the} idea to increase the bulk viscosity $\xi$, as suggested by Dellar \cite{Dellar02}. As discussed in Appendix \ref{appD}, instead of the standard $f_i^{(e)}$ (see Eq. (\ref{d2q9_equilibrium_compact}) in the Appendix \ref{appEQ}) in Eq. (\ref{ACM2.0}), {we} consider a modified set of functions, namely $f_i^{(e*)}=f_i^{(e*)}(\gamma)$, where $\gamma$ is a free parameter related to the bulk viscosity $\xi$: $\xi=2\rho_0\nu\,(1+2\,\gamma)$. This strategy {enables} to increase the bulk viscosity {and represents a valuable} example, {showing how to use moments of the updated distribution function for the local computation of derivatives} (see the Appendix \ref{appD} for details). {The above argument is a further confirmation that the present LW-ACM can easily incorporate technologies originally developed for LBM.}

As a concluding remark, it is worth to point out that, even though LW-ACM can solve this under-resolved test case by means of previously discussed improvements, the average and peak values of the velocity field divergence remain slightly larger than those computed by MRT-LBM. For sake of completeness, optimized ACM \cite{ohwada11} yields much smaller values of velocity field divergence than those of MRT-LBM. However, it is important to keep in mind that the Minion \& Brown flow is a very severe test due to the small initial perturbation $\delta$, which realizes a very sharp boundary layer. Hence this test is a multi-scale problem and some of the regularity assumptions used in deriving the numerical schemes may not hold completely.%}

\begin{figure}[ht]
\begin{center}
\includegraphics[width=10cm]{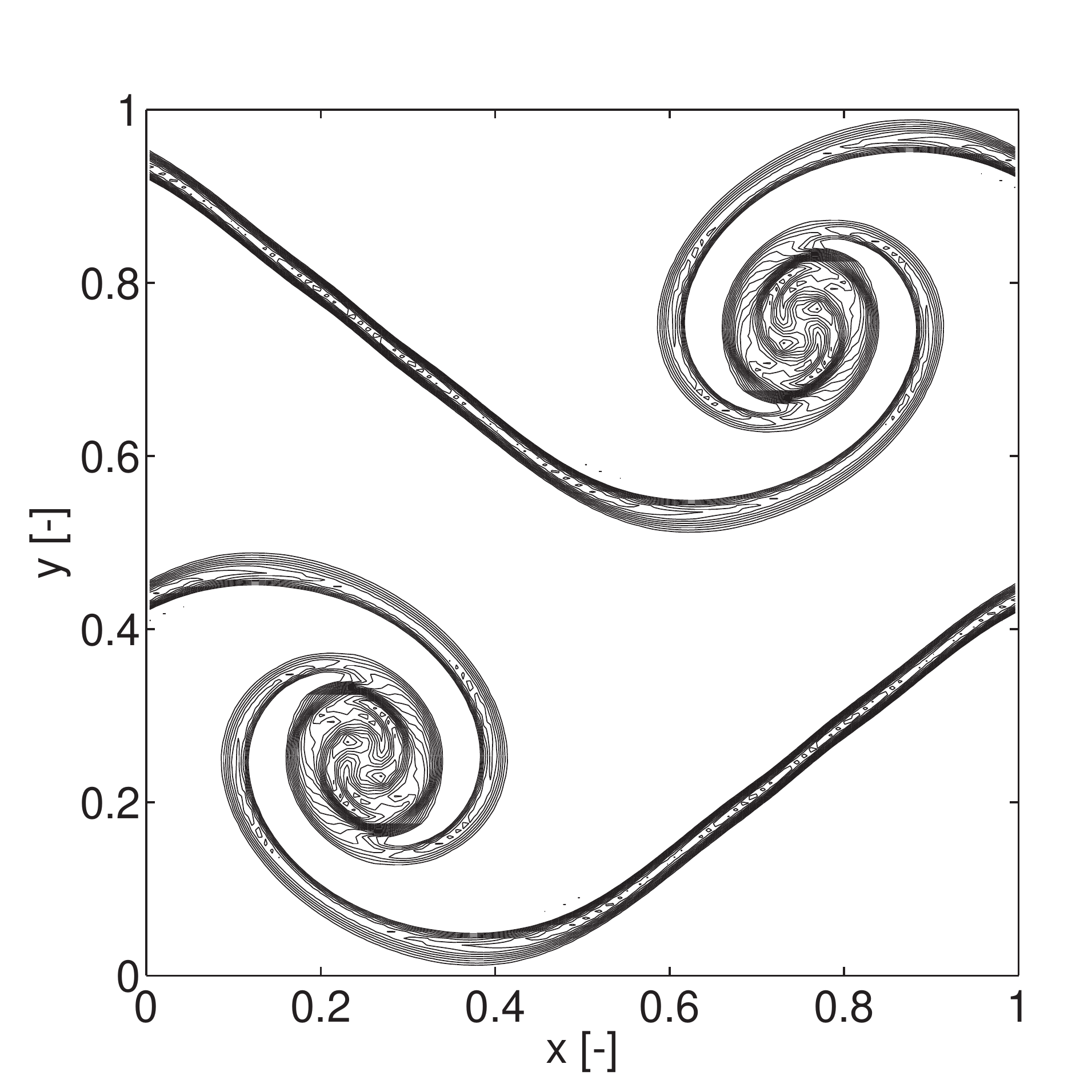}
\end{center}
  \caption{Contours of vorticity at $t = 1$ from the link-wise ACM with the both suggested improvements ($\nu_p = 170\,\nu$, see Section \ref{INSE}, and $\gamma=0.4$, see Section \ref{appD}) on a $128\times128$ grid with $\mbox{Ma} = 0.04$ and $\mbox{Re} = 10000$.}
\label{fig:minion2}
\end{figure}

\subsection{\label{WBC}Link-wise wall boundary conditions}
%One of the most remarkable features of {LW-ACM} consists in its link-wise formulation, i.e. along a single direction given by the $i$-th lattice velocity $\hat{\mathbf{v}}_i$. 
{The link-wise formulation of the proposed method offers significant advantages when dealing with wall boundary conditions. First, let us consider simple structured boundaries, where walls are aligned along the mesh (for general cases, the reader can refer to the next section).} In particular, let us suppose that walls are located halfway between two consecutive nodes in an ideal step-wise geometry. Instead of using Eqs. (\ref{ACM2.0b}), {it proves convenient focusing on the} quantities $f_i^*$ (after pre-combining) streaming out of a single point $\mathbf{x}$ (``push'' formulation)
\begin{equation}\label{ACM2.0cbb}
f_i^*(\hat{\mathbf{x}},\hat{t})=
f_i^{(e)}(\hat{\mathbf{x}},\hat{t})
-2\left(\frac{\omega-1}{\omega}\right)
f_i^{(e,o)}(\hat{\mathbf{x}},\hat{t}).
\end{equation}
Let us suppose that $\hat{\mathbf{x}}$ is a fluid node close to a complex wall boundary at rest such that $\hat{\mathbf{x}}+\hat{\mathbf{v}}_i$ is a wall node. In an ideal step-wise geometry, the wall location is assumed halfway between $\hat{\mathbf{x}}$ and $\hat{\mathbf{x}}+\hat{\mathbf{v}}_i$. Hence, during the streaming step, the information stored in the discrete distribution function pointing towards the wall are reflected along the same link by the wall (according to the {popular} {\em bounce-back} rule), namely 
\begin{equation}\label{ACM2.0ccbb}
f_{BB(i)}^{**}(\hat{\mathbf{x}},\hat{t}+1)=f_i^*(\hat{\mathbf{x}},\hat{t}),
\end{equation}
where $BB(i)$ is the bounce-back operator giving the lattice link opposite to $i$-th. Finally the post-combining step can be performed in the usual way, namely
\begin{equation}\label{ACM2.0cccbb}
f_{BB(i)}(\hat{\mathbf{x}},\hat{t}+1)=
f_{BB(i)}^{**}(\hat{\mathbf{x}},\hat{t}+1)
+2\left(\frac{\omega-1}{\omega}\right)\,f_{BB(i)}^{(e,o)}(\hat{\mathbf{x}},\hat{t}).
\end{equation}
{Considering that $\hat{\mathbf{v}}_i=-\hat{\mathbf{v}}_{BB(i)}$ and $f_i^{(e,o)}=-f_{BB(i)}^{(e,o)}$, the combination of previous steps yields}
\begin{equation}\label{ACM2.0cccbb2}
f_{BB(i)}(\hat{\mathbf{x}},\hat{t}+1)=
f_i^{(e)}(\hat{\mathbf{x}},\hat{t})
+\left(2-\frac{2}{\omega}\right)\,2 f_{BB(i)}^{(e,o)}(\hat{\mathbf{x}},\hat{t}).
\end{equation}

In case of moving complex boundary with velocity $\mathbf{u}_w$, the procedure reported in \cite{bouzidi01} (here reformulated in terms of $f_{BB(i)}^{(e,o)}$) suggests {the inclusion of} the additional term
\begin{equation}\label{ACM2.0movebb}
\delta f_{BB(i)}(\rho_0,\mathbf{u}_w)=2 f_{BB(i)}^{(e,o)}(\rho_0,\mathbf{u}_w),
\end{equation}
where $f_{BB(i)}^{(e,o)}$ is given by Eq. (\ref{feqo}) and $\rho_0$ is the averaged density over the whole computational domain (see Appendix \ref{appA} for details). For sake of simplicity, let us {consider} the diffusive scaling: $\Delta x=\epsilon\ll 1$, $\Delta t=\epsilon^2$. Hence, the incompressible limit implies $\rho(\hat{\mathbf{x}})=\rho_0+O(\epsilon^2)$. Moreover, the point $\hat{\mathbf{x}}$ is only $\epsilon/2$ away from the moving wall. Combining the previous conditions, the following approximation holds $\mathbf{u}(\hat{\mathbf{x}})=\mathbf{u}_w+O(\epsilon^2)$ and 
\begin{equation}\nonumber
f_{BB(i)}^{(e,o)}(\hat{\mathbf{x}},t)=f_{BB(i)}^{(e,o)}(\rho_0,\mathbf{u}_w)+O(\epsilon^2), 
\end{equation}
{meaning that the {rightmost} term of Eq. (\ref{ACM2.0cccbb2}) produces a similar effect to the correction (\ref{ACM2.0movebb}).} Hence, the suggested correction for LBM will be multiplied by a scaling factor {(complement to one of the factor multiplying the last term in Eq. (\ref{ACM2.0cccbb2}))} in link-wise ACM, namely
\begin{equation}\label{ACM2.0ccccbb}
f^w_{BB(i)}(\hat{\mathbf{x}},\hat{t}+1)=f_{BB(i)}(\hat{\mathbf{x}},\hat{t}+1)
+\left(\frac{2}{\omega}-1\right)\delta f_{BB(i)}(\rho_0,\mathbf{u}_w),
\end{equation}
where $f^w_{BB(i)}$ is the proper boundary condition in case of moving boundary.

Link-wise formulation is also very useful in computing hydrodynamical forces acting on bodies. A popular approach in LBM literature is based on the so-called momentum exchange algorithm (MEA), originally proposed in \cite{Ladd1994} and {lately} improved in \cite{Lorenz2009} (see \cite{Caiazzo2007} for a complete discussion). In LBM, at every time step, the momentum 
\begin{equation}\label{MEA}
\mathbf{p}_i = \hat{\mathbf{v}}_i f_i^{\text{post}}(\hat{\mathbf{x}},\hat{t})-\hat{\mathbf{v}}_{BB(i)} f_{BB(i)}(\hat{\mathbf{x}},\hat{t}+1) = \hat{\mathbf{v}}_i\left[f_i^{\text{post}}(\hat{\mathbf{x}},\hat{t})+f_{BB(i)}(\hat{\mathbf{x}},\hat{t}+1)\right],
\end{equation}
is transferred from the fluid to the solid body ($f_i^{\text{post}}$ is the post-collisional distribution function). For sake of simplicity, here we restrict ourselves on bodies at rest. In link-wise ACM, the quantity $f_i^{\text{post}}$ is purposely defined in order to get rid of the last term in the post-combining step, namely
\begin{equation}\label{MEA2}
f_i^{\text{post}}(\hat{\mathbf{x}},\hat{t}) = f_i^*(\hat{\mathbf{x}},\hat{t})-2\left(\frac{\omega-1}{\omega}\right)\,f_{BB(i)}^{(e,o)}(\hat{\mathbf{x}},\hat{t}).
\end{equation}
Introducing the previous definition in Eq. (\ref{MEA}) yields
\begin{equation}\label{MEA3}
\mathbf{p}_i = \hat{\mathbf{v}}_i\left[f_i^*(\hat{\mathbf{x}},\hat{t})+f_{BB(i)}^{**}(\hat{\mathbf{x}},\hat{t}+1)\right].
\end{equation}
The previous expression for link-wise ACM is general. In case of step-wise geometries (as those considered in this section), {the expression (\ref{ACM2.0ccbb}) holds and Eq. (\ref{MEA3}) can be recast as: $\mathbf{p}_i = 2 \hat{\mathbf{v}}_i f_i^*(\hat{\mathbf{x}},\hat{t})$.} The force exerted on a body is computed by a summation of the contributions (\ref{MEA3}) over all the boundary links {surrounding} its surface: 
\begin{equation}\label{MEA4}
\mathbf{\mathcal{F}}=\sum_{i\in \textsf{S}}\mathbf{p}_i,
\end{equation}
where $\textsf{S}$ is the set of links {starting from all surrounding nodes intersecting the body.} 

{Conversion of the force (\ref{MEA4})} from lattice units to physical units requires {substraction of} the hydrostatic {component} $\mathbf{\mathcal{F}}_0$ {generated by} the averaged density field $\rho_0$ (see also the Appendix \ref{appA} for details). Computing the hydrostatic force on {partial boundaries of a body by Eq. (\ref{MEA4} and \ref{MEA3}) can be accomplished after exclusion of the velocity-dependent components of $f_i^*$ and $f_{BB(i)}^{**}$.} If the force is computed on {the entire} body surface, the hydrostatic force is null. The remaining quantity $\mathbf{\mathcal{F}}-\mathbf{\mathcal{F}}_0$ scales as the second order tensor: in case of diffusive scaling $\mathbf{\mathcal{F}}-\mathbf{\mathcal{F}}_0\sim\epsilon^2$. However the number of points in the set $\textsf{S}$ increases proportionally to $1/\epsilon$. Consequently the following scaling holds {(see also Table \ref{tab:scaling})}
\begin{equation}\label{MEA6}
\Delta\bar{\mathbf{\mathcal{F}}} = (\mathbf{\mathcal{F}}-\mathbf{\mathcal{F}}_0)/\epsilon^2 = \left(-\mathbf{\mathcal{F}}_0/\epsilon+\sum_{i\in \textsf{S}(1/\epsilon)}\mathbf{p}_i\right)/\epsilon.
\end{equation}

\subsubsection{2D lid driven cavity flow}
%
%\red{[PIETRO] We should mention the following paper: Li-Shi Luo, Wei Liao, Xingwang Chen, Yan Peng and Wei Zhang, Numerics of the lattice Boltzmann method: Effects of collision models on the lattice Boltzmann simulations, PHYSICAL REVIEW E 83, 056710 (2011); Maybe in the table?}

In this section, the effective enhanced stability of the present LW-ACM method is tested by means of the classical two-dimensional (2D) lid driven cavity problem (see also \cite{Bordeaux2006,Sahin03}). Such a benchmark has been considered owing to a singularity of the pressure in the lid corners, which makes it a severe test for robustness of numerical schemes, especially starting from moderately high Reynolds numbers. In all simulations, we consider a square domain $(x,y)\in[0,L]\times[0,L]$, with $L=1$. %Moreover the simulation time is $t\in [0,t_0]$, where $t_0=100$, which is considered enough to reach the steady state conditions for the considered Reynolds numbers (see next). 
Such a domain is discretized by uniform collocated grid with $N\times N$ points. The boundaries are located half-cell away from the computational nodes. Let us denote $\hat{\mathbf{x}}_b$ the generic boundary computational node. In all inner nodes ($\hat{\mathbf{x}}\neq\hat{\mathbf{x}}_b$),  Eq. (\ref{ACM2.0}) holds for any lattice velocity $\hat{\mathbf{v}}_i$. In an arbitrary boundary node $\hat{\mathbf{x}}_b$, Eq. (\ref{ACM2.0}) holds for any lattice velocity $\hat{\mathbf{v}}_i$ such that $\hat{\mathbf{x}}_b+\hat{\mathbf{v}}_i$ is still a computational node. In case $\hat{\mathbf{x}}_b+\hat{\mathbf{v}}_i$ falls out of the computational grid, the boundary condition (\ref{ACM2.0ccccbb}) is applied, with $\mathbf{u}_w$ being the boundary velocity (imposed half-cell away from the boundary node $\hat{\mathbf{x}}_b$). In the following numerical simulations, $\mathbf{u}_w=(u_L,0)^T$ at the lid wall, where $u_L$ is the lid velocity, and $\mathbf{u}_w=\bm{0}$ for all the other walls. At the lid corners, the lid velocity is imposed, while for other corners the boundary conditions (\ref{ACM2.0ccccbb}) are adopted.

%%%%%%%%%Elio%%%%%%%%%%%%%%%%%%
\begin{figure}[ht]
\begin{center}
\includegraphics[trim=32mm 10mm 40mm 8mm,clip,width=1\textwidth]{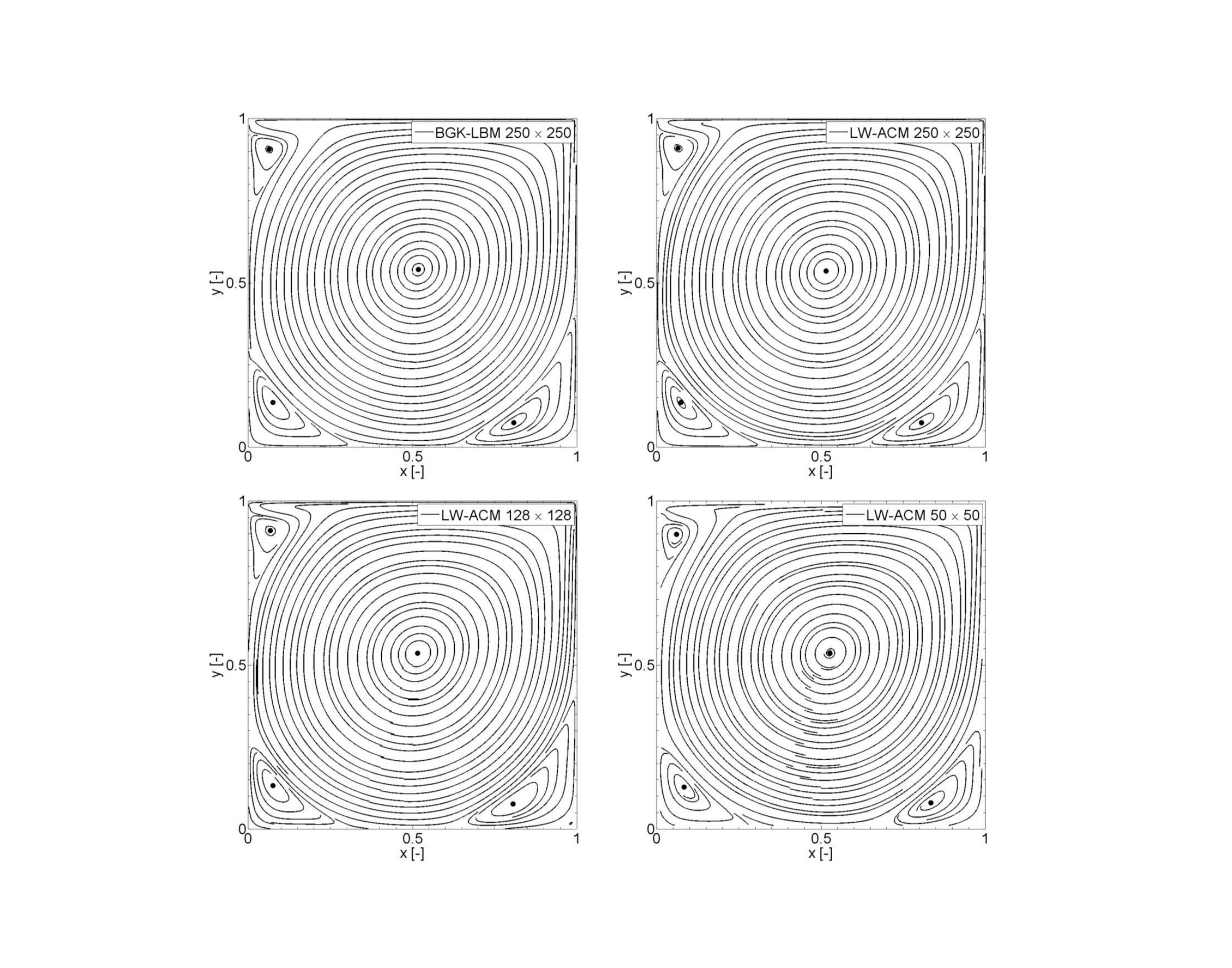}
\end{center}
  \caption{Streamlines for the lid driven cavity flow with $\mbox{Re}=5000$. Different numerical methods and different meshes are compared. The location of the four minima of the stream-function is denoted by filled circles.}
\label{several_streamlines_vel}
\end{figure}
%
%\begin{figure}[ht]
%\begin{center}
%\includegraphics[width=10cm]{BGK-250x250_LWACM-050x050}
%\end{center}
%  \caption{Streamlines for the lid driven cavity flow with $\mbox{Re}=5000$ and $t_0=100$. Different numerical methods and different meshes are compared.}
%\label{fig:lid-50}
%\end{figure}
%%%%%%%Elio%%%%%%%%%%%%%%%%%%%%%
%
\begin{figure}[ht]
\centering
	   \includegraphics[width=\textwidth]{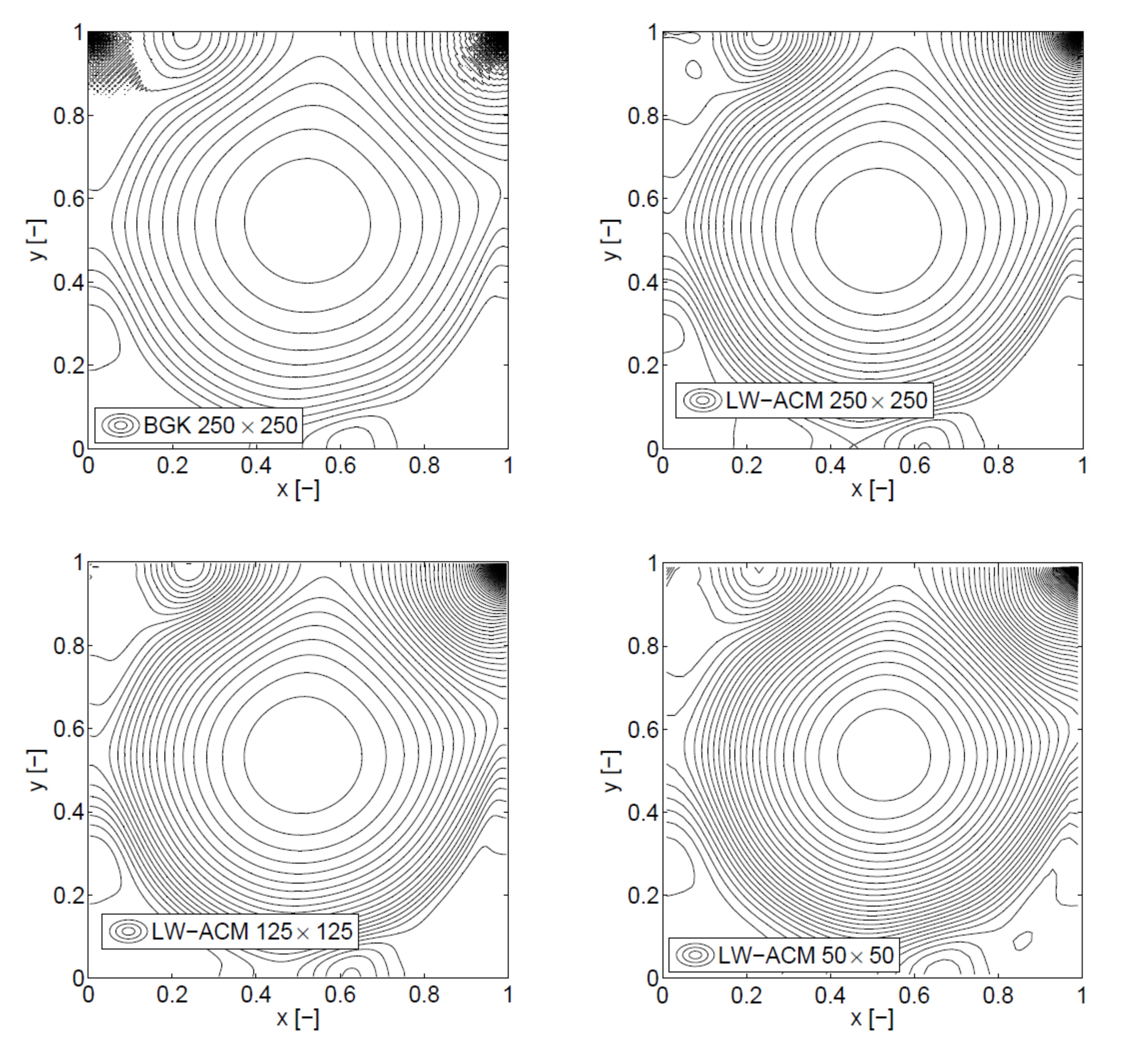}       
\caption{Pressure contours for the lid driven cavity flow with $\mbox{Re}=5000$. Different numerical methods and different meshes are compared.}
\label{several_streamlines_P}
\end{figure}
In Figs. \ref{several_streamlines_vel} and \ref{several_streamlines_P}, numerical results corresponding to several grids and methods are reported for Reynolds number $\mbox{Re}=u_L L/ \nu = 5000$. In this case, it is known that the flow is characterized by four main vortexes, whose actual {\em centers} can be found by searching for local extrema of the stream function $\psi$ defined as:
\begin{equation}
\label{streamf}
u=\frac{\partial\psi}{\partial y},\qquad
v=-\frac{\partial\psi}{\partial x},
\end{equation}
with $u$ and $v$ being the horizontal and vertical component of the velocity field, respectively. 

%%%%%%%%%%%%%Elio
\begin{figure}[ht]
  \centering
  %\begin{tabular}{cc}
     \includegraphics[width=\textwidth]{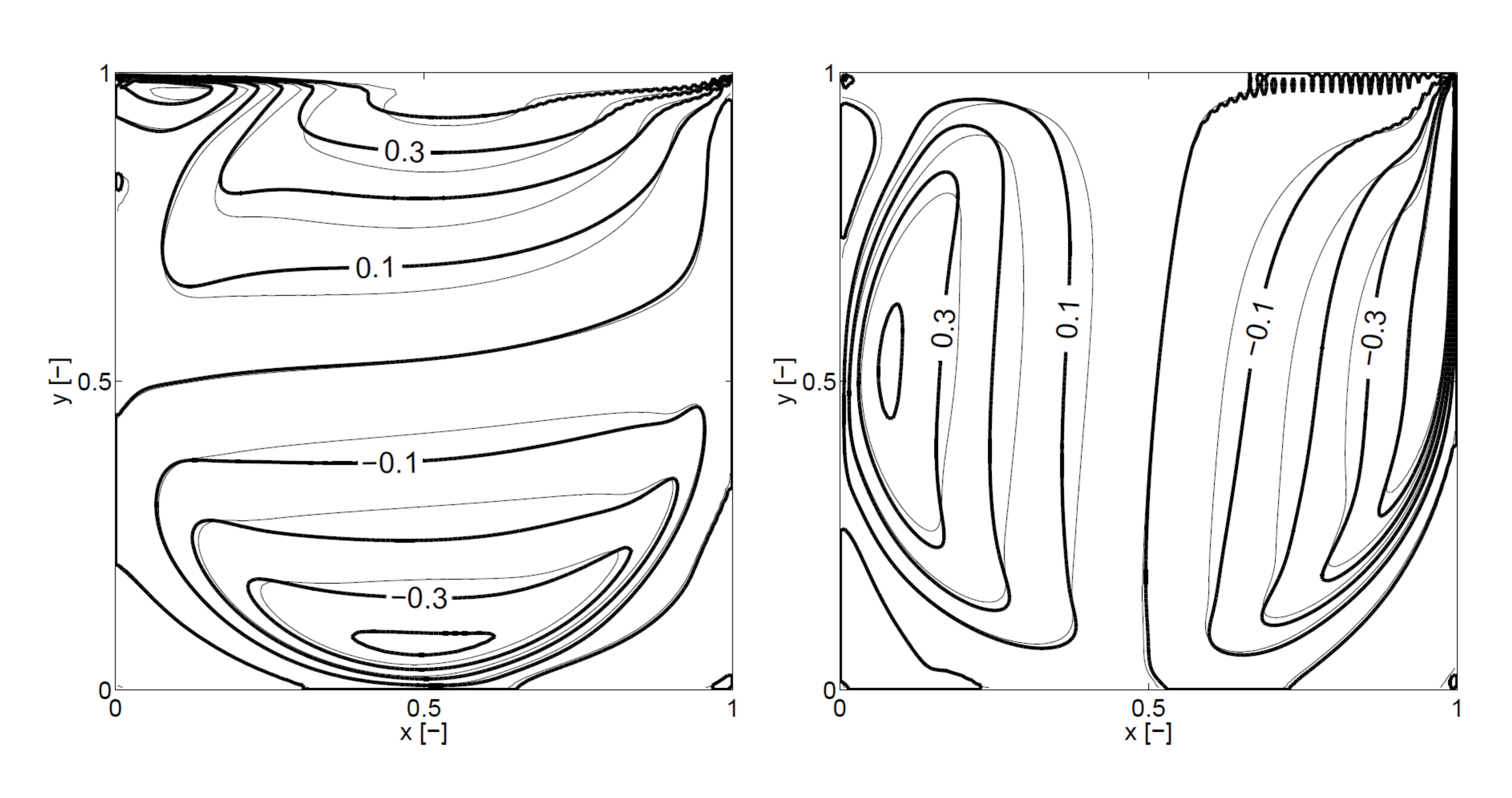} 
	%\end{tabular}
\caption{Comparison of the velocity field for the lid driven cavity flow at $\mbox{Re}=5000$ with $128 \times 128$ grids: horizontal component $u$ and vertical component $v$ are reported on the left and right hand side, respectively. Thin and bold lines denote the present LW-ACM (with $\nu_p=\nu$) and optimized ACM solution \cite{ohwada10}, respectively. The latter method is based on a second order accurate scheme in time, and fourth order accurate scheme in space (bulk fluid).}\label{lwacmAcmCOMvel.128}
\end{figure}
%%%%%%%%%%%%%Elio
%%%%%%%%%%%%%
%%%%%%%%%%%%%Elio
\begin{figure}[ht]
  \centering
  %\begin{tabular}{cc}
     \includegraphics[width=0.7\textwidth]{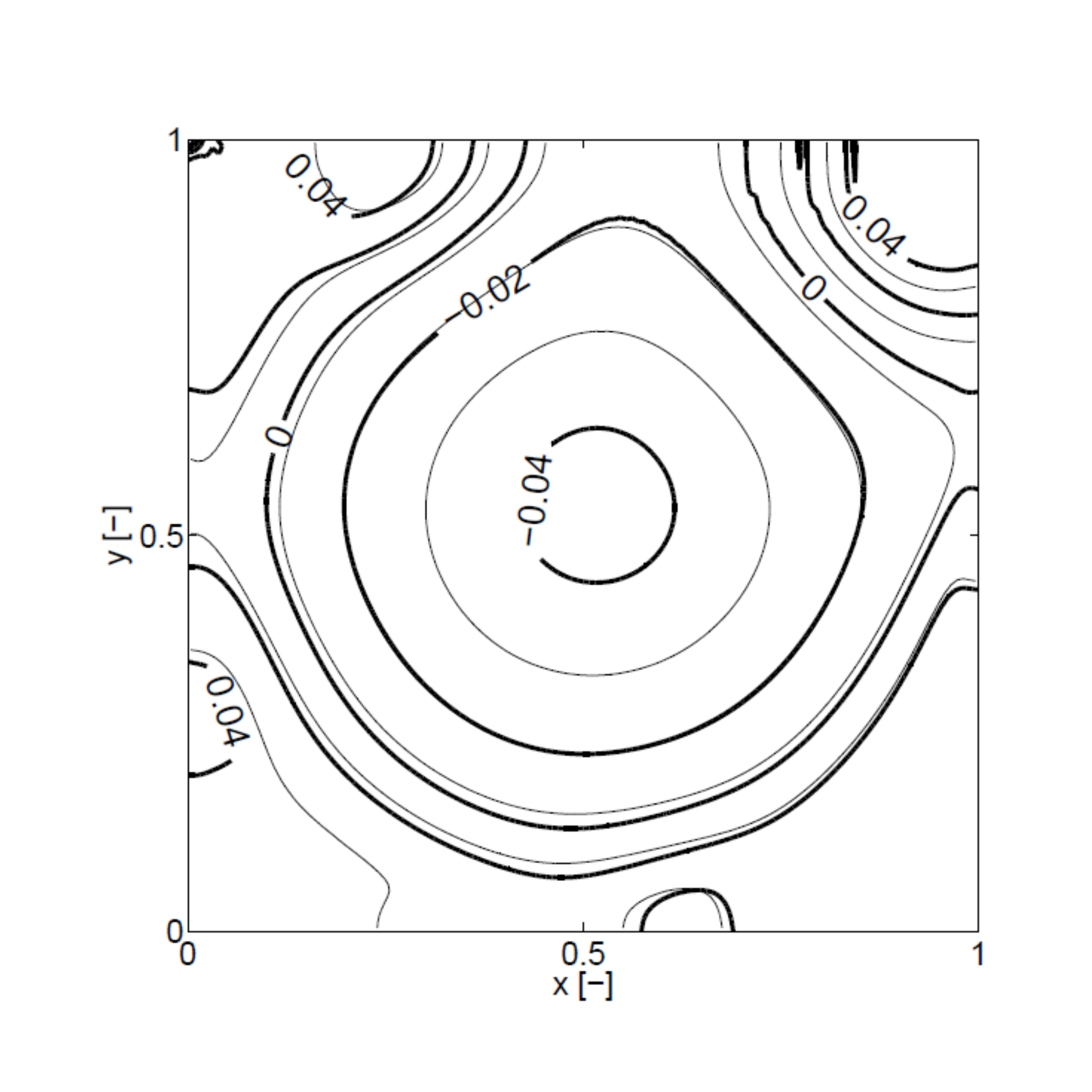} 
	%\end{tabular}
\caption{Comparison of the pressure field for the lid driven cavity flow at $\mbox{Re}=5000$ with $128 \times 128$ grids. Thin and bold lines denote the present LW-ACM (with $\nu_p=\nu$) and optimized ACM solution \cite{ohwada10}, respectively. The latter method is based on a second order accurate scheme in time, and fourth order accurate scheme in space (bulk fluid).}\label{lwacmAcmCOMP.128}
\end{figure}
%%%%%%%%%%%%%Elio
%%%%%%%%%%%%%

For standard Lattice Boltzmann method \cite{qian92} with BGK collisional operator and $D2Q9$ lattice, the coarsest grid which ensures numerical stability was found to be $250\times 250$. On the other hand, the present LW-ACM method can be safely adopted with $125\times 125$ grid, and reasonably accurate results are found as reported in Figs. \ref{several_streamlines_vel} and \ref{several_streamlines_P}. In addition, LW-ACM shows stability even with $50\times 50$ grid ($1/25$ the total number of nodes needed by BGK-LBM method), where LW-ACM is still able to describe the main features of the 2D cavity flow at $\mbox{Re}= 5000$.

In Fig. \ref{several_streamlines_P} the pressure contours for the same test are reported. As visible in the upper-left part of Fig. \ref{several_streamlines_P} the BGK solution shows some {\em checkerboard} pressure distribution at the top corners of the cavity. The mesh resolution is still enough to overcome the checkerboard instability mechanism: however this comes at the price of a very large computational domain (larger than $250 \times 250$). On the other hand, no such a problem was noticed with LW-ACM, even for quite coarse grids (down to $50 \times 50$). The absence of spurious oscillations in the numerical solutions by the artificial compressibility method (ACM) for this test case has been already pointed out \cite{ohwada10}. Hence, the previous numerical evidences demonstrate that also the present link-wise formulation of ACM inherits the same feature.

It is worth stressing that numerical stability on coarse grids, yet with poor accuracy, is a highly desirable feature in several engineering problems, e.g. where a loose grid resolution of details of no interest (for the overall flow phenomena) should not prevent global convergence of the code.

%\begin{figure}[ht]
%\centering
%  \begin{tabular}{cc}
%	   \includegraphics[width=0.5\textwidth]{ACM-LWACM-u-250x250}  &
%       \includegraphics[width=0.5\textwidth]{ACM-LWACM-v-250x250} \\
%	\end{tabular}
%\caption{Comparison of flow field in the problem of the lid-driven cavity flow ($\mbox{Re}=5000$): $u$ field (left) and $v$ field (right). The black line indicate the ACM solution %\cite{ohwada10} and the red lines indicate the solution obtained by LW-ACM.}
%\label{fig:acm-lwacm}
%\end{figure}
%%%%%%%%%%%%%Elio
\begin{figure}[ht]
  \centering
  %\begin{tabular}{cc}
     \includegraphics[width=\textwidth]{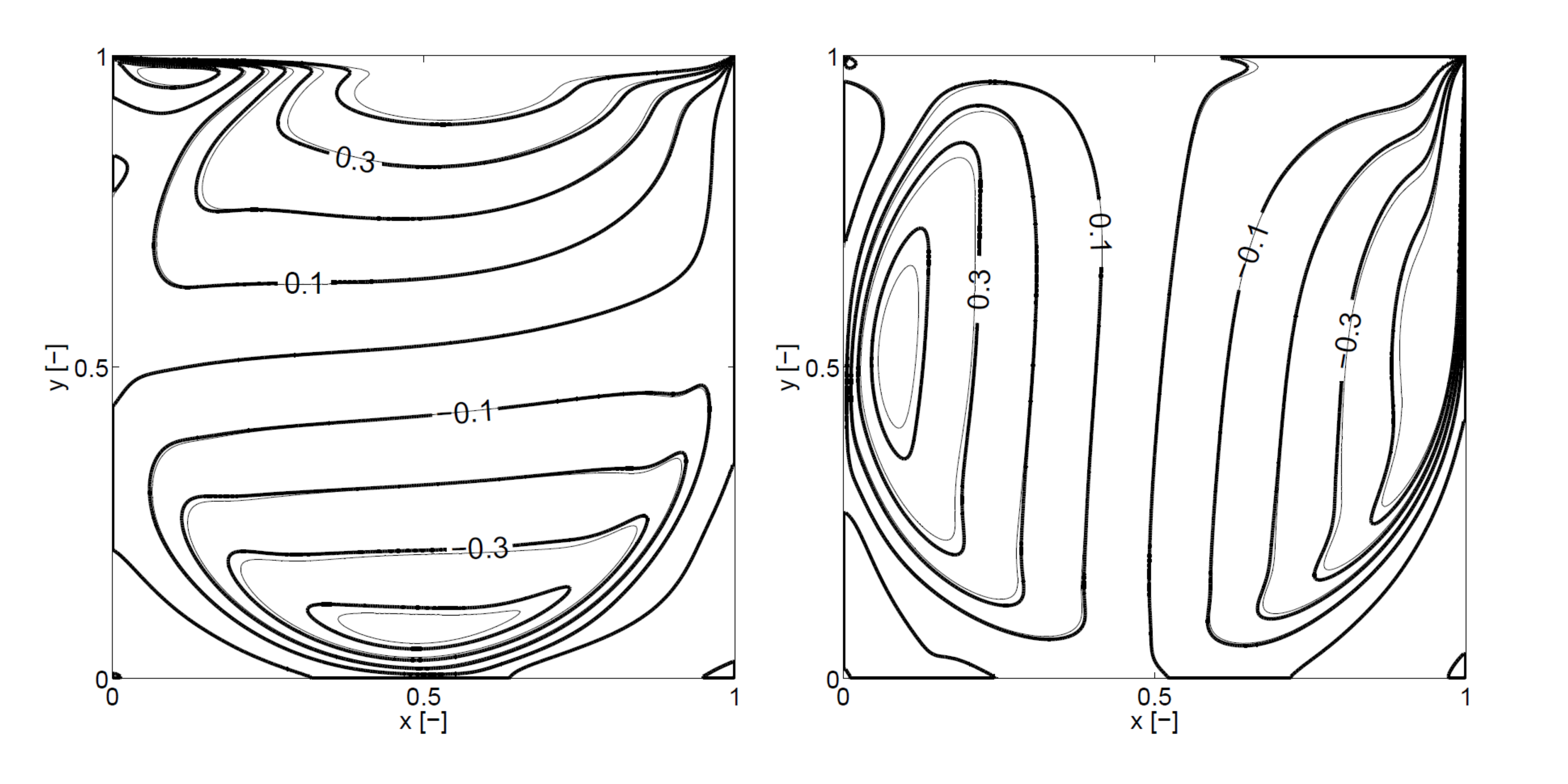} 
	%\end{tabular}
\caption{Comparison of the velocity field for the lid driven cavity flow at $\mbox{Re}=5000$ with $256 \times 256$ grids: horizontal component $u$ and vertical component $v$ are reported on the left and right hand side, respectively. Thin and bold lines denote the present LW-ACM (with $\nu_p=\nu$) and ACM solution \cite{ohwada10}, respectively. The latter method is based on a second order accurate scheme in time, and fourth order accurate scheme in space.}\label{lwacmAcmCOMvel}
\end{figure}
%%%%%%%%%%%%%Elio
%%%%%%%%%%%%%
%%%%%%%%%%%%%Elio
\begin{figure}[ht]
  \centering
  %\begin{tabular}{cc}
     \includegraphics[width=0.65\textwidth]{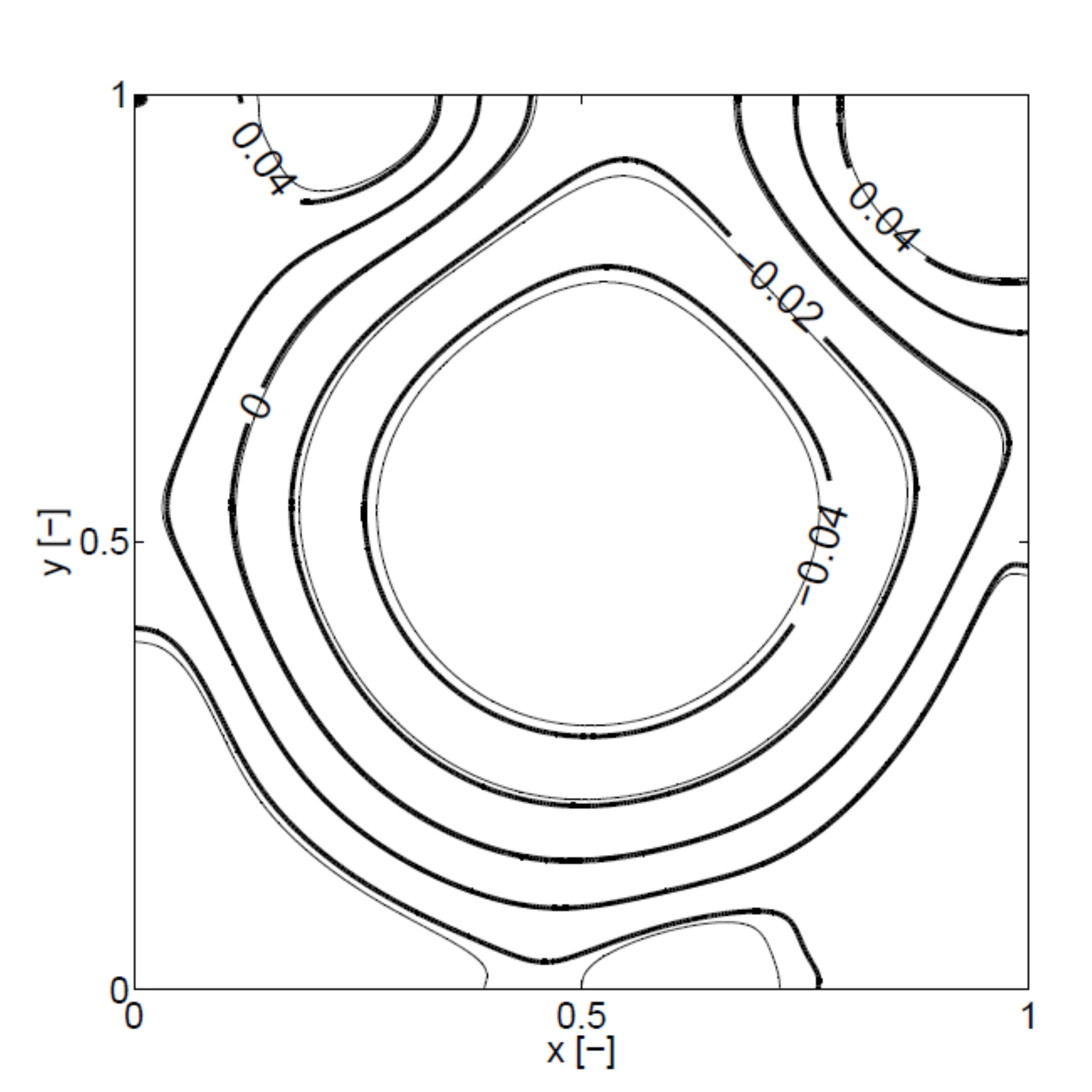} 
	%\end{tabular}
\caption{Comparison of the pressure field for the lid driven cavity flow at $\mbox{Re}=5000$ with $256 \times 256$ grids. Thin and bold lines denote the present LW-ACM (with $\nu_p=\nu$) and ACM solution \cite{ohwada10}, respectively. The latter method is based on a second order accurate scheme in time, and fourth order accurate scheme in space.}\label{lwacmAcmCOMP}
\end{figure}
%%%%%%%%%%%%%Elio
%%%%%%%%%%%%%

{Clearly the numerical schemes should be compared in terms of the actual accuracy as well. From the very beginning, it is worth to point out that all considered methods (i.e. BGK-LBM, MRT-LBM, ACM, LW-ACM) are based on artificial compressibility and even steady state solutions depend on the numerical Mach number (in particular, the pressure gradients depend on the Mach number, as well as the number of time steps). In particular, reducing the numerical Mach number improves the quality of the results. See Appendix \ref{appA} for details. Hence a fair comparison among different methods requires using the same numerical Mach number.}

The flow fields of a 2D lid-driven cavity problem with $\mbox{Re}=5000$ and $128 \times 128$ grid, as predicted by the optimized ACM method \cite{ohwada10} and the present LW-ACM (with Mach number $\mbox{Ma}=0.2$ and $\nu_p=\nu$), have been compared. Here optimized ACM means that (a) high wave numbers are damped for the suppression of the checkerboard instability and (b) the Richardson extrapolation in the Mach number (except around top singular corners) is employed \cite{ohwada10}. As reported in Figs. \ref{lwacmAcmCOMvel.128} and \ref{lwacmAcmCOMP.128}, LW-ACM shows both a smoother and more accurate behavior (see also Table \ref{LW-ACMvsBordeauxTab}), beside a remarkably simpler implementation. 

%\blue{[PIETRO,TAKU: Maybe you want to add a few lines describing the optimized/no-optimized ACM!!!]} 

Moreover, in Figs. \ref{lwacmAcmCOMvel} and \ref{lwacmAcmCOMP}, we report the flow fields of a 2D lid-driven cavity problem with $\mbox{Re}=5000$ and $256 \times 256$ grid as predicted by the optimized ACM \cite{ohwada10} and the present LW-ACM (with Mach number $\mbox{Ma}=0.2$ and $\nu_p=\nu$), where a small mismatch between the two solutions is observed.  This time the optimized ACM method is able to reproduce the reference results \cite{Bordeaux2006} with excellent accuracy (despite the use of a much coarser grid: $256^2$ vs. $2048^2$), although minor pressure oscillations are still visible at the lid corners and this affects, e.g., the prediction of entrophy (see Eq. (\ref{global.quantities}) and Table \ref{LW-ACMvsBordeauxTab}). Differences between ACM and LW-ACM are mainly due to: 1)different accuracy of the two schemes (second and first order accuracy in time for ACM and the present LW-ACM respectively, whereas fourth and second order accuracy in space for ACM and the present LW-ACM respectively), and 2) slightly different treatment of boundaries. ACM imposes boundary conditions on the computational nodes, while in LW-ACM, analogously to LBM, the wall boundary conditions are imposed half cell away from the computational node. This allows one to avoid singularities, which appear in the top corners for this test case. Despite all this, it is fair to say that the agreement between the above two solutions is quite good.

%\blue{[ELIO: LW-ACM vs. Bordeaux]}
%%%%%%%%%%%%%Elio
\begin{figure}[ht]
  \centering
%  \begin{tabular}{cc}
     \includegraphics[width=0.95\textwidth]{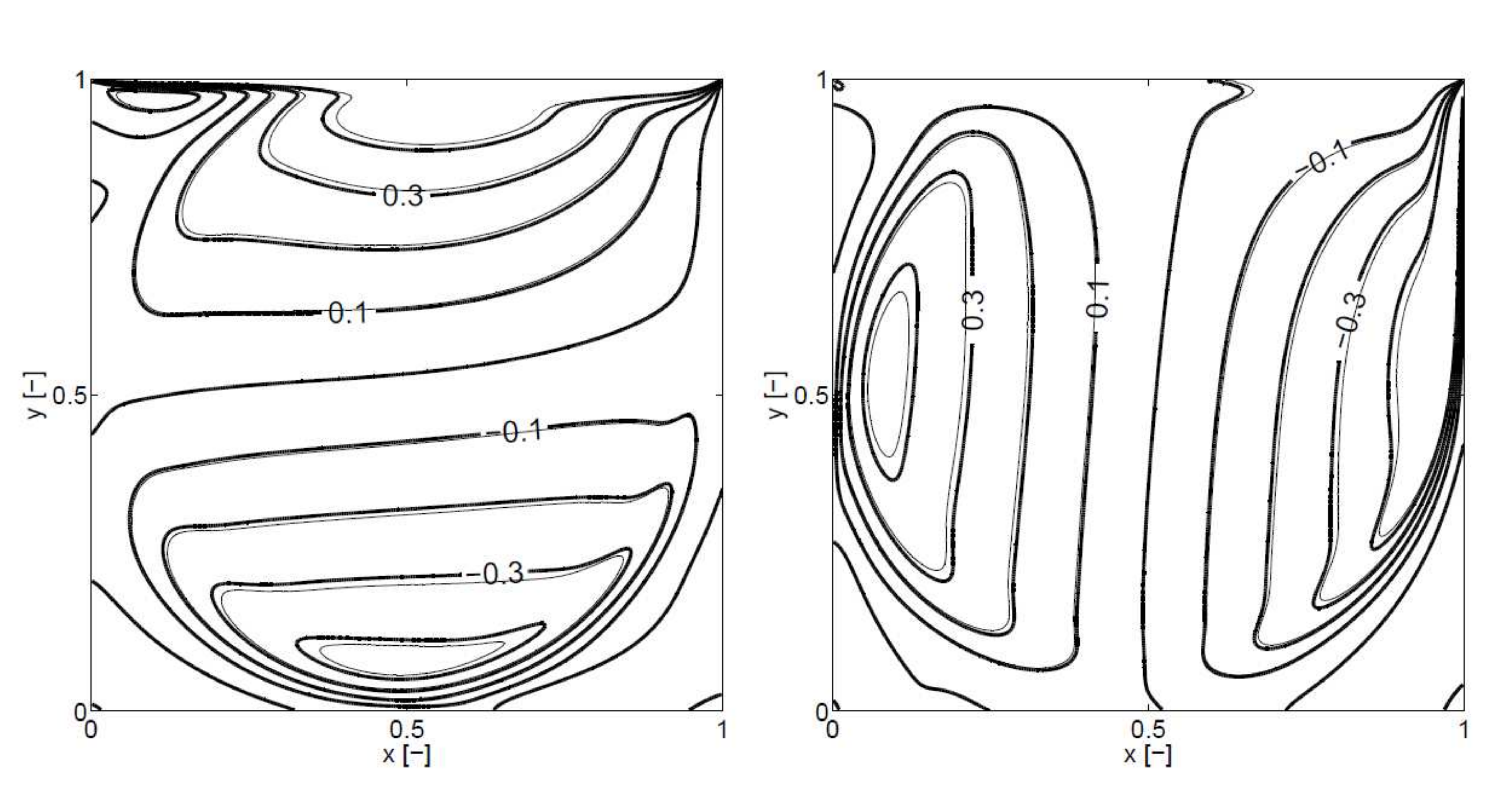} 
%	\end{tabular}
\caption{Comparison of the velocity field for the lid driven cavity flow at $\mbox{Re}=5000$: horizontal component $u$ and vertical component $v$ are reported on the left and right hand side, respectively. Thin and bold lines denote the present LW-ACM ($256 \times 256$ grid, with $\nu_p=\nu$) and the reference solution \cite{Bordeaux2006} ($2048 \times 2048$ grid), respectively.}\label{lwacmBorCOMvel}
\end{figure}
%%%%%%%%%%%%%Elio
%%%%%%%%%%%%%
%%%%%%%%%%%%%Elio
\begin{figure}[ht]
  \centering
%  \begin{tabular}{cc}
     \includegraphics[width=0.65\textwidth]{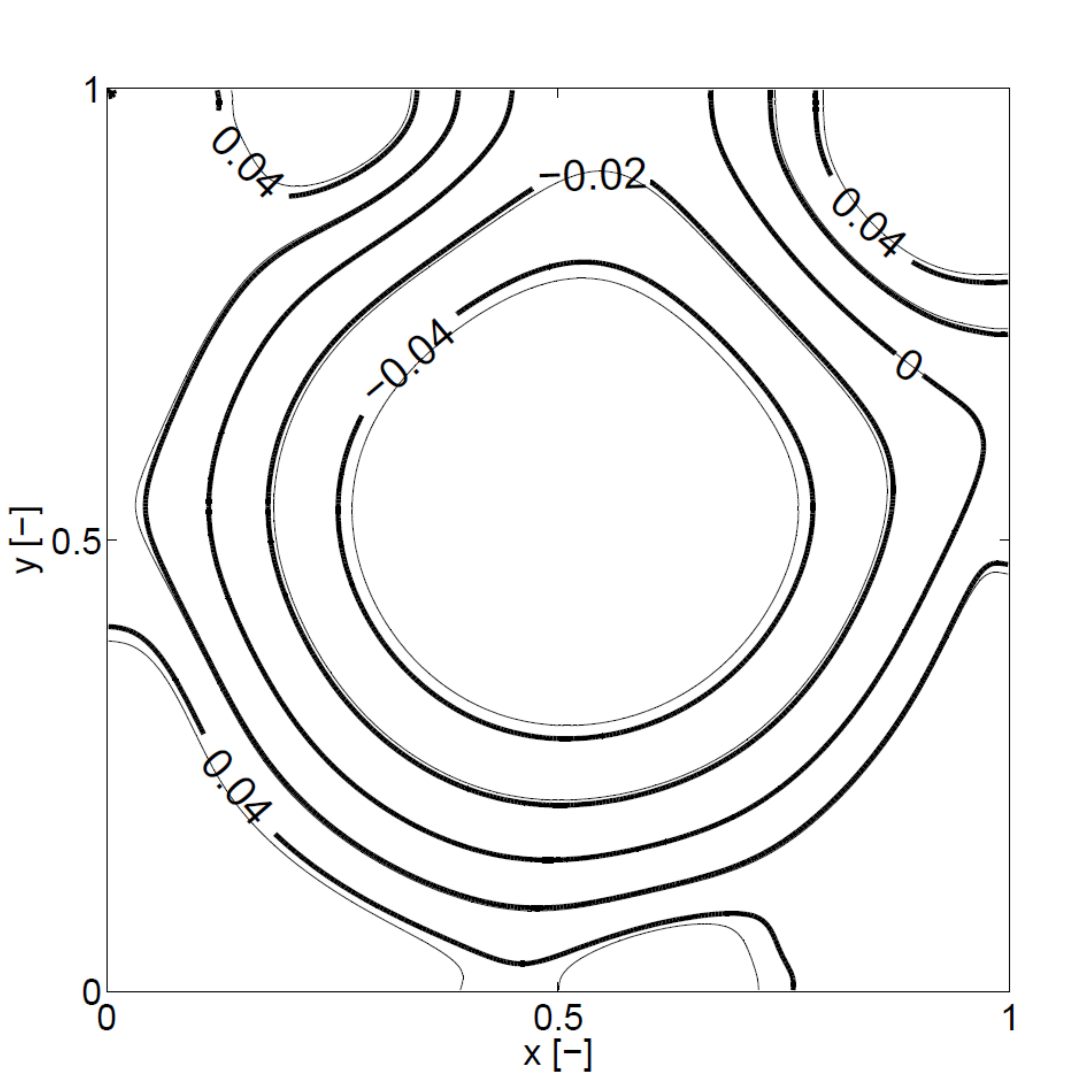} 
%	\end{tabular}
\caption{Comparison of the pressure field for the lid driven cavity flow at $\mbox{Re}=5000$. Thin and bold lines denote the present LW-ACM ($256 \times 256$ grid) and the reference solution \cite{Bordeaux2006} ($2048 \times 2048$ grid), respectively.}\label{lwacmBorCOMP}
\end{figure}
%%%%%%%%%%%%%Elio
%%%%%%%%%%%%%

In Figs. \ref{lwacmBorCOMvel} and \ref{lwacmBorCOMP}, the flow field computed by the LW-ACM with $\mbox{Re}=5000$ and $256 \times 256$ grid is compared to a reference solution from the literature \cite{Bordeaux2006}, where a state of the art code based on finite differences is used with a remarkably fine grid ($2048 \times 2048$). Despite a significant disparity in the number of computational nodes (LW-ACM makes use of $1/64$ the nodes adopted for the reference solution), an excellent agreement is found. It is worth stressing that, the above problem was also simulated by the multiple relaxation time lattice Boltzmann method (MRT-LBM) with $256 \times 256$ grid. Comparison between MRT-LBM and the reference solution is also very good, however the issue of spurious pressure oscillations in the upper-left corner of the cavity could not be avoided. Further comparisons between the LW-ACM and other methods are proposed in Table \ref{LW-ACMvsBordeauxTab}. Here, coordinates of the primary vortex center ($x_p,y_p$), coordinates of the lower-right vortex center ($x_{lr},y_{lr}$), total kinetic energy $E$ and enstrophy $Z$ are reported for several schemes and grids, where the latter two quantities are computed as follows:
%%%%%%%%%%%%%
\begin{equation}\label{global.quantities}
\begin{split}
E = \frac{1}{2}\int_\Omega \left\| \mathbf{u}\right\|^2 d \Omega \approx \frac{1}{2} \Delta x \Delta y \sum_{i,j}\left( u_{i,j}^2+v_{i,j}^2\right), \\ \; Z=\frac{1}{2}\int_\Omega \left\| \omega \right\|^2 d \Omega \approx \frac{1}{2} \Delta x \Delta y \sum_{i,j} \omega_{i,j}^2,
\end{split}
\end{equation}
%%%%%%%%%%%%%%%%%%%%%%%%%%%%%%%%%%%%%
%
with $\omega=\partial_x v - \partial_y u$, $\Delta x$ and $\Delta y$ being the vorticity and the grid spacings respectively. In our study, we notice that one consequence of spurious pressure oscillations in the solution of classical lattice Boltzmann schemes is that both BGK-LBM and MRT-LBM show remarkable inaccuracy in recovering the enstrophy value predicted by the reference \cite{Bordeaux2006}, whereas the present LW-ACM overcomes the above issue. 

Finally, based on Figs. \ref{several_streamlines_vel}, \ref{several_streamlines_P}, \ref{lwacmAcmCOMvel.128}, \ref{lwacmAcmCOMP.128}, \ref{lwacmAcmCOMvel}, \ref{lwacmAcmCOMP}, \ref{lwacmBorCOMvel}, \ref{lwacmBorCOMP}, on comparisons of local and global quantities proposed in Table \ref{LW-ACMvsBordeauxTab}, we can conclude that LW-ACM represents an excellent alternative in terms of simplicity, stability and accuracy. 

%\blue{[PIETRO: Do you want to spend a few words on the importance of the Mach number on the accuracy of the solution???]}

%\blue{$^1$This simulation was performed by the present LW-ACM method with Mach number $Ma=0.04$ and fictitious viscosity $\nu_p=170 \nu$ (see section \ref{minion})}.
\vspace{0.5cm}
\begin{table}[ht]
\caption{2D lid driven cavity flow at $\mbox{Re}=5000$: Comparison between the present LW-ACM (with Mach number $\mbox{Ma}=0.2$, $\nu=\nu_p$) and alternative solvers for INSE from literature \cite{ohwada10,Bordeaux2006,Succi2001,dHumieres2002,Luo2011pre}. {In artificial compressibility methods (LW-ACM, BGK-LBM, MRT-LBM), even steady state solutions depend on the numerical Mach number: Hence a fair comparison among different methods requires using the same numerical Mach number ($\mbox{Ma}=0.2$ in this case).} $^\dagger$This is the optimized version of the ACM method proposed in \cite{ohwada10} where (a) high wave numbers are damped for the suppression of the checkerboard instability and (b) the Richardson extrapolation in the Mach number (except around top singular corners) is employed. %a special treatment for spurious pressure oscillations is implemented, while second and fourth order accuracy in time and space is ensured in the bulk fluid, respectively \blue{[PIETRO, TAKU: Maybe you wish to add a few comments on that! E.G., What about accuracy at the boundaries???]}. 
$^*$Owing to both the accuracy of the scheme and the size of meshes adopted, these results are considered as a reference for the present study. However, since enstrophy $Z$ for 2D lid-driven cavity goes to infinity as the grid spacing goes to zero \cite{Bordeaux2006}, a meaningful comparison for $Z$ is among similar grids.}
\label{LW-ACMvsBordeauxTab}
\vspace{0.5cm}
\begin{center}
\begin{tabular}{cccccc}
%\hline
%\multicolumn{5}{c}{2D lid driven cavity flow at $\mbox{Re}=5000$}\\
\hline
Scheme & Grid & ($x_p,y_p$) & ($x_{lr},y_{lr}$) & Energy & Enstrophy\\
\hline
Present & $128 \times 128$ & ($0.51652,0.53754$) & ($0.81081,0.079079$) & $0.039845$ & $29.247$\\
%\blue{Present$^1$} & $128 \times 128$ & ($0.51652,0.53353$) & ($0.80981,0.074074$) & $0.043016$ & $30.452$\\
\cite{ohwada10} & $128 \times 128$ & ($0.52052,0.53954$) & ($0.82883,0.071071$) & $0.027430$ & $41.249$\\
\cite{ohwada10}$^\dagger$ & $128 \times 128$ & ($0.51652,0.53854$) & ($0.80981,0.072072$) & $0.038371$ & $37.704$\\
BGK-LBM & $128 \times 128$ & unstable & unstable & unstable & unstable\\
MRT-LBM & $128 \times 128$ & ($0.51652,0.53554$) & ($0.80881,0.075075$) & $0.043600$ & $37.404$\\
\cite{Bordeaux2006}$^*$ & $128 \times 128$ & ($0.51562,0.53906$) & ($0.80469,0.070313$) & $0.043566$ & $30.861$\\
\hline
Present & $256 \times 256$ & ($0.51552,0.53554$) & ($0.80581,0.074074$) & $0.044391$ & $34.821$\\
\cite{ohwada10}& $256 \times 256$ & ($0.51652,0.53654$) & ($0.80881,0.072072$) & $0.040896$ & $43.198$\\
\cite{ohwada10}$^\dagger$ & $256 \times 256$ & ($0.51451,0.53654$) & ($0.80380,0.072072$) & $0.048114$ & $42.290$\\
BGK-LBM & $250 \times 250$ & ($0.51752,0.54054$) & ($0.80781,0.074074$) & $0.041614$ & $40.455$\\
MRT-LBM & $256 \times 256$ & ($0.51552,0.53554$) & ($0.80681,0.074074$) & $0.045222$ & $40.833$\\
\cite{Bordeaux2006}$^*$ & $256 \times 256$  & ($0.51562,0.53516$) & ($0.80859,0.074219$) & $0.046204$ & $34.368$\\
\hline
\cite{Bordeaux2006}$^*$ & $2048 \times 2048$ & ($0.51465,0.53516$) & ($0.80566,0.073242$) & $0.047290$ & $40.261$\\
\end{tabular}
\end{center}
\end{table}
%%%%%%%%%%%%%%%%%%%%%%%%%%%%%%%%%%%%%
%%%%%%%%%%%%%%%%%%%%%%%%%%%%%%%%%%%%%
%\blue{[ELIO: Periodic solutions to be added!!!]}
%%%%%%%%%%%%%%%%%%%%%%%%%%%%%%%%%%%%%
%%%%%%%%%%%%%%%%%%%%%%%%%%%%%%%%%%%%%
%\red{[PIETRO: some comments about both Table \ref{LW-ACMvsBordeauxTab} and Figure \ref{fig:LWpreiodic8100} must be added.]}

{\em Remark-}In this study, towards the end of making an extensive comparison among state of the art INSE solvers, simulations are performed by different methods and grids as reported in Table \ref{LW-ACMvsBordeauxTab}. Since boundaries may be located differently for different methods (e.g. unlike ACM \cite{ohwada10} where boundaries coincide with computational nodes, in LW-ACM boundaries are half cell away from computational nodes), upon convergence, all the fields are first interpolated (by cubic spline interpolation) on a shifted grid (same size as the one used for fluid dynamic computations) having the boundaries located on the computational nodes. The values of global kinetic energy and enstrophy given by Eq. (\ref{global.quantities}) are based on the latter shifted grids. 

Moreover, if one performs the calculation of streamlines and vortex locations on the basis of the same nodes of the fluid dynamic grid, the final accuracy will depends on both the accuracy of the numerical solution and the grid itself. Hence, results on coarse grids are penalized twice. Therefore, towards the end of computing the coordinates ($x_p,y_p$) and ($x_{lr},y_{lr}$), all the hydrodynamic fields (as computed by the several schemes and different meshes) are first interpolated (by cubic spline interpolation) on a larger mesh, and thereafter the stream function and the vortex locations are computed. The latter post-processing procedure is composed of the following subsequent steps: 1) interpolation of the results on fixed fine grid ($1000^2$ in Table \ref{LW-ACMvsBordeauxTab}); 2) computation of the stream function and its local extrema denoting the vortex centers.

\subsubsection{3D diagonally driven cavity flow}
%%%%%%%%%%%
\begin{figure}[ht]
  \centering
  %\begin{tabular}{cc}
     \includegraphics[width=0.9\textwidth]{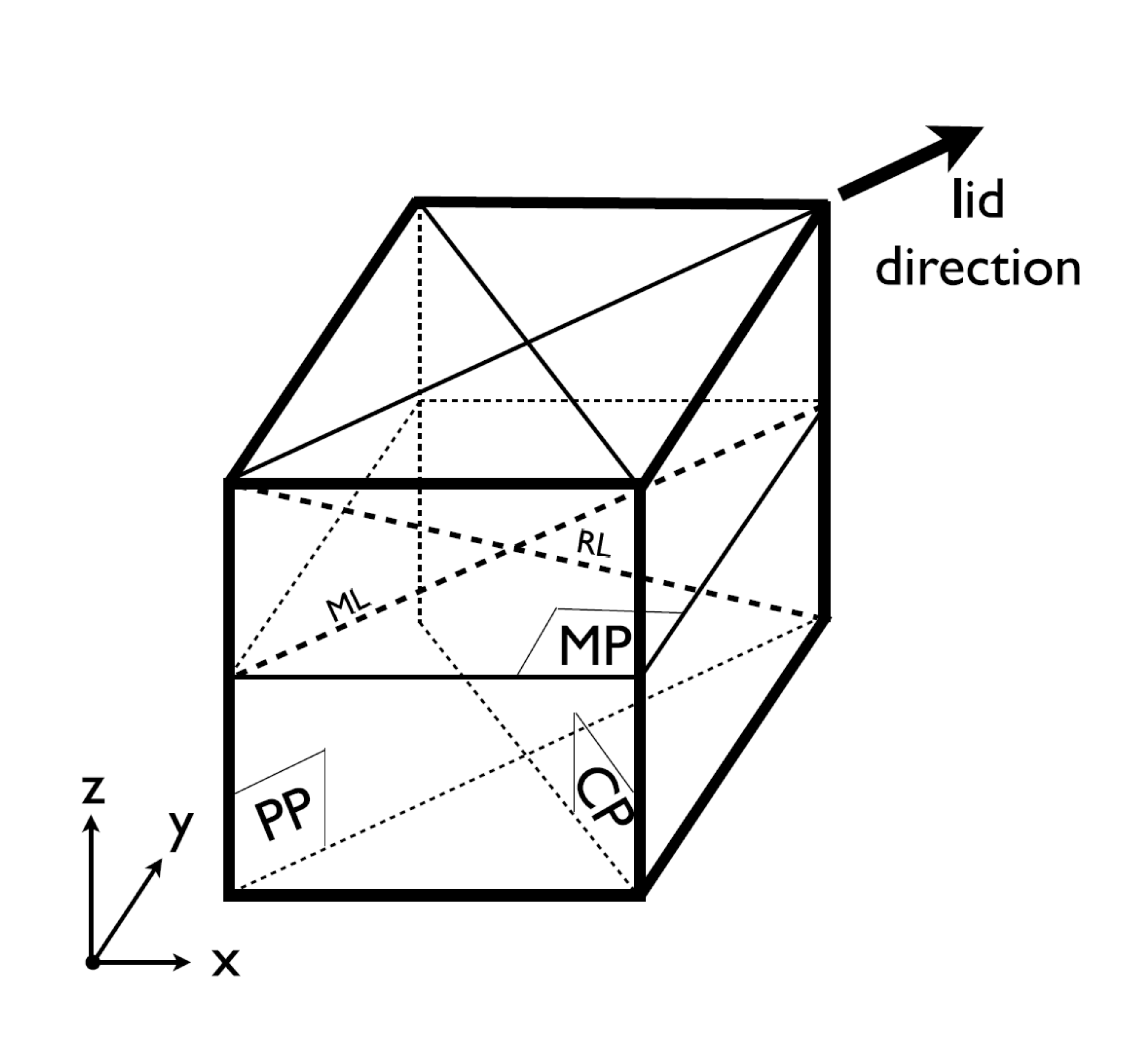} 
	%\end{tabular}
\caption{Cavity flow with the lid moving along its diagonal. Velocity components along axes $x$, $y$ and $z$ are denoted $u$, $v$ and $w$ respectively.}\label{3dlidcartoon}
\end{figure}
%%%%%%%%%%%
One of the main advantage of the proposed link-wise formulation of ACM consists in its independence on the space dimensionality (as far as the considered equilibrium satisfies the constraints required by the target equations: see Appendix \ref{appA} for details). As a result, the extension of LW-ACM to three-dimensional flows is straightforward. In the following calculations, the D3Q19 lattice \cite{qian92} will be used: even though that is not a Hermitian lattice (such as D3Q27), the former lattice allows to satisfy the constraints required by the Navier-Stokes equations (in particular Eq. (\ref{third_def})).

Here, we have chosen the three-dimensional (3D) diagonally lid-driven cavity flow, which is a classical benchmark for numerical solvers of the incompressible Navier-Stokes equations (see also \cite{Povitsky2001,dHumieres2002,Povitsky2005}). The cavity is a cubic box with unit edge as schematically sketched in Fig. \ref{3dlidcartoon}. The boundary condition at the top plane $(x,y,1)$ is $\mathbf{u}_L=(\sqrt{2},\sqrt{2},0)/20$ so that $u_L=\left\|\mathbf{u}_L\right\|=1/10$, whereas the remaining five walls are subject to no-slip boundary conditions. The computational domain is discretized by a uniform collocated grid with $N^3$ nodes, with boundaries located half-cell away from the computational nodes. Towards the end of making a comparison with data from literature, calculations have been performed by the LW-ACM at two Reynolds numbers studied in \cite{Povitsky2001,Povitsky2005} and \cite{dHumieres2002} ($\mbox{Re}=700$, $\mbox{Re}=2000$), and two grids: $N=48$ and $N=60$. Let us denote $\hat{\mathbf{x}}_w$ the generic boundary computational node. In all inner computational nodes ($\hat{\mathbf{x}}\neq\hat{\mathbf{x}}_w$), Eq. (\ref{ACM2.0}) holds for any lattice velocity $\hat{\mathbf{v}}_i$.

In this test case, numerical stability is significantly affected by boundary conditions. For that reason, in Ref. \cite{dHumieres2002}, Authors suggest to use equilibrium-based boundary conditions for the sliding wall at relatively high Reynolds numbers on small computational grids. Nevertheless, as pointed out \cite{dHumieres2002}, this implementation imposes an incorrect constant pressure at the boundary, with the momentum transfer significantly weakened in the direction perpendicular to the lid. Moreover, in Ref. \cite{dHumieres2002}, the ``node'' bounce-back boundary conditions are applied to the remaining five walls for imposing no-slip boundary conditions. Although such an approach reduces oscillations caused by the parity invariance and thus enhances the numerical stability, the several simplifications discussed above were necessary to simulate the 3D cavity flow with $\mbox{Re}=2000$, $D3Q15$ lattice and $52^3$ grid.

On the contrary, the present LW-ACM method does not need to resort to the above simplifications any longer. At an arbitrary boundary node $\hat{\mathbf{x}}_w$ Eq. (\ref{ACM2.0ccccbb}) holds, with $\mathbf{u}_w$ being the boundary velocity (imposed half-cell away from the boundary computational node $\hat{\mathbf{x}}_w$). This increases the accuracy in treating the boundaries (compared to \cite{dHumieres2002}) and, most importantly, it makes problems in three-dimensions just a straightforward extension of the ones in two-dimensions (see previous section).
%
%%%%%%%%%%%%%%%Elio
%\begin{figure}[ht]
%\centering
%  \begin{tabular}{cc}
%	   \includegraphics[width=0.5\textwidth]{LID3D_ref}  &
%       \includegraphics[width=0.5\textwidth]{LID3D_ACM} \\
%	\end{tabular}
%\caption{Velocity vectors for the diagonally driven cavity flow for $\mbox{Re}=2000$ at the mid-plane ($z=0.5$): reference solution obtained by commercial code FLUENT %\cite{Povitsky2001} (left) and present LW-ACM (right).}
%\label{fig:lid3d-flow}
%\end{figure}
%%%%%%%%%%%%%%%Elio

\begin{figure}[ht]
\centering
	     \includegraphics[width=1\textwidth]{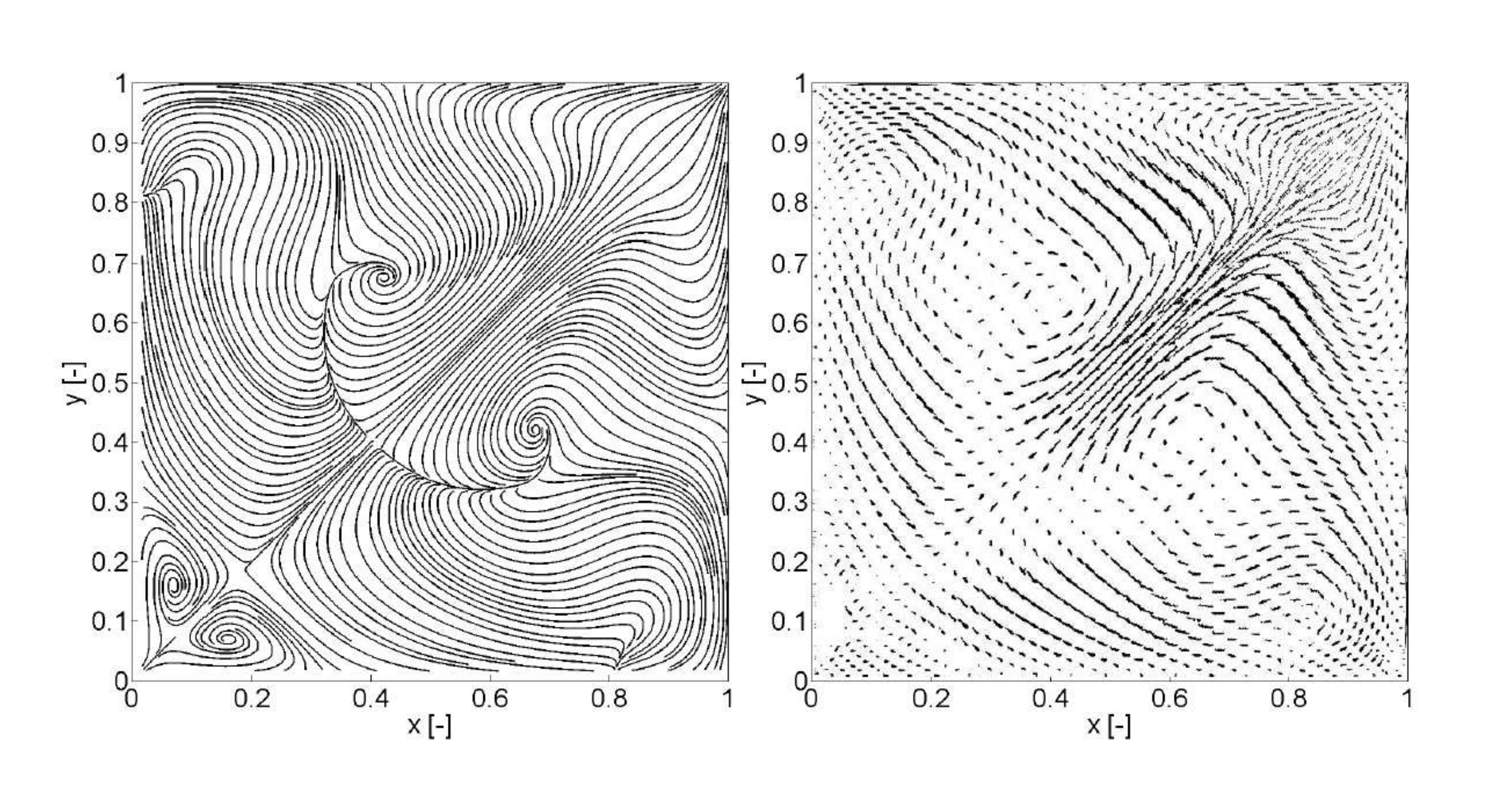}
\caption{Flow in the middle plane $z=0.5$ of the 3D diagonally driven cavity (MP plane in Fig. \ref{3dlidcartoon}) at $\mbox{Re}=700$. Comparison between the LW-ACM with $60^3$ grid (left) and a reference solution \cite{Povitsky2001} obtained by the commercial code FLUENT (right) with $68^3$ total number of grid nodes.\label{comparison_3d_midflow}}
\end{figure}

\begin{figure}[ht]
\centering
	     \includegraphics[width=0.7\textwidth]{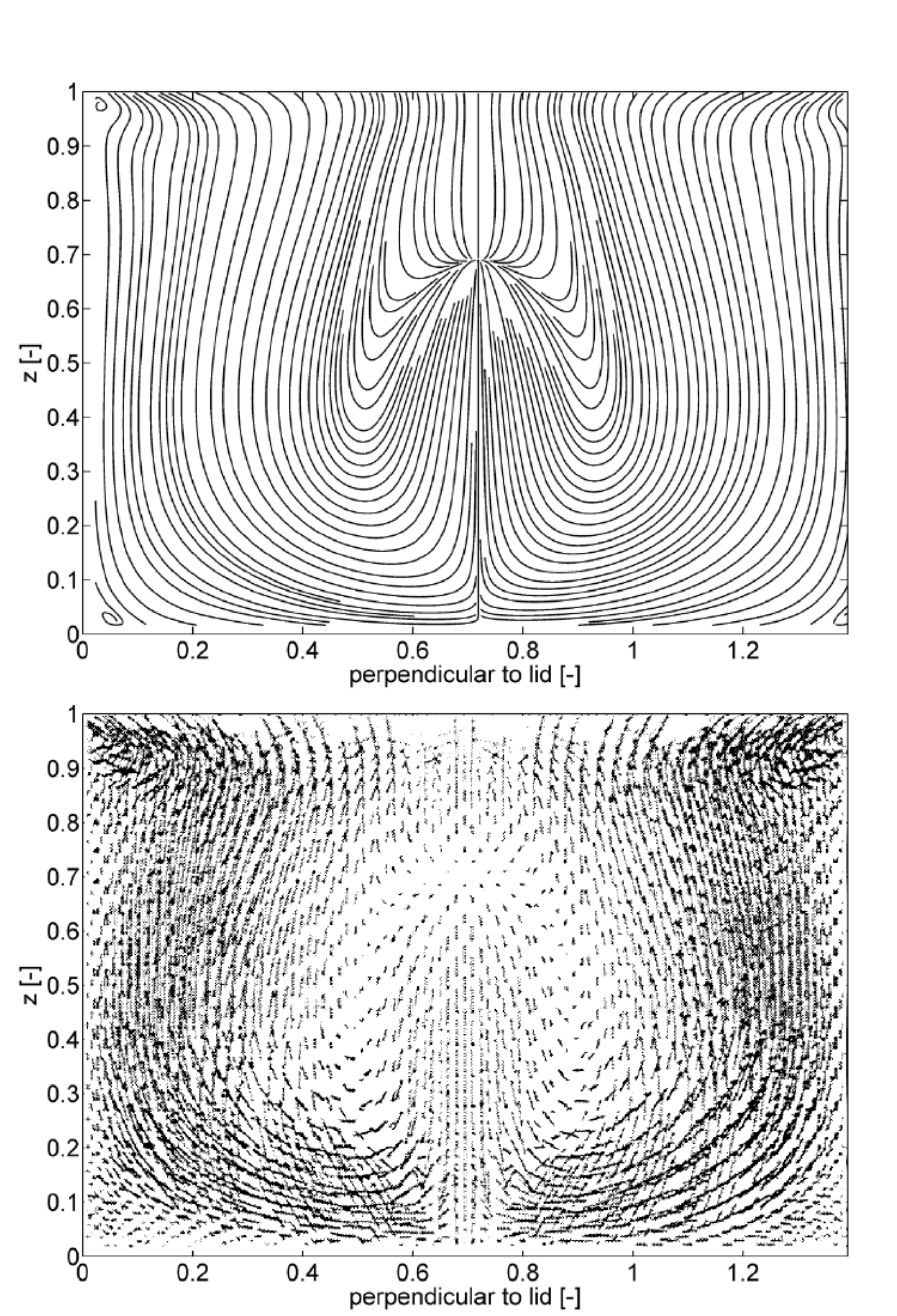}
\caption{Flow in the plane perpendicular to the direction of the lid (CP plane in Fig. \ref{3dlidcartoon}) at $\mbox{Re}=700$. Comparison between the LW-ACM with $60^3$ Cartesian grid (top) and a reference solution \cite{Povitsky2001} obtained by the commercial code FLUENT (bottom) with $68^3$ total number of grid nodes. At the centerline, a stagnation point is observed at $z=0.68$ (present), and $z=0.74$ (FLUENT).\label{comparison_3d_ppflow}}
\end{figure}

\begin{figure}[ht]
\centering
	     \includegraphics[width=0.7\textwidth]{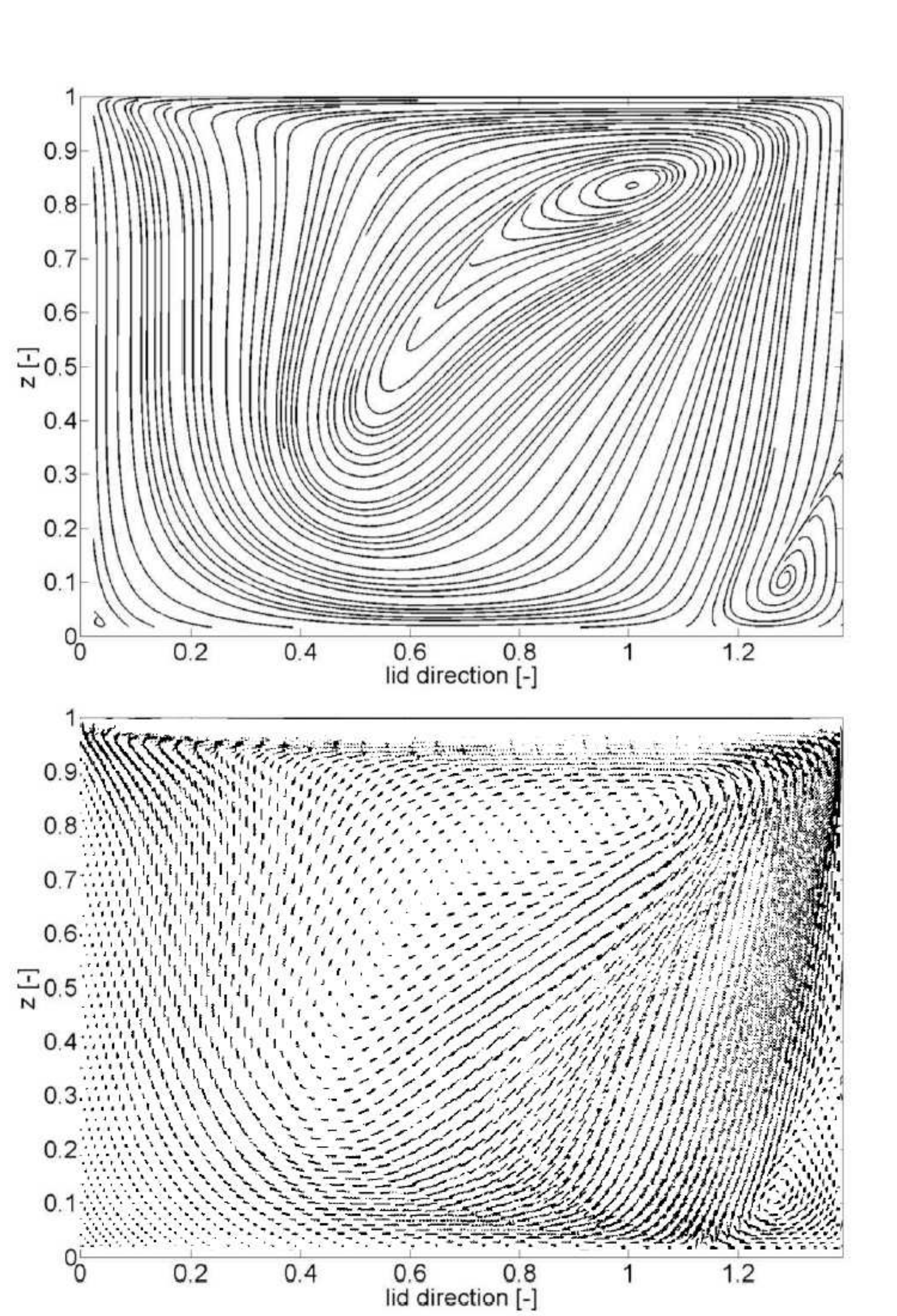}
\caption{Flow in the plane parallel to the direction of the lid (PP plane in Fig. \ref{3dlidcartoon}) at $\mbox{Re}=700$. Comparison between the LW-ACM with $60^3$ Cartesian grid (top) and a reference solution \cite{Povitsky2001} obtained by the commercial code FLUENT (bottom) with $68^3$ total number of grid nodes.\label{comparison_3d_cpflow}}
\end{figure}

\begin{figure}[ht]
\centering
	     \includegraphics[width=1\textwidth]{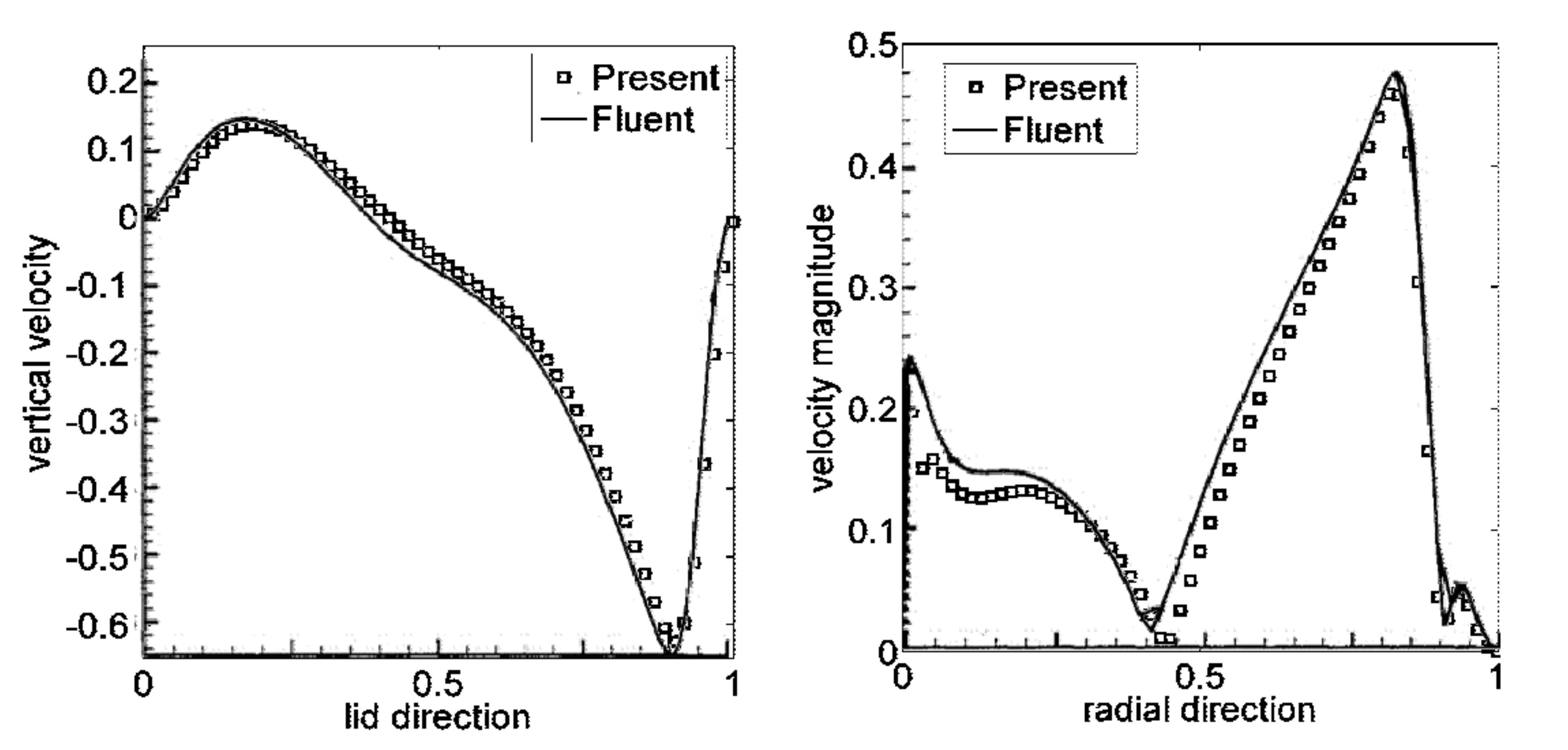}
\caption{Velocity profile along the line ML (left) and the line RL (right) at $\mbox{Re}=700$ (see also Fig. \ref{3dlidcartoon}). Comparison between the present LW-ACM with $60^3$ Cartesian grid and a reference solution \cite{Povitsky2001} obtained by the commercial code FLUENT with $68^3$ total number of grid nodes.\label{comparison_3d_quant}}
\end{figure}

In Figs. \ref{comparison_3d_midflow}, \ref{comparison_3d_ppflow} and \ref{comparison_3d_cpflow} we report a comparison between the velocity fields (in the MP, CP and PP planes of Fig. \ref{3dlidcartoon}) by both the commercial code FLUENT (non-uniform $68^3$ grid) \cite{Povitsky2001,Povitsky2005}, here considered as a reference, and the present LW-ACM ($60^3$ uniform grid) at Reynolds $\mbox{Re}=700$: All the flow structures are correctly reproduced by LW-ACM. Moreover, based on a comparison of both local quantities in Fig. \ref{comparison_3d_quant} and parallel/perpendicular global momenta $M_\|$, $M_\bot$:
\begin{equation}\label{3Dmomenta}
\begin{split}
M_\|=\frac{1}{2} \int_{V} (u+w)^2 d{V} \approx \frac{1}{2} \Delta x \Delta y \Delta z \sum_{i,j,k}\left( u_{i,j,k}+w_{i,j,k}\right)^2, \\
M_\bot=\frac{1}{2} \int_{V} (u-w)^2 d{V} \approx \frac{1}{2} \Delta x \Delta y \Delta z \sum_{i,j,k}\left( u_{i,j,k}-w_{i,j,k}\right)^2
\end{split}
\end{equation}
reported in Table \ref{global.quant.3D}, we can conclude that the present LW-ACM is indeed able to recover the reference solution with significant accuracy.
%Some numerical results and comparisons are reported in Figs. \ref{fig:lid3d-flow} and \ref{fig:lid3d-p}. In particular, Figure \ref{fig:lid3d-flow} compares the numerical results obtained by means of the commercial code FLUENT \cite{Povitsky2001} with the present LW-ACM. Clearly LW-ACM is able to catch the main features of the flow field and all the secondary vortexes: in particular, the top-left and bottom-right vortexes are quite difficult to catch.

\begin{figure}[ht]
\centering
       \includegraphics[width=1\textwidth]{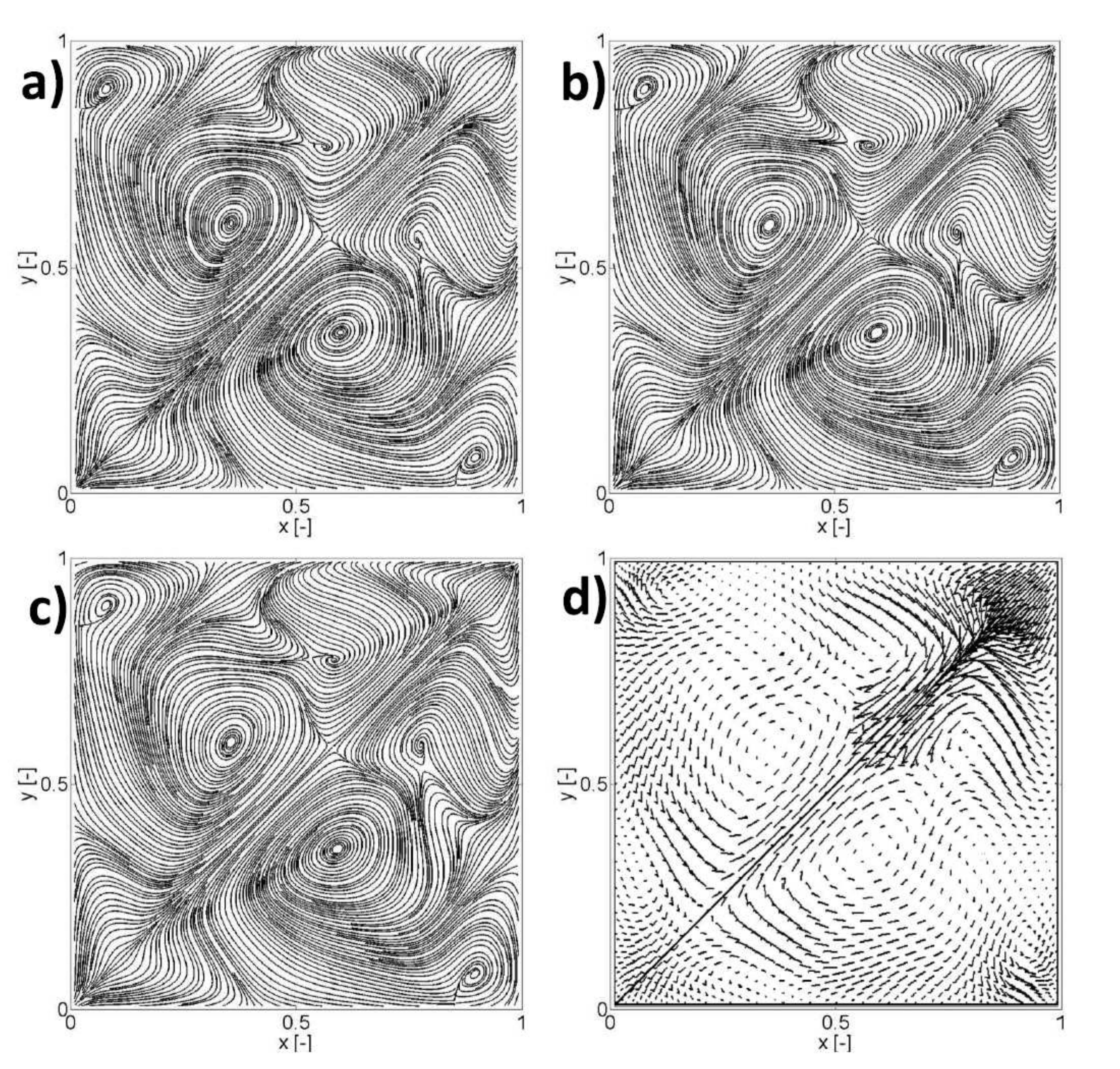} 
\caption{Flow in the middle plane $z=0.5$ of the 3D diagonally driven cavity (MP plane in Fig. \ref{3dlidcartoon}) at $\mbox{Re}=2000$. a) LW-ACM with $48^3$ uniform grid. b)  LW-ACM with $60^3$ uniform grid. c) LW-ACM with $68^3$ uniform grid. d) Reference solution by the commercial code FLUENT with $68^3$ non-uniform grid \cite{Povitsky2005}.\label{flowlid3d-lw-mrt}}
\end{figure}

\begin{figure}[ht]
\centering
       \includegraphics[width=1\textwidth]{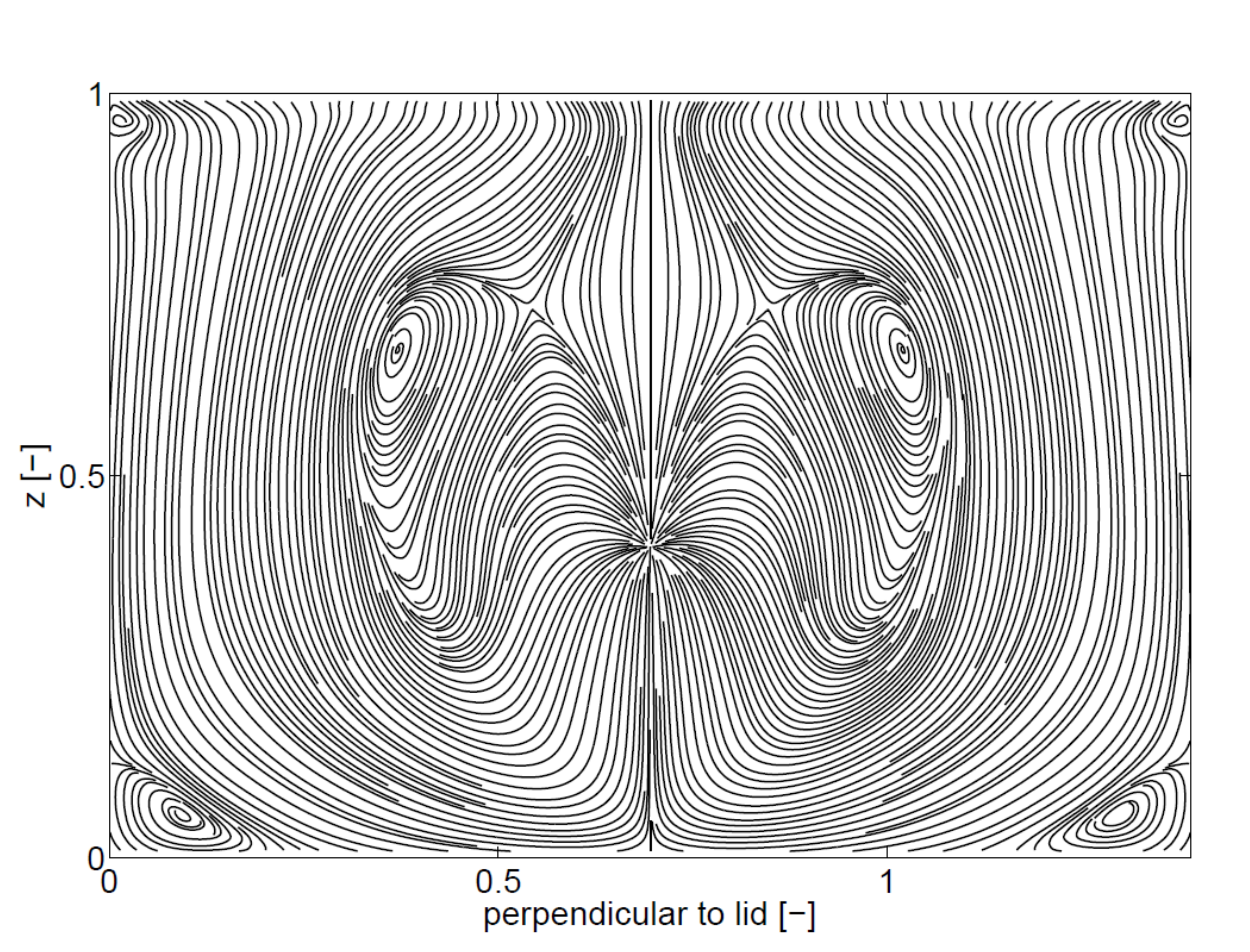} 
\caption{Flow in the plane perpendicular to the direction of the lid (CP plane in Fig. \ref{3dlidcartoon}) at $\mbox{Re}=2000$ obtained by the LW-ACM with $60^3$ uniform grid.  At the centerline, a stagnation point is observed at $z=0.405$.\label{cpflowre2000}}
\end{figure}

\begin{figure}[ht]
\centering
       \includegraphics[width=1\textwidth]{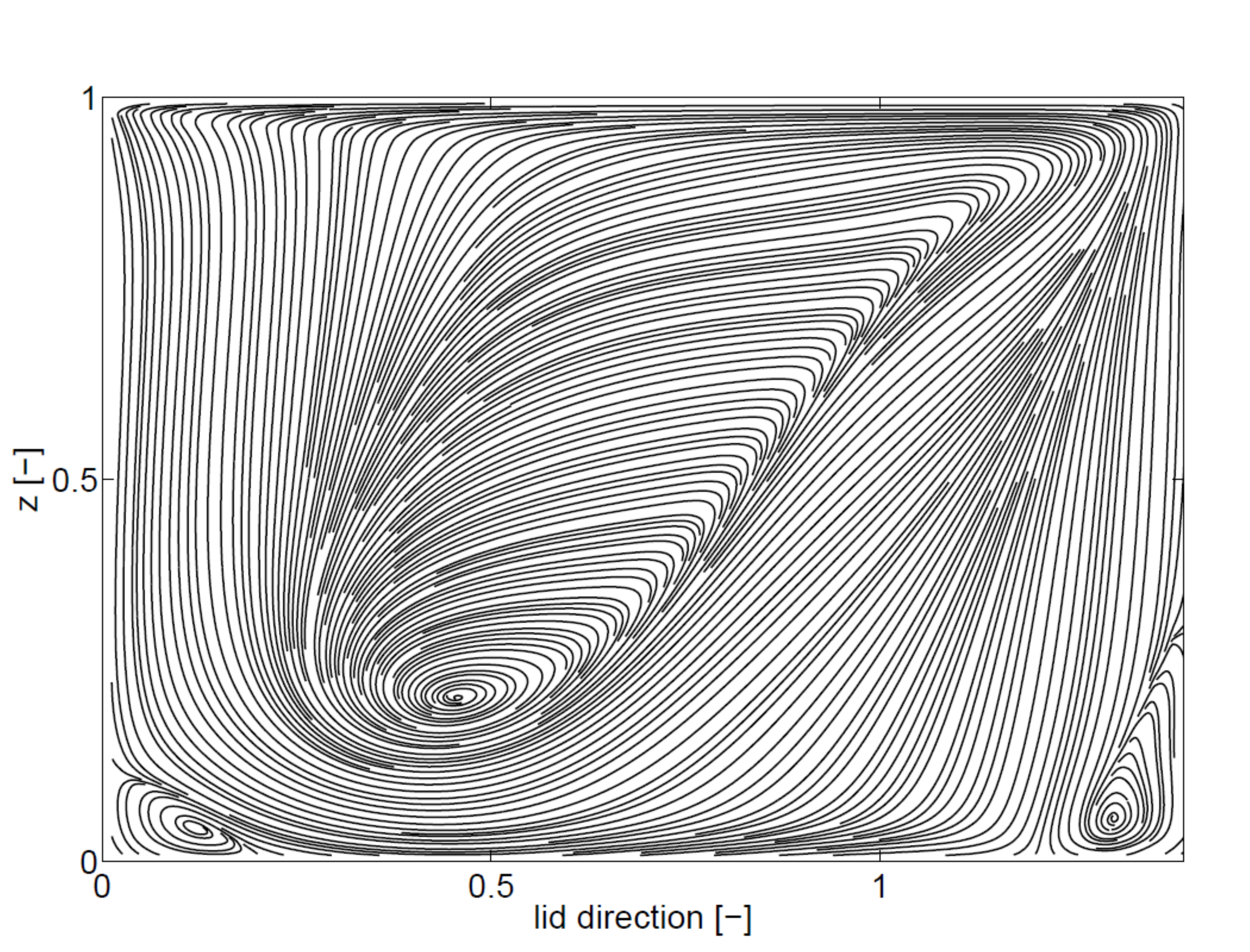} 
\caption{Flow in the plane parallel to the direction of the lid (PP plane in Fig. \ref{3dlidcartoon}) at $\mbox{Re}=2000$ obtained by the LW-ACM with $60^3$ uniform grid.\label{ppflowre2000}}
\end{figure}

In Figs. \ref{flowlid3d-lw-mrt}, \ref{cpflowre2000} and \ref{ppflowre2000} results of the 3D lid-driven cavity flow at higher Reynolds number, $\mbox{Re}=2000$, are reported. It is worth stressing that, here all the main structures of the flow are correctly described by LW-ACM even with grids coarser than the one adopted in the reference solution \cite{Povitsky2001,Povitsky2005}. We stress that, describing secondary vortexes in this case is known to be a severe test for numerical schemes (in particular catching top-left and bottom-right secondary vortexes).

For the sake of completeness, we also notice as the Reynolds number increases larger deviations of the LW-ACM solution from the reference are observed in terms of the parallel/perpendicular global momenta $M_\|$, $M_\bot$ (see Table \ref{global.quant.3D}).

\begin{figure}[ht]
\centering
       \includegraphics[width=1\textwidth]{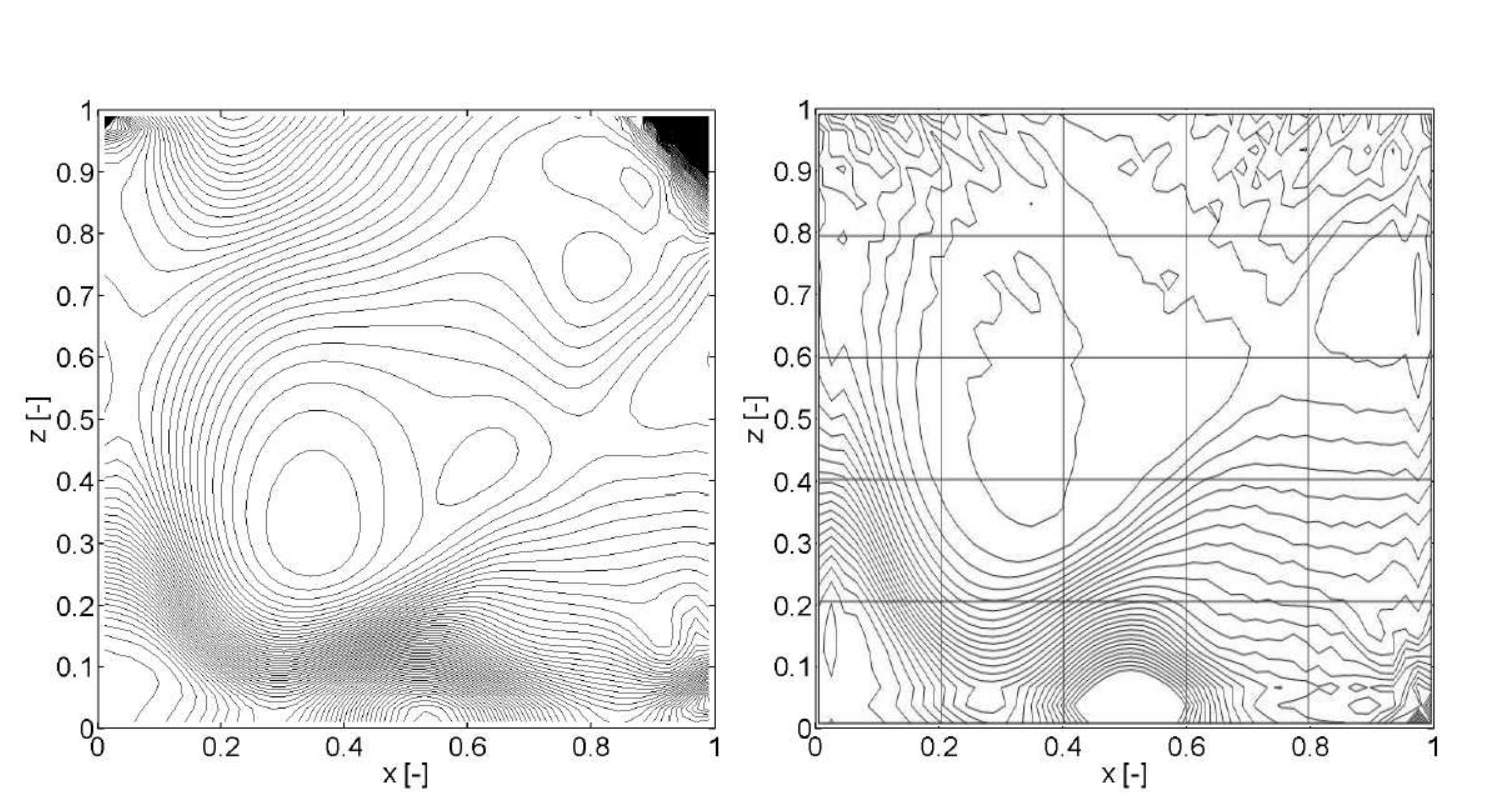} 
\caption{Pressure contours for the diagonally driven cavity flow for $\mbox{Re}=2000$ at the lateral mid-plane ($y=0.5$): Solution obtained by present LW-ACM with $48^3$ uniform grid and $D3Q19$ lattice (left), and MRT-LBM method with $52^3$ uniform grid and $D3Q15$ lattice \cite{dHumieres2002} (right).\label{Plid3d-lw-mrt}}
\end{figure}

Finally, in Fig. \ref{Plid3d-lw-mrt} the numerical results obtained by both LW-ACM and the MRT-LBM \cite{dHumieres2002} are shown. These two simulations are not perfectly comparable each other. In fact, Authors in Ref. \cite{dHumieres2002} were forced by stability issues to implement some simplifications when dealing with the boundary conditions (mainly, equilibrium-based boundary conditions for imposing the lid velocity and ``node'' bounce-back for the no-slip boundary conditions). Those simplifications reduce the accuracy with regards to that recovered by Eq. (\ref{ACM2.0ccccbb}).

Other minor difference is that \cite{dHumieres2002} and the present study were obtained by the D3Q15 lattice with $52^3$ grid, and D3Q19 lattice with $48^3$ grid respectively. We notice that, in this case, MRT-LBM makes use of a larger number of degrees of freedom compared to LW-ACM: $52^3 \times 15 > 48^3 \times 19$. In spite of this, it is quite clear by Fig. \ref{Plid3d-lw-mrt} that the pressure field recovered by the present LW-ACM is remarkably smoother than the one obtained by MRT-LBM. More specifically, a crucial difference is that LW-ACM predicts smooth pressure increase at the top-right corner, while the MRT-LBM results are affected by oscillations around the imposed constant pressure at the top plane.

%%%%%%%%%%%%%%%%%%%%%%%%%%%%%%%%%%%%%%%%%%%%%%%%%%%%%%%%%%%%%%%%%%%%
%%%%%%%%%%%%%%%%%%%%%%%%%%%%%RE=700%%%%%%%%%%%%%%%%%%%%%%%%%%%%%%%%%
%%%%%%%%%%%%%%%%%%%%%%%%%%%%%%%%%%%%%%%%%%%%%%%%%%%%%%%%%%%%%%%%%%%%
%%%%%%%%%%%%%%%%%%%
%%%%
%%%%
%%%%%%%%%%%%%%%%%%%
%%%%%%%%%%%%%%%%%%
%%%%%%%%%%%%%%%%%%
%%%%%%%%%%%%%%%%%%
%%%%%%%%%%%%%%%%%%
%%%%%%%%%%%%%%%%%%%%%%%%%%%%%%%%%%%%%%%%%%%%%%%%%%%%%%%%%%%%%%%%%%%%
%%%%%%%%%%%%%%%%%%%%%%%%%%%%%RE=700%%%%%%%%%%%%%%%%%%%%%%%%%%%%%%%%%
%%%%%%%%%%%%%%%%%%%%%%%%%%%%%%%%%%%%%%%%%%%%%%%%%%%%%%%%%%%%%%%%%%%%
%%%%%%%%%%%
%%%%%%%%%%%%%%%%%%%%%%%%%%%%%%%%%%%%%%%%%%%%%%%%%%%%%%%%%%%%%%%%%%%%
%%%%%%%%%%%%%%%%%%%%%%%%%%%%%RE=2000%%%%%%%%%%%%%%%%%%%%%%%%%%%%%%%%
%%%%%%%%%%%%%%%%%%%%%%%%%%%%%%%%%%%%%%%%%%%%%%%%%%%%%%%%%%%%%%%%%%%%
%%%%%%%%%%%
%%%%%%%%%%%
%%%%%%%%%%%
%%%%%%%%%%%
%%%%%%%%%%%
%%%%%%%%%%%%%%%%%%%%%%%%%%%%%%%%%%%%%%%%%%%%%%%%%%%%%%%%%%%%%%%%%%%%
%%%%%%%%%%%%%%%%%%%%%%%%%%%%%RE=2000%%%%%%%%%%%%%%%%%%%%%%%%%%%%%%%%
%%%%%%%%%%%%%%%%%%%%%%%%%%%%%%%%%%%%%%%%%%%%%%%%%%%%%%%%%%%%%%%%%%%%
%
%%%%%%%%%%%%%%%%%%%%%%%%%%%%%%%%%%%%%%%%%%%%%%%%%%%%%%%%%%%%%%%%%%%%
\vspace{0.5cm}
\begin{table}[ht]
\caption{3D diagonally driven cavity: Volume integral of momentum flux. Comparisons are carried out between the present LW-ACM method and the reference solution in \cite{Povitsky2001} adopting $60^3$ and $68^3$ total number of grid nodes, respectively.\label{global.quant.3D}}
\vspace{0.5cm}
\begin{center}
\begin{tabular}{ccccc}
\hline
  & Present, $\mbox{Re}=700$  & \cite{Povitsky2001,Povitsky2005}, $\mbox{Re}=700$ & Present, $\mbox{Re}=2000$ & \cite{Povitsky2001,Povitsky2005}, $\mbox{Re}=2000$\\
  \hline
$\int_V{M_\|}$ & $0.203 \times 10^{-1}$ & $0.216\times 10^{-1}$ & $0.134 \times 10^{-1}$ & $0.163 \times 10^{-1}$\\
$\int_V{M_\bot}$ & $0.232 \times 10^{-2}$ & $0.283\times 10^{-2}$ & $0.174 \times 10^{-2}$ & $0.239 \times 10^{-2}$\\
\end{tabular}
\end{center}
\end{table} 
%%%%%%%%%%%%%%%%%%%%%%%%%%%%%%%%%%%%%%%%%%%%%%%%%%%%%%%%%%%%%%%%%%%%

\subsubsection{\label{moving}Circular Couette flow}
\begin{figure}[ht]
\centering
       \includegraphics[width=1\textwidth]{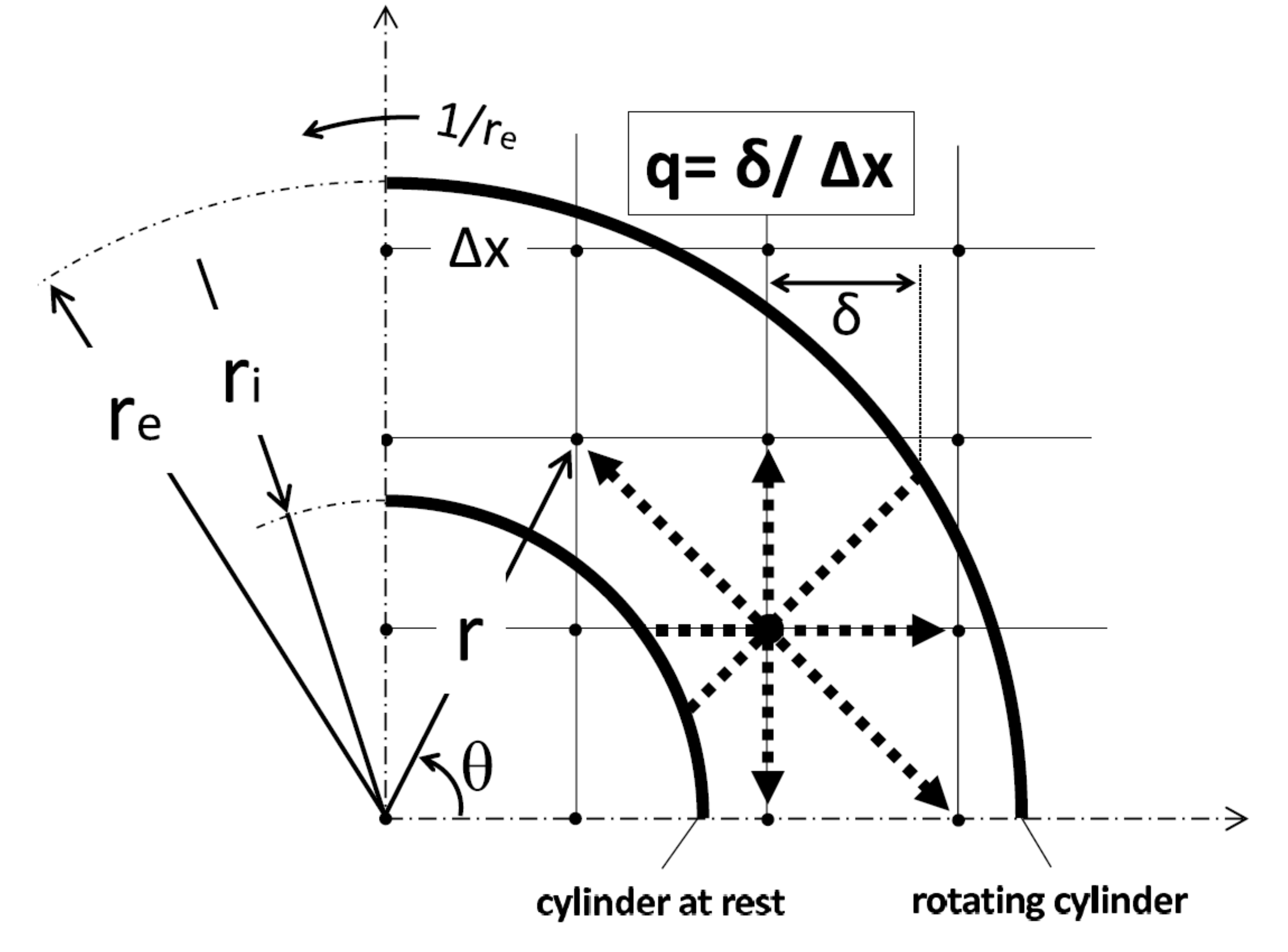} 
\caption{Set-up of the circular Couette flow, where one quarter of the domain is reported. For the sake of clarity, a significantly coarse grid is represented.}\label{circularcouette}
\end{figure}
Dealing with moving complex boundaries is very important in many applications: for example, particle suspensions, granular flows and active (bio-)agents immersed in the flow. In these cases, the essential issue is to reduce as much as possible the computational demand by avoiding re-meshing every time that the considered objects move in the flow. Taking into account moving objects is also complicated by the need of re-initializing the portions of the flow field which are filled again by the fluid after the motion of the objects. The latter feature is neglected here, because it is a general issue, not peculiar of the link-wise methods.

First of all, we extend the wall boundary treatment discussed in the previous sections. Let us suppose that $\hat{\mathbf{x}}$ is a fluid node close to a complex wall boundary at rest such that $\hat{\mathbf{x}}+\hat{\mathbf{v}}_i$ is a wall node. Let us focus on the intersection between the $i$-th lattice link and the wall. The distance between the latter intersection and the fluid node, divided by the mesh spacing $\Delta x$, gives the normalized distance $0\leq q\leq 1$ (above we considered only the case: $q=1/2$). In this case, the streaming step can be performed following the same procedure provided for LBM in \cite{bouzidi01} (instead of using Eq. (\ref{ACM2.0cccbb})), namely
\begin{eqnarray}\label{ACM2.0cc}
f_{BB(i)}^{**}(\hat{\mathbf{x}},\hat{t}+1)=\left\{
\begin{array}{ll}
2qf_i^*(\hat{\mathbf{x}},\hat{t})+(1-2q)f_i^*(\hat{\mathbf{x}}-\epsilon\,\hat{\mathbf{v}}_i,\hat{t}),
 & q<1/2,\\
\frac{1}{2q}f_i^*(\hat{\mathbf{x}},\hat{t})+\left(1-\frac{1}{2q}\right)f_{BB(i)}^*(\hat{\mathbf{x}},\hat{t}),
 & q\geq1/2,\\
\end{array}\nonumber
\right.
\end{eqnarray}
where $BB(i)$ is the bounce-back operator giving the lattice link opposite to $i$-th. Finally the post-combining step can be performed in the usual way, namely by means of Eq. (\ref{ACM2.0cccbb}).

In case of moving complex boundary with velocity $\mathbf{u}_w$, the procedure reported in \cite{bouzidi01} suggests to consider an additional term, namely
\begin{equation}\label{ACM2.0move}
\delta f_{BB(i)}(\rho_0,\mathbf{u}_w)=\left\{
\begin{array}{ll}
2f_{BB(i)}^{(e,o)}(\rho_0,\mathbf{u}_w),
 & q<1/2,\\
\frac{1}{q}f_{BB(i)}^{(e,o)}(\rho_0,\mathbf{u}_w),
 & q\geq1/2,\\
\end{array}
\right.
\end{equation}
where $f_{BB(i)}^{(e,o)}$ is given by Eq. (\ref{feqo}) and $\rho_0$ is the average value of the density over the whole computational domain (see Appendix \ref{appA} for details). Similarly to what we did in the previous sections, in case of diffusive scaling, the suggested correction for LBM will be multiplied by a scaling factor in link-wise ACM. Hence $f^w_{BB(i)}$ given by Eq. (\ref{ACM2.0ccccbb}) is the proper boundary condition in case of moving complex boundary.
%%%%%%%%%%%%%Elio
%\begin{figure}[ht]
%\begin{center}
%\includegraphics[width=\textwidth]{CirCouette}
%\end{center}
%  \caption{General setup for the circular Couette flow and some details about fitting curved walls on the Cartesian mesh.}
%\label{fig:circouette}
%\end{figure}

\vspace{0.5cm}
\begin{table}[ht]
\caption{The $L^1$ norm of the error versus $\epsilon\equiv\Delta x\equiv\mbox{Ma}$ at $t=20$ in the problem of the circular Couette flow for $\nu=0.07$.}
\label{table:circouette}
\vspace{0.5cm}
\begin{center}
\begin{tabular}{cccc}
  \hline
\multicolumn{4}{c}{Link-wise ACM} \\ 
  \hline
 $\epsilon\equiv\Delta x$ & Error $L^1[\bar{u}_\theta]$ & Error $L^1[\bar{p}]$ & Error $L^1[\left|{\mathcal{\bar T}}\right|]$ \\
  \hline
$1/20$ & $1.53627\times 10^{-3}$ & $8.81208\times 10^{-4}$ & $4.20562\times10^{-4}$\\
$1/40$ & $3.50537\times 10^{-4}$ & $3.58432\times 10^{-4}$ & $1.98687\times10^{-4}$\\
$1/80$ & $1.77257\times 10^{-4}$ & $1.97650\times 10^{-4}$ & $9.13289\times10^{-5}$\\
$1/160$& $3.42570\times 10^{-5}$ & $6.16474\times 10^{-5}$ & $3.66160\times10^{-5}$\\
  \hline\hline
\multicolumn{4}{c}{MRT-LBM} \\ 
  \hline
 $\epsilon\equiv\Delta x$ & Error $L^1[\bar{u}_\theta]$ & Error $L^1[\bar{p}]$ & Error $L^1[\left|{\mathcal{\bar T}}\right|]$ \\
  \hline
$1/20$ & $4.66795\times 10^{-3}$ & $2.52316\times 10^{-3}$ & $7.58091\times10^{-5}$\\
$1/40$ & $1.52864\times 10^{-3}$ & $8.47929\times 10^{-4}$ & $1.39351\times10^{-4}$\\
$1/80$ & $3.08607\times 10^{-4}$ & $2.99584\times 10^{-4}$ & $6.98541\times10^{-5}$\\
$1/160$& $7.99695\times 10^{-5}$ & $1.09817\times 10^{-4}$ & $3.39760\times10^{-5}$\\
\end{tabular}
\end{center}
\end{table}
%
%\begin{figure}[ht]
%\begin{center}
%\includegraphics[width=14cm]{circular-couette}
%\end{center}
%\caption{Normalized error analysis for circular Couette flow as a function of the grid spacing $\Delta x$ (left) and the kinematic viscosity $\nu$ (right).}
%\label{circular-couette}
%\end{figure}

In this section, numerical results are reported for the circular Couette flow, where a viscous fluid is confined in the gap between two concentric rotating cylinders. In our study, we assume the inner cylinder (with radius $r_i$) at rest while the outer cylinder (with radius $r_e$) rotates at a constant angular velocity $1/r_e$. The latter flow admits the following exact solution: %\blue{[ELIO: add a case test description] Details about this test case can be found in standard textbooks.} 

%\blue{
\begin{eqnarray}\label{circcouttesol}
\bar{u}(t,r,\theta) &=& -C\left( \frac{r}{r_i}-\frac{r_i}{r} \right)\sin(\theta),\\
\bar{v}(t,r,\theta) &=&  C\left( \frac{r}{r_i}-\frac{r_i}{r} \right)\cos(\theta),\\
\bar{p}(t,r,\theta) &=&  \bar p_i+C^2\ln\left(\frac{r_i^2}{r^2}\right)-\frac{C^2}{2}\left(\frac{r_i^2}{r^2}-\frac{r^2}{r_i^2}\right),\\
\bar{\mathcal{T}}   &=& 4 \pi C\,\nu\,r_i,
\end{eqnarray}%}
where
\begin{equation}\label{circcouttesolC}
C = \frac{1}{r_e/r_i-r_i/r_e}.
\end{equation}%}
%
%\blue{
%\begin{equation}\label{circcouttesolT}
%\bar{t} = 4\pi C\,\nu\,r_i.
%\mathbf{\mathcal{T}} = 4\pi C\,\nu\,r_i.
%\mathbf{\mathcal{F}}
%\end{equation}%}
%
Here, $\bar u$, $\bar v$, $\bar p$ and $\mathbf{\mathcal{\bar T}}$ denote horizontal velocity, vertical velocity, pressure and the torque on the inner cylinder respectively, with $\theta$ being the angle between radial direction and the horizontal axis. A schematic representation of this setup is reported in Figure \ref{circularcouette}. Diffusive scaling is considered for this test case, where the velocity field is scaled on meshes with different sizes, keeping fixed the relaxation frequency (see Appendix \ref{appA} for details). The latter scaling ensures second order convergence in the accuracy, as reported in Table \ref{table:circouette}. %The numerical results are plotted in Fig. \ref{circular-couette} for making more evident the order of convergence and discussing the error structure. Moreover the numerical results obtained by standard Lattice Boltzmann method are reported as well for comparison.}

%\blue{
Moreover, the torque exerted by the fluid on the inner cylinder is computed. To this end, Eq. (\ref{MEA3}) is applied in combination with the boundary conditions provided by (\ref{ACM2.0cc}). The computation of $\mathbf{\mathcal{\bar T}}$ was finally performed by a summation of the contributions (\ref{MEA3}) over all the boundary links around its surface, namely%}
\begin{equation}\label{MEA5}
\mathbf{\mathcal{T}}=\sum_{i\in \textsf{S}}\left(\hat{\mathbf{x}}-\hat{\mathbf{x}}_c\right)\times\mathbf{p}_i,
\end{equation}
%
%\blue{
where $\hat{\mathbf{x}}_c$ is the center of the cylinders and $\textsf{S}$ is the set of links starting from all nodes surrounding the body and intersecting the body itself. Similarly to what has been done for scaling the force exerted on a body, the above torque (assumed acting on the whole inner cylinder) must be converted from lattice units to physical units (see Table \ref{tab:scaling} for details). More specifically, the formula is the following%}
\begin{equation}\label{MEA7}
{\mathcal{\bar T}}=\sum_{i\in \textsf{S}(1/\epsilon)}\left(\hat{\mathbf{x}}-\hat{\mathbf{x}}_c\right)\times\mathbf{p}_i,
\end{equation}
because $\left|\hat{\mathbf{x}}-\hat{\mathbf{x}}_c\right|\sim 1/\epsilon$ and this automatically takes into account the force scaling reported in Eq. (\ref{MEA6}). 

In Table \ref{table:circouette}, the numerical results for the circular Couette flow are reported. As expected, the numerical solution in terms of velocity and pressure shows almost second order convergence rate. On the other hand, the modified MEA for link-wise ACM in computing $\left|\mathbf{\mathcal{\bar T}}\right|$ shows first order convergence rate (similarly to the original MEA for LBM).

It is important to highlight that, by taking into account the curvature correctly (e.g. by finite-volume ACM using body-fitted cell system), the accuracy is dramatically improved. The reason is that, in link-wise ACM and in LBM, the boundary condition for curved wall is based on one dimensional interpolation, but this method is not accurate for computing stresses (depending on local spatial derivatives). Hence for accurate solving boundary layers, the finite-volume ACM using body-fitted cell system is preferable. However, in the present paper, we used link-wise boundary conditions because they are extremely simple to be generalized in three dimensions and they have potential when dealing with moving complex obejcts (e.g. particles) for avoiding re-meshing. Hence, which formulation to chose is a matter of the considered application.

\section{\label{conclusions}Conclusions}

%\blue{[PIETRO,ELIO: To be re-formulated and integrated!!!]}
%\blue{
In the present work, a novel method for low Mach number fluid dynamic simulations is proposed, taking inspiration from the best features of both the Lattice Boltzmann Method (LBM) and more classical computational fluid dynamic (CFD) techniques such as the Artificial Compressibility Method (ACM). The main advantage is the possibility of exploiting well established technologies originally developed for LBM and classical CFD, with special emphasis on finite differences (at least in the present paper), at the cost of minor changes. For instance, like LBM, it is possible to use simple Cartesian structured meshes, eventually recursively refined in the vicinity of solid walls, and there is no need of solving Poisson equations for pressure. On the other hand, any boundary condition designed for finite difference schemes can be easily included.%}

As far as solving incompressible Navier-Stokes equations - INSE - (by minimal amount of unknowns) is the only concern, the pseudo-kinetic heritages of LBM represent a severe limitation to several aspects such as designing flexible boundary conditions, introducing tunable forcing terms and analyzing consistency of the numerical scheme (asymptotics). On the contrary, the suggested method has no such pseudo-kinetic heritages. Or in other words, following the standard LBM nomenclature, the present LW-ACM requires no high-order moments beyond hydrodynamics (often referred to as {\em ghost moments}) and no kinetic expansion such as Chapman-Enskog, Hilbert, van Kampen. Like finite difference schemes, only standard Taylor expansion is needed for analyzing consistency. Beside the above aspects, numerical evidences reported in this work suggest that LW-ACM represents an excellent alternative in terms of simplicity, stability and accuracy. Hence, in this framework (solving INSE by minimal amount of unknowns), the utility of high-order moments is questionable.

Finally, preliminary efforts towards optimal implementations have shown that LW-ACM is capable of similar computational speed as optimized (BGK-) LBM. In addition, the memory demand is significantly smaller than (BGK-) LBM. In our opinion, there is still room for improvement according to the performance model (based on assuming either infinitely fast memory or infinitely fast compute units). Importantly, with an efficient implementation, this algorithm may be one of the few which is compute-bound and not memory-bound. The latter {observation} is of particular interest for General-Purpose computing on Graphics Processing Units (GPGPU).
%The artificial compressibility method for the incompressible Navier-Stokes equations was (link-wise) reformulated by a finite set of discrete directions (links) on a regular Cartesian mesh, in analogy with the lattice Boltzmann method. This novel formulation allows to increase further the analogies between the artificial compressibility method and the lattice Boltzmann method. In fact, in addition to the standard formulation in terms of macroscopic variables (see Appendix \ref{appB}), the alternative link-wise formulation is based on the local equilibrium of the discrete distribution function (see Eq. (\ref{ACM2.0})), which allows one to use some of the techniques previously proposed for the lattice Boltzmann method. In this paper, how to deal with moving complex boundaries is discussed in details. Further examples of porting lattice Boltzmann technology in the context of the artificial compressibility method are currently under investigation, in particular with regards to thermal flows.

\section*{Acknowledgments}

The authors are grateful to Professor Charles-Henri Bruneau of Universit\'e Bordeaux I, for providing us the numerical data of lid-driven cavity flow. PA would like to thank Dr. Paul J. Dellar for pointing out the Minion \& Brown flow test case, Dr. Martin Geier for stressing the importance of having techniques able to compute hydrodynamical stresses locally, Dr. Salvador Izquierdo for critical discussions about LBM boundary conditions. Moreover PA would like to thank Dr. Thomas Zeiser for suggestions about efficient coding and for measuring the performance reported in Table \ref{performance} (based on elementary codes developed by PA). Finally, PA and EC would like to acknowledge the support of the FIRB project ``Revolutionary surface coatings by carbon nanotubes for high heat transfer efficiency - THERMALSKIN'' 2011-2014.

%*****************************************************************

\appendix

\section{\label{appEQ}Appendix: Equilibrium distribution functions}

The quantities $f_i^{(e)}$ are designed in order to recover the incompressible isothermal fluid dynamics. For sake of completeness, we report here the explicit expressions of the equilibrium functions for some popular lattices.

The D2Q9 lattice \cite{qian92}, suitable for two dimensional problems ($D=2$), consists of the following discrete velocities (${Q}=9$): $\hat{\mathbf{v}}_0 = (0,\, 0)$, $\hat{\mathbf{v}}_i = (\pm 1,\, 0)$ and $(0,\, \pm 1)$, for $i =$ 1--4, and $\hat{\mathbf{v}}_i = (\pm 1,\, \pm 1)$, for $i =$ 5--8, where the $i$-th equilibrium distribution function $f_i^{(e)}$ reads
\begin{equation}\label{d2q9_equilibrium_compact}
f_i^{(e)}=w_i\rho\left[1+3\hat{\mathbf{v}}_i\cdot\mathbf{u}+\frac{9}{2}\left(\hat{\mathbf{v}}_i\cdot\mathbf{u}\right)^2
-\frac{3}{2}\mathbf{u}^2\right],
\end{equation}
with $\rho$ the fluid density, and $w_i$ the weights
\begin{equation}\label{weight_factors}
w_i=\left\{
\begin{array}{lll}
&4/9&\;\;\;i=0,\\
&1/9&\;\;\;i=\text{1--4},\\
&1/36&\;\;\;i=\text{5--8}.\\
\end{array}\right.
\end{equation}
More explicitly, the complete set of equilibria takes the form:
\begin{eqnarray}\label{d2q9_equilibrium}
{f^{(e)}}=\left[\begin{array}{c}
4/9\,{\rho}-2/3\, \rho u^2-2/3\, \rho v^2,\\
1/9\,{\rho}{+1/3\, \rho u}+1/3\, \rho u^2-1/6\, \rho v^2,\\
1/9\,{\rho}{+1/3\, \rho v}+1/3\, \rho v^2-1/6\, \rho u^2,\\
1/9\,{\rho}{-1/3\, \rho u}+1/3\, \rho u^2-1/6\, \rho v^2,\\
1/9\,{\rho}{-1/3\, \rho v}+1/3\, \rho v^2-1/6\, \rho u^2,\\
1/36\,{\rho}{+1/12\,\rho( u+ v)}+1/8\,\rho( u+ v)^2-1/24\,\rho( u^2+ v^2),\\
1/36\,{\rho}{-1/12\,\rho( u- v)}+1/8\,\rho(- u+ v)^2-1/24\,\rho( u^2+ v^2),\\
1/36\,{\rho}{-1/12\,\rho( u+ v)}+1/8\,\rho(- u- v)^2-1/24\,\rho( u^2+ v^2),\\
1/36\,{\rho}{+1/12\,\rho( u- v)}+1/8\,\rho( u- v)^2-1/24\,\rho( u^2+ v^2)
\end{array}\right]\nonumber
,\end{eqnarray}
where $u$ and $v$ are the velocity components, i.e. $(u,v)^T=\mathbf{u}$, with pressure being $p=\rho / 3$. The above equations (\ref{d2q9_equilibrium_compact}) and (\ref{d2q9_equilibrium}) can be generalized as follows \cite{Asinari2009}
\begin{eqnarray}\label{d2q9_equilibrium_gen}
{f^{(g)}}\left(\Pi_{xx},\Pi_{yy}\right)=\left[\begin{array}{c}
\rho\left(1-\Pi_{xx}\right)\left(1-\Pi_{yy}\right),\\
{\rho}\left(\Pi_{xx}+u\right)\left(1-\Pi_{yy}\right)/2,\\
{\rho}\left(\Pi_{xx}-u\right)\left(1-\Pi_{yy}\right)/2,\\
{\rho}\left(1-\Pi_{xx}\right)\left(\Pi_{yy}+v\right)/2,\\
{\rho}\left(1-\Pi_{xx}\right)\left(\Pi_{yy}-v\right)/2,\\
{\rho}\left(\Pi_{xx}+u\right)\left(\Pi_{yy}+v\right)/4,\\
{\rho}\left(\Pi_{xx}-u\right)\left(\Pi_{yy}+v\right)/4,\\
{\rho}\left(\Pi_{xx}-u\right)\left(\Pi_{yy}-v\right)/4,\\
{\rho}\left(\Pi_{xx}+u\right)\left(\Pi_{yy}-v\right)/4,
\end{array}\right]
\end{eqnarray}
with the equation (\ref{d2q9_equilibrium_compact}) being a special case of (\ref{d2q9_equilibrium_gen}): If one assumes $\Pi_{xx}=1/3+u^2$ and $\Pi_{yy}=1/3+v^2$, then ${f^{(g)}}\left(1/3+u^2,1/3+v^2\right)={f^{(e)}}$ (if third order terms with respect to velocity components are neglected). However, it is possible to introduce more involved functions depending on additional parameters. For instance, a quasi-equilibrium function which is useful for tuning bulk viscosity of both lattice Boltzmann and link-wise ACM schemes can be expressed as
\begin{equation}\label{quasi-equilibrium}
{f^{(qe)}}\left(\rho,\mathbf{u},\text{Tr}\right)=
{f^{(g)}}\left(\frac{\text{Tr}+u^2-v^2}{2},\frac{\text{Tr}-u^2+v^2}{2}\right),
\end{equation}
where $\text{Tr}$ is an additional tunable parameter (usually corresponding to the trace of the second order tensor $\mathbf{\Pi}=\sum_i\hat{\mathbf{v}}_i\hat{\mathbf{v}}_i f_i$ normalized by density (see also the Appendix \ref{appD}).

%Similarly analytical expressions hold for three dimensional computational domains. 
The D3Q19 lattice, which is suitable for three dimensional problems ($D=3$), consists of the following discrete velocities (${Q}=19$): $\hat{\mathbf{v}}_0 = (0,\, 0,\, 0)$; $\hat{\mathbf{v}}_i = (\pm 1,\, 0,\, 0)$ and $(0,\, \pm 1,\, 0)$ and $(0,\, 0,\, \pm 1)$, for $i =$ 1--6; $\hat{\mathbf{v}}_i = (\pm 1,\, \pm 1,\, 0)$ and $(\pm 1,\, 0,\, \pm 1)$ and $(0,\, \pm 1,\, \pm 1)$, for $i =$ 7--18. Here, the the $i$-th function $f_i^{(e)}$ is formally identical to (\ref{d2q9_equilibrium_compact}), with the following weights
\begin{equation}\label{weight_factors_3d}
w_i=\left\{
\begin{array}{lll}
&1/3&\;\;\;i=0,\\
&1/18&\;\;\;i=\text{1--6},\\
&1/36&\;\;\;i=\text{7--18}.\\
\end{array}\right.
\end{equation}

\section{\label{appPHYS}Appendix: Physical derivation}

The Boltzmann equation is the fundamental equation in kinetic theory of gases, describing time evolution of the distribution function of gas molecules as a function of time, space coordinates, and molecular velocity. The Bhatnagar-Gross-Krook (BGK) model equation inherits the main features of the original Boltzmann equation, with the fluid-dynamic description of the BGK solution for small Knudsen numbers being much simpler to obtain. Hence, owing to a remarkably less demanding effort, it come advantageous the employment of the BGK equation at the heart of kinetic methods for solving INSE.
%Its computational efforts are also much less than the original Boltzmann equation, and therefore, it is quite natural and advantageous to employ the BGK equation as the basis of kinetic method for incompressible Navier-Stokes equation. 
A well known drawback of the BGK equation is that the recovered Prandtl number is unity, while the original Boltzmann equation yields a value near to 2/3. However, since most of the LBM schemes do not consider the energy equation, the issue of inaccurate thermal diffusivity can be often neglected. At the same time, it is allowed to employ the isothermal BGK with a constant collision frequency for this purpose \cite{asinari09}. 

A crucial {\em ingredient} of any lattice Boltzmann scheme is a finite set of microscopic velocities, called lattice. The generic lattice velocity is identified by the subscript $i$, where $0\leq i\leq Q-1$. The LBM simulates the time evolution of a weakly compressible gas flow in nearly continuum regime by solving a kinetic equation on the lattice and yields the solution of the incompressible Navier-Stokes equation as its leading order. Hence, the relaxation frequency in the BGK equation can be expressed as a function of the kinematic viscosity $\nu$. The dimensionless form of the simplified BGK equation on a lattice takes the form
\begin{equation}\label{ilbe}
\frac{\partial f_i''}{\partial \hat{t}}+\hat{\mathbf{v}}_i\cdot\hat{\nabla} f_i'' = \frac{1}{3\nu}\left(f_i^{(e)}-f_i''\right),
\end{equation}
where $\hat{\mathbf{x}}$, $\hat t$, and $\hat{\mathbf{v}}_i$ are the (dimensionless) space coordinates, time, and molecular velocity components, respectively; $f_i''$ is the distribution function of gas molecules for the $i$-th velocity on the lattice; $f_i^{(e)}$ is the equilibrium distribution function. Let us suppose that $\nu\ll 1$: then it is possible to find an approximated solution of (\ref{ilbe}) by singular regular expansion, where:
\begin{equation}\label{approx}
f_i'' = f_i^{(e)}-3\nu\,\hat{\mathbf{v}}_i\cdot\hat{\nabla} f_i^{(e)}+O(\nu^2).
\end{equation}
Introducing the above approximation in the advection term of Eq. (\ref{ilbe}), it yields
\begin{equation}\label{ilbe2}
\frac{\partial f_i''}{\partial \hat{t}}=-\hat{\mathbf{v}}_i\cdot\hat{\nabla} f_i^{(e)} 
+3\nu\,(\hat{\mathbf{v}}_i\cdot\hat{\nabla})^2 f_i^{(e)}+\frac{1}{3\nu}\left(f_i^{(e)}-f_i''\right)+O(\nu^2).
\end{equation}
Here, the goal is to derive an algorithm formulated in terms of only hydrodynamic quantities, i.e. the statistical macroscopic moments of $f_i''$ corresponding to microscopic quantities conserved by the collisional operator (right hand side of Eq. (\ref{ilbe})). In particular, the local equilibrium $f_i^{(e)}$ is defined such that it has the same hydrodynamic quantities of $f_i''$. Hence, as far as the computation of the hydrodynamic quantities is concerned, the collisional operator in (\ref{ilbe2}) is unessential. Removing the latter term determines a modification in the model equation, though there is no effect on the hydrodynamic quantities. Let us define a new model equation by removing the collisional term and neglecting terms $O(\nu^2)$ in (\ref{ilbe2}), which can be re-formulated with respect to the new distribution function $f_i'$ as follows:
\begin{equation}\label{ilbe3}
\frac{\partial f_i'}{\partial \hat{t}}=-\hat{\mathbf{v}}_i\cdot\hat{\nabla} f_i^{(e)} 
+3\nu\,(\hat{\mathbf{v}}_i\cdot\hat{\nabla})^2 f_i^{(e)}.
\end{equation}
The above model equation (\ref{ilbe3}) can be recast in the equivalent form
\begin{equation}\label{ilbe4}
\frac{\partial f_i'}{\partial \hat{t}}=
-\eta_2\,\left(\hat{\mathbf{v}}_i\cdot\hat{\nabla} f_i^{(e,e)} 
-\eta_4/\eta_2\,(\hat{\mathbf{v}}_i\cdot\hat{\nabla})^2 f_i^{(e,e)}\right)
-\eta_1\,\left(\hat{\mathbf{v}}_i\cdot\hat{\nabla} f_i^{(e,o)}
-\eta_3/\eta_1\,(\hat{\mathbf{v}}_i\cdot\hat{\nabla})^2 f_i^{(e,o)}\right),
\end{equation}
where the odd part of the equilibrium distribution function is defined by (\ref{feqo}), while the even part is $f_i^{(e,e)}=f_i^{e}-f_i^{(e,o)}$, $\eta_1=\eta_2=1$ and $\eta_3=\eta_4=3\nu$. Recalling that 
\begin{equation}\label{ACM2.0taylor}
\hat{\mathbf{v}}_i\cdot\hat{\nabla} f_i^{(e,o/e)} 
-\frac{1}{2}\,(\hat{\mathbf{v}}_i\cdot\hat{\nabla})^2 f_i^{(e,o/e)}\approx
f_i^{(e,o/e)}(\hat{\mathbf{x}}-\hat{\mathbf{v}}_i,\hat{t})-f_i^{(e,o/e)}(\hat{\mathbf{x}},\hat{t}),
\end{equation}
we modify once more the model equation by setting $\eta_1=6\nu$ and $\eta_4=1/2$ (while other parameters remain unchanged, namely $\eta_2=1$ and $\eta_3=3\nu$). By doing so, $\eta_4/\eta_2=\eta_3/\eta_1=1/2$ which enables to use the approximation (\ref{ACM2.0taylor}). By means of the above set of parameters, Eq. (\ref{ilbe4}) becomes
\begin{equation}\label{ACM2.0nu2}
\frac{\partial f_i'}{\partial \hat{t}}=
-\left(f_i^{(e,e)}(\hat{\mathbf{x}},\hat{t})
-f_i^{(e,e)}(\hat{\mathbf{x}}-\hat{\mathbf{v}}_i,\hat{t})\right)
-6\nu\left(
f_i^{(e,o)}(\hat{\mathbf{x}},\hat{t})-
f_i^{(e,o)}(\hat{\mathbf{x}}-\hat{\mathbf{v}}_i,\hat{t})\right).
\end{equation}
As common in LBM, we apply the forward Euler rule for approximating first order time derivatives:
\begin{equation}\label{ACM2.0nu3}
f_i'(\hat{\mathbf{x}},\hat{t}+1)=
f_i'(\hat{\mathbf{x}},\hat{t})
-\left(f_i^{(e,e)}(\hat{\mathbf{x}},\hat{t})
-f_i^{(e,e)}(\hat{\mathbf{x}}-\hat{\mathbf{v}}_i,\hat{t})\right)
-6\nu\left(
f_i^{(e,o)}(\hat{\mathbf{x}},\hat{t})-
f_i^{(e,o)}(\hat{\mathbf{x}}-\hat{\mathbf{v}}_i,\hat{t})\right).
\end{equation}
As far as the computation of hydrodynamic quantities is concerned, the first term of the right hand side in (\ref{ACM2.0nu3}) can be substituted by $f_i^{(e)}(\hat{\mathbf{x}},\hat{t})$ (they have same hydrodynamic moments). The novel model equation can thus be re-formulated in terms of the distribution function $f_i$ as follows
\begin{equation}\label{ACM2.0nu4}
f_i(\hat{\mathbf{x}},\hat{t}+1)=
f_i^{(e)}(\hat{\mathbf{x}},\hat{t})
-\left(f_i^{(e,e)}(\hat{\mathbf{x}},\hat{t})
-f_i^{(e,e)}(\hat{\mathbf{x}}-\hat{\mathbf{v}}_i,\hat{t})\right)
-6\nu\left(
f_i^{(e,o)}(\hat{\mathbf{x}},\hat{t})-
f_i^{(e,o)}(\hat{\mathbf{x}}-\hat{\mathbf{v}}_i,\hat{t})\right),
\end{equation}
or equivalently
\begin{equation}\label{ACM2.0nu}
f_i(\hat{\mathbf{x}},\hat{t}+1)=
f_i^{(e)}(\hat{\mathbf{x}}-\hat{\mathbf{v}}_i,\hat{t})
+\left(1-6\nu\right)\left(
f_i^{(e,o)}(\hat{\mathbf{x}},\hat{t})-
f_i^{(e,o)}(\hat{\mathbf{x}}-\hat{\mathbf{v}}_i,\hat{t})\right).
\end{equation}
Eq. (\ref{ACM2.0}) can be recovered, upon substitution of (\ref{viscosity2}) into (\ref{ACM2.0nu}).

\section{\label{appA}Appendix: Asympthotic analysis}

In (\ref{ACM2.0}), with $\hat{\mathbf{x}}=\mathbf{x}/\Delta x$ and $\hat{t}=t/\Delta t$, both $\Delta t$ and $\Delta x$ must approach zero, though it is not clear the value of the limit: $\lim_{\Delta x \to 0} \Delta t/\Delta x$. In order to clarify this point, we apply Taylor expansion to (\ref{ACM2.0})
\begin{eqnarray}\label{AA}
f_i(t+\Delta t)&&=
f_i^{(e)}
-\Delta x\,\hat{\mathbf{v}}_i\cdot\nabla f_i^{(e)}
+\frac{\Delta x^2}{2}(\hat{\mathbf{v}}_i\cdot\nabla)^2 f_i^{(e)}
-\frac{\Delta x^3}{6}(\hat{\mathbf{v}}_i\cdot\nabla)^3 f_i^{(e)}+\dots\nonumber\\
&&+\left(2-\frac{2}{\omega}\right)\left(
\Delta x\,\hat{\mathbf{v}}_i\cdot\nabla f_i^{(e,o)}
-\frac{\Delta x^2}{2}(\hat{\mathbf{v}}_i\cdot\nabla)^2 f_i^{(e,o)}
+\frac{\Delta x^3}{6}(\hat{\mathbf{v}}_i\cdot\nabla)^3 f_i^{(e,o)}\right)+\dots,\nonumber\\
\end{eqnarray}
where all the quantities are computed in the same point $\mathbf{x}$ and hence this is no more explicitly reported. A summation of Eqs. (\ref{AA}) over $i$ yields:
\begin{eqnarray}\label{AArho}
\frac{\partial\rho}{\partial t}
+\left(\frac{2}{\omega}-1\right)
\frac{\Delta x}{\Delta t}\,\nabla\cdot (\rho\mathbf{u})
&+&\frac{1}{6}\left(\frac{2}{\omega}-1\right)
\frac{\Delta x^3}{\Delta t}\,(\nabla\cdot)^3 \mathsf{Q}^{(e)}=\nonumber\\
&&\frac{\Delta x^2}{2\,\Delta t}(\nabla\cdot)^2\mathbf{\Pi}^{(e)}
+O\left(\frac{\Delta x^4}{\Delta t}(\nabla\cdot)^2\mathbf{\Pi}^{(e)}\right),
\end{eqnarray}
with $\mathsf{Q}^{(e)}=\sum_i\hat{\mathbf{v}}_i\hat{\mathbf{v}}_i\hat{\mathbf{v}}_i f_i^{(e)}$. In the rightmost term of the above expression (\ref{AArho}), we rely upon the fact that spatial derivatives of the fourth-order moments have the same growth rate of spatial derivatives of the second-order moments. Moreover, the diverge is raised to a power which is selected for consistency with the units of other terms. This is unessential, as far as the order of magnitude of the term is concerned. Multiplying (\ref{AA}) by $\hat{\mathbf{v}}_i$ and summing over $i$, it yields
\begin{equation}\label{AAu}
\frac{\partial(\rho\mathbf{u})}{\partial t}
+\frac{\Delta x}{\Delta t}\,\nabla\cdot \mathbf{\Pi}^{(e)}=
\frac{\Delta x^2}{\Delta t}\left(\frac{1}{\omega}-\frac{1}{2}\right)
(\nabla\cdot)^2 \mathsf{Q}^{(e)}+O\left(\frac{\Delta x^3}{\Delta t}\nabla\cdot\mathbf{\Pi}^{(e)}\right).
\end{equation}
Similar considerations as Eq. (\ref{AArho}) apply to the rightmost term in the above equation (\ref{AAu}). Concerning the relationship between $\Delta t$ and $\Delta x$, (\ref{AAu}) suggests two possible strategies:
\begin{subeqnarray}\label{scaling}
\Delta t&\propto\Delta x,\qquad&\mbox{acoustic scaling},\\
\Delta t&\propto\Delta x^2,\qquad&\mbox{diffusive scaling}.
\end{subeqnarray}
Sometimes, the dimensionless mesh spacing $\Delta x=\Delta x'/L$ is referred to as spacing ($\Delta x\equiv h$) in the literature on finite difference method or even numerical Knudsen number ($\Delta x\equiv\mbox{Kn}$) in the Lattice Boltzmann literature. Similarly, it is possible to introduce a numerical Mach number: $\mbox{Ma}=U/(\Delta x'/\Delta t')$. In this way, the dimensionless time step $\Delta t=\Delta t'/(L/U)$ can be expressed as $\Delta t\equiv \mbox{Kn}\,\mbox{Ma}$. Hence, the acoustic scaling corresponds to constant $\mbox{Ma}$, while the diffusive scaling to $\mbox{Ma}\propto \Delta x$.
%%%%%%%%%%%%%%%%%%%%%%%%%%

Regardless of the adopted strategy, the numerical scheme must converge towards the physical solution of the incompressible Navier-Stokes equations. The physical solution is identified by the Reynolds number, which is the reciprocal of the factor multiplying the second-order spatial derivatives in Eq. (\ref{AAu}), namely
\begin{equation}\label{Re}
\frac{\Delta x^2}{\Delta t}\left(\frac{1}{\omega}-\frac{1}{2}\right)\propto\frac{1}{\mbox{Re}}.
\end{equation}
Hence, according to (\ref{Re}), in acoustic scaling $\omega$ needs to be tuned in order to guarantee a constant Reynolds number, while in diffusive scaling a fixed $\omega$ already ensures a constant Reynolds number. Moreover, in acoustic scaling, the smaller $\Delta x$ the smaller $\omega$ for keeping fixed the Reynolds number. The latter case can be problematic in the present LW-ACM due to the heuristic stability domain, $1\leq \omega<2$, thus diffusive scaling is generally preferable.

Similarly to standard LBM, in LW-ACM, the definition of $\Delta t$ and $\Delta x$ is implicit (and this is sometimes a source of confusion). In fact, the end-user can select both $\mbox{Kn}\propto 1/N$ (with $N$ the number of mesh points along the flow characteristic length) and $\mbox{Ma}$ by a scaling factor for the velocity field. Hence, the acoustic scaling requires the tuning of the relaxation frequency $\omega$ on different meshes, keeping constant the computed velocity field. On the other hand, the diffusive scaling corresponds to a scaling of the velocity field (see also the section below) on different meshes, keeping fixed the relaxation frequency.

\subsection{Diffusive scaling}

Substituting $\Delta x=\epsilon\ll 1$ and $\Delta t=\epsilon^2$ into (\ref{AAu}), it yields
\begin{equation}\label{AAuds}
\frac{\partial(\rho\mathbf{u})}{\partial t}
+\frac{1}{\epsilon}\,\nabla\cdot \mathbf{\Pi}^{(e)}=
\left(\frac{1}{\omega}-\frac{1}{2}\right)
(\nabla\cdot)^2 \mathsf{Q}^{(e)}+O\left(\epsilon\nabla\cdot\mathbf{\Pi}^{(e)}\right).
\end{equation}
Due to the presence of a mesh-dependent parameter in (\ref{AAuds}), upon convergence, all moments need to be scaled via the following post-process:
%The problem with the previous equation is that there is a mesh-dependent parameter, which one would like to get rid of. The solution consists in scaling all moments. Once the numerical solution is available, let us suppose to apply the following post-processing:
%
\begin{equation}\label{scaling2}
\bar{\mathbf{u}}=(\mathbf{u}-\mathbf{u}_0)/\epsilon,\qquad \bar{\mathbf{\Pi}}^{(e)}=(\mathbf{\Pi}^{(e)}-\mathbf{\Pi}_0)/\epsilon^2,\qquad \bar{\mathsf{Q}}^{(e)}=(\mathsf{Q}^{(e)}-\mathbf{Q}_0)/\epsilon, 
\end{equation}
where $\mathbf{u}_0$, $\mathbf{\Pi}_0$ and $\mathbf{Q}_0$ are constants, with the odd quantities $\mathbf{u}_0$ and $\mathbf{Q}_0$ equal to zero. On the contrary, the even quantity $\mathbf{\Pi}_0=p_0\mathbf{I}$, where $p_0$ is the average pressure over the whole computational domain (the average even moment $\mathbf{\Pi}^{(e)}$ over the whole computational domain depends mainly on the amount of mass and only slightly on the flow field).

As a result, the normalized pressure field is defined as $\bar{p}=(p-p_0)/\epsilon^2$ and the normalized density field as $\bar{\rho}=(\rho-\rho_0)/\epsilon^2$ or equivalently $\rho=\rho_0+\epsilon^2\,\bar{\rho}$. Upon substitution of the latter quantities into (\ref{AArho}), it follows
\begin{equation}\label{AArho2}
%\epsilon^2\frac{\partial\bar{\rho}}{\partial t}
%+\left(\frac{2}{\omega}-1\right)\rho_0
%\nabla\cdot\bar{\mathbf{u}}=O(\epsilon^2),
\nabla\cdot\bar{\mathbf{u}}=O(\epsilon^2).
\end{equation}
Recalling that $\mathbf{\Pi}^{(e)}=\rho\mathbf{u}\mathbf{u}+p\,\mathbf{I}$ and introducing the scaled quantities, we obtain
\begin{equation}\label{AAu2}
\rho_0\frac{\partial\bar{\mathbf{u}}}{\partial t}
+\rho_0\bar{\mathbf{u}}\cdot\nabla\bar{\mathbf{u}}
+\nabla\bar{p}=
\left(\frac{1}{\omega}-\frac{1}{2}\right)
(\nabla\cdot)^2 \bar{\mathsf{Q}}^{(e)}+O(\epsilon^2).
\end{equation}
In order to recover the incompressible isothermal fluid dynamics, the quantities $f_i^{(e)}$ are designed \cite{qian92} such that
\begin{equation}\label{third_def}
{Q}_{ijk}^{(e)}=\frac{\rho}{3}\left({u}_i\delta_{jk}+{u}_j\delta_{ik}+{u}_k\delta_{ij}\right).
\end{equation}
%
%or equivalently
%
%\begin{equation}\label{third_def2}
%\bar{Q}_{ijk}^{(e)}=\frac{\rho}{3}\left(\bar{u}_i\delta_{jk}+\bar{u}_j\delta_{ik}+\bar{u}_k\delta_{ij}\right).
%\end{equation}
%
Consequently 
\begin{equation}\label{third}
\nabla\cdot\nabla\cdot \bar{\mathsf{Q}}^{(e)}=\frac{\rho_0}{3}\nabla^2\bar{\mathbf{u}}+\frac{2\,\rho_0}{3}\nabla\,\nabla\cdot\bar{\mathbf{u}}+O(\epsilon^2)=
\frac{\rho_0}{3}\nabla^2\bar{\mathbf{u}}+O(\epsilon^2),
\end{equation}
where the last result is due to Eq. (\ref{AArho2}). Introducing the previous assumption in Eq. (\ref{AAu2}), it yields
\begin{equation}\label{AAu3}
\boxed{\frac{\partial\bar{\mathbf{u}}}{\partial t}
+\bar{\mathbf{u}}\cdot\nabla\bar{\mathbf{u}}
+\frac{1}{\rho_0}\nabla\bar{p}=\nu\nabla^2\bar{\mathbf{u}}+O(\epsilon^2).}
\end{equation}
where
\begin{equation}\label{viscosity}
\nu=\frac{1}{3}\left(\frac{1}{\omega}-\frac{1}{2}\right).
\end{equation}
Introducing the previous expression into Eq. (\ref{AArho}), it yields
\begin{equation}\label{AArho3}
\frac{\epsilon^2}{\rho_0}\frac{\partial\bar{\rho}}{\partial t}
+6\nu\nabla\cdot\bar{\mathbf{u}}=
\frac{\epsilon^2}{2}\nabla\cdot
\left(\bar{\mathbf{u}}\cdot\nabla\bar{\mathbf{u}}
+\frac{1}{\rho_0}\nabla\bar{p}\right)+O(\epsilon^4).
\end{equation}
Taking into account the new definition given by (\ref{viscosity}), Eq. (\ref{AArho2}) should be expressed more rigorously as $\nabla\cdot\bar{\mathbf{u}}=O(\epsilon^2/\nu)$. Combining the latter equation and Eq. (\ref{AAu3}), it follows
\begin{equation}\label{AAu4}
\nabla\cdot
\left(\bar{\mathbf{u}}\cdot\nabla\bar{\mathbf{u}}
+\frac{1}{\rho_0}\nabla\bar{p}\right)=O(\epsilon^2/\nu).
\end{equation}
Introducing the above expression into Eq. (\ref{AArho3}), it yields
\begin{equation}\label{AArho4}
\boxed{\frac{\epsilon^2}{6\rho_0\nu}\frac{\partial\bar{\rho}}{\partial t}
+\nabla\cdot\bar{\mathbf{u}}=O(\epsilon^4/\nu^2).}
\end{equation}
The divergence-free condition for the velocity field requires that $\epsilon^2/\nu\ll 1$, which is consistent with (\ref{AAu3}) as well.

\subsection{Acoustic scaling}

Assuming $\Delta x=\epsilon\ll 1$ and $\Delta t=\epsilon$, Eq. (\ref{AAu}) yields
\begin{equation}\label{AAu_as}
\frac{\partial(\rho\mathbf{u})}{\partial t}
+\nabla\cdot \mathbf{\Pi}^{(e)}=
\epsilon\left(\frac{1}{\omega}-\frac{1}{2}\right)
(\nabla\cdot)^2 \mathsf{Q}^{(e)}+O(\epsilon^2).
\end{equation}
This time, there is no need to scale all the moments and a proper tuning of $\omega$ is sufficient instead (see below). Leaving the moments unscaled and substituting the above assumptions in Eq. (\ref{AArho}), we obtain
\begin{equation}\label{AArho_as}
\frac{\partial\rho}{\partial t}
+\left(\frac{2}{\omega}-1\right)\,\nabla\cdot (\rho\mathbf{u})=O(\epsilon).
\end{equation}
The above equation (\ref{AArho_as}) proves that acoustic scaling should not be used in link-wise ACM, when dealing with transient simulations, while for steady state flows we get:
\begin{equation}\label{AArho_as2}
\nabla\cdot (\rho\mathbf{u})=O(\epsilon/\nu),
\end{equation}
showing that, if the density (pressure) gradients are small (consistently with the incompressible limit), the above equation provides indeed an accurate divergence-free velocity field. However, in acoustic (unlike diffusive scaling) the compressibility error cannot be reduced by mesh refinement. Taking into account Eq. (\ref{third_def}) and Eq. (\ref{AArho_as2}), we get
\begin{equation}\label{third_as}
\nabla\cdot\nabla\cdot {\mathsf{Q}}^{(e)}=\frac{1}{3}\nabla^2(\rho{\mathbf{u}})+O(\epsilon/\nu),
\end{equation}
hence
\begin{equation}\label{AAu_as2}
\frac{\partial(\rho\mathbf{u})}{\partial t}
+\nabla\cdot (\rho\mathbf{u}\mathbf{u})+\nabla p=
\bar{\nu}\nabla^2(\rho{\mathbf{u}})+O(\epsilon),
\end{equation}
where $\bar{\nu}=\epsilon\,\nu$. If the density (and pressure) gradients are small, the solution to the system formed by (\ref{AArho_as2}) and (\ref{AAu_as2}) provides a reasonable approximation of the Navier-Stokes solution in the incompressible limit. However, the latter system does not asymptotically converge towards the incompressible Navier-Stokes solution, as the mesh get finer and finer (at least, as far as the discretization error is smaller than the compressibility error). This is the main reason why the diffusive scaling is preferred in this work.

\subsection{Forcing}

Let us consider the forcing step described by Eq. (\ref{force}), with $\Delta x=\epsilon\ll 1$, $\Delta t=\epsilon^{\beta+1}$ (or equivalently $\mbox{Ma}=\mbox{Kn}^\beta$) where $\beta$ is a free parameter ($\beta=0$ and $\beta=1$ denote acoustic and diffusive scaling, respectively). The correction due to (\ref{force}) leads to an additional term in (\ref{AAu}):
\begin{equation}\label{AAu_for}
\frac{\partial(\rho\mathbf{u})}{\partial t}
+\frac{\Delta x}{\Delta t}\,\nabla\cdot \mathbf{\Pi}^{(e)}=
\frac{\Delta x^2}{\Delta t}\left(\frac{1}{\omega}-\frac{1}{2}\right)
(\nabla\cdot)^2 \mathsf{Q}^{(e)}+O\left(\frac{\Delta x^3}{\Delta t}\nabla\cdot\mathbf{\Pi}^{(e)}\right)+\frac{1}{\Delta t}\,\rho\mathbf{g}.
\end{equation}
From the definition $\Delta t\equiv \mbox{Kn}\,\mbox{Ma}$ and $\mathbf{u}=\mbox{Ma}\,\bar{\mathbf{u}}$ (see the previous section), the proper scaling for the forcing term follows:
\begin{equation}\label{force_scaling}
\bar{\mathbf{g}}=\frac{1}{\mbox{Kn}\,\mbox{Ma}^2}\,\mathbf{g}=\frac{1}{\epsilon^{2\beta+1}}\,\mathbf{g}.
\end{equation}
A certain physical acceleration $\bar{\mathbf{g}}$ (fixed for a given problem), can be imposed in the numerical code through the mesh-dependent acceleration $\mathbf{g}=\epsilon^{2\beta+1}\,\bar{\mathbf{g}}$.

\section{\label{appD}Appendix: Computing derivatives locally}

LW-ACM allows a straightforward local computation of spatial derivatives. In this respect, a good example is provided by the following strategy for tuning bulk viscosity. Since the LW-ACM (like LBM and ACM) is an artificial compressibility scheme, bulk viscosity can be regarded as a free parameter (if the incompressible limit is the only concern).

One possible strategy for tuning the bulk viscosity is described below. Instead of standard $f_i^{(e)}$ (see Eq. (\ref{d2q9_equilibrium_compact}) in Appendix \ref{appEQ}) in Eq. (\ref{ACM2.0}), we consider a modified set of functions, namely $f_i^{(e*)}$, defined as
\begin{equation}\label{quasi-equilibrium2}
f_i^{(e*)} = 
{f_i^{(qe)}}\left(\rho,\mathbf{u},\text{Tr}^{(e)}+\gamma\,(\text{Tr}^{(+)}-\text{Tr}^{(e)})\right),
\end{equation}
where ${f^{(qe)}}\left(\rho,\mathbf{u},\text{Tr}\right)$ is given by Eq. (\ref{quasi-equilibrium2}), $\gamma$ is a free parameter, $\text{Tr}^{(e)}$ is the trace of the tensor $\mathbf{\Pi}^{(e)}$ normalized by the density and $\text{Tr}^{(+)}$ is the trace of the tensor $\mathbf{\Pi}^{(+)}=\sum_i\hat{\mathbf{v}}_i\hat{\mathbf{v}}_i f_i(t+\Delta t)$ normalized again by the density. Let us consider the two dimensional case ($D=2$): by definition, $\text{Tr}^{(e)}=2/3+\mathbf{u}^2$. Substituting the latter into Eq. (\ref{quasi-equilibrium2}) and taking into account the definition of $f_i^{(qe)}$ given by Eq. (\ref{quasi-equilibrium}), it is possible to compute the second order tensor of $f_i^{(e*)}$, namely
\begin{equation}\label{second-quasi-equilibrium2}
\mathbf{\Pi}^{(e*)}=\mathbf{\Pi}^{(e)}+\frac{\gamma}{2}\,
(\text{Tr}^{(+)}-\text{Tr}^{(e)})\,\mathbf{I}.
\end{equation}
At the leading order, the modified equilibrium differs from the standard one only due to even moments.

In order to simplify the last term of the Eq. (\ref{second-quasi-equilibrium2}), by Eqs. (\ref{AA}), we compute the following quantities
\begin{equation}\label{AA2}
\mathbf{\Pi}^{(+)}=
\mathbf{\Pi}^{(e)}
-6\nu\,\Delta x\,\nabla\cdot\mathsf{Q}^{(e)}
+O\left(\Delta x^2(\nabla\cdot)^2\mathbf{\Pi}^{(e)}\right),
\end{equation}
where $\mathbf{\Pi}^{(+)}=\sum_i\hat{\mathbf{v}}_i\hat{\mathbf{v}}_i f_i(t+\Delta t)$. Recalling the definition in (\ref{third_def}), it yields
\begin{equation}\label{AA3}
\mathbf{\Pi}^{(+)}=
\mathbf{\Pi}^{(e)}
-2\nu\,\Delta x\,\left(\nabla(\rho\mathbf{u})+\nabla(\rho\mathbf{u})^T+\nabla\cdot(\rho\mathbf{u})\,\mathbf{I}\right)+O\left(\Delta x^2(\nabla\cdot)^2\mathbf{\Pi}^{(e)}\right).
\end{equation}
with its trace taking the form:
\begin{equation}\label{trAA3}
\text{Tr}^{(+)}=
\text{Tr}^{(e)}
-8\nu\,\Delta x\,\nabla\cdot(\rho\mathbf{u})+O\left(\Delta x^2(\nabla\cdot)^2\mathbf{\Pi}^{(e)}\right).
\end{equation}
Considering the diffusive scaling, namely $\Delta x=\epsilon\ll 1$ and $\Delta t=\epsilon^2$ (see Appendix \ref{appA} for further details about the diffusive scaling), the previous expression can be recast as:
\begin{equation}\label{trAA3scaled}
\bar{\text{Tr}}^{(+)}=
\bar{\text{Tr}}^{(e)}
-8\rho_0\nu\,\nabla\cdot\bar{\mathbf{u}}+O(\epsilon^2).
\end{equation}
Introducing the previous expression into Eq. (\ref{second-quasi-equilibrium2}) and applying the scaling to the remaining terms, it reads:
\begin{equation}\label{second-quasi-equilibrium3}
\bar{\mathbf{\Pi}}^{(e*)}=\bar{\mathbf{\Pi}}^{(e)}-4\rho_0\nu\,\gamma\,\nabla\cdot\bar{\mathbf{u}}\,\mathbf{I}+O(\epsilon^2).
\end{equation}
Substituting $\bar{\mathbf{\Pi}}^{(e*)}$ instead of $\bar{\mathbf{\Pi}}^{(e)}$ into Eq. (\ref{AAu2}) and taking into account Eq. (\ref{third}), it yields
\begin{equation}\label{AAu2my}
\rho_0\frac{\partial\bar{\mathbf{u}}}{\partial t}
+\rho_0\bar{\mathbf{u}}\cdot\nabla\bar{\mathbf{u}}
+\nabla\bar{p} = \nu\,\nabla^2 \bar{\mathbf{u}}+\xi\,\nabla\,\nabla\cdot\bar{\mathbf{u}}+O(\epsilon^2),
\end{equation}
where $\xi=2\rho_0\nu\,(1+2\,\gamma)$ is related to the bulk viscosity. The previous equation is consistent with Eq. (\ref{AAu3}), because the gradient of the divergence of the velocity field is as large as the leading error (hence it is not spoiling the consistency). However, the range $\gamma\geq 0$ is usually beneficial to numerical stability. This strategy enables to increase the bulk viscosity by using the updated distribution function for computing locally all required derivatives (involved in the divergence of the velocity field).

\section{\label{appC}Appendix: Equivalent finite-difference formulas}

Here, we provide some finite-difference formulas fully equivalent to Eq. (\ref{ACM2.0}) for a chosen lattice. Let us consider the popular D2Q9 lattice \cite{qian92} for two dimensional problems ($D=2$), and consisting of nine discrete velocities (${Q}=9$). The quantities $f_i^{(e)}$ can be explicitly defined for this lattice \cite{qian92}. They allow to recover the incompressible Euler equations (with $p=\rho/3$), and they are consistent with the property given by (\ref{third}), which is essential for recovering Navier-Stokes equations. The numerical algorithm is fully defined, upon substitution of $f_i^{(e)}$ into the Eq. (\ref{ACM2.0}).

According to the finite-difference literature, we define the generic computational stencil by means of cardinal directions. The generic point $P$ with a position vector $\hat{\mathbf{x}}=(n,m)^T$ (the superscript $T$ denotes transposition) is identified by a pair of integers $n$ and $m$. By means of the subscripts $E$ and $W$, we denote the neighboring points $(n\pm 1,m)^T$, respectively. Similarly, by means of the subscripts $N$ and $S$, we mean the neighboring points $(n,m\pm 1)^T$, respectively. Two types of subscripts may be used concurrently for identifying the diagonal points. 

Concerning time levels, if not otherwise stated, all quantities are intended as computed at the generic time level $\hat{t}$, with the superscript ``$+$'' meaning a quantity at the new time level $\hat{t}+1$. The unknown quantities are given by the velocity components $\mathbf{u}=(u,v)^T$ and the pressure $p$ (the density for this model is given by $\rho=3\,p$). Hence the equivalent finite-difference formulas must provide a way to compute $u_P^+$, $v_P^+$ and $p_P^+$.

Applying the definitions of hydrodynamic quantities to Eq. (\ref{ACM2.0}), it follows:  
\begin{eqnarray}\label{FDjx}
 p_P^+\,u_P^+ &=& p_E (- 6 u_E^2 + 6 u_E + 3 v_E^2 - 2)/18+ p_W (6 u_W^2 + 6 u_W - 3 v_W^2 + 2)/18 \nonumber\\
 		&&- p _{NE} (3 u _{NE}^2 + 9 u _{NE} v _{NE} - 3 u _{NE} + 3 v _{NE}^2 - 3 v _{NE} + 1)/36\nonumber\\
 		&&+ p _{NW} (3 u _{NW}^2 - 9 u _{NW} v _{NW} + 3 u _{NW} + 3 v _{NW}^2 - 3 v _{NW} + 1)/36\nonumber\\
 		&&- p _{SE} (3 u _{SE}^2 - 9 u _{SE} v _{SE} - 3 u _{SE} + 3 v _{SE}^2 + 3 v _{SE} + 1)/36\nonumber\\
 		&&+ p _{SW} (3 u _{SW}^2 + 9 u _{SW} v _{SW} + 3 u _{SW} + 3 v _{SW}^2 + 3 v _{SW} + 1)/36\nonumber\\
 		&&+ 2/3(1/\omega - 1) \left[p_E u_E-3 p_P u_P + p_W u_W\right.\nonumber\\
 		&&+ (p _{NE} u _{NE} + p _{NW} u_{NW}  + p _{SE} u _{SE}+ p _{SW} u _{SW})/4\nonumber\\
 		&&\left.+(p_{NE} v_{NE}- p_{NW} v_{NW}- p_{SE} v_{SE} + p_{SW} v_{SW})/4\right], 		
\end{eqnarray}
\begin{eqnarray}\label{FDjy}
 p_P^+\,v_P^+ &=& p_N (3 u_N^2 - 6 v_N^2 + 6 v_N - 2)/18+ p_S (- 3 u_S^2 + 6 v_S^2 + 6 v_S + 2)/18  \nonumber\\
    &&- p_{NE} (3 u_{NE}^2 + 9 u_{NE} v_{NE} - 3 u_{NE} + 3 v_{NE}^2 - 3 v_{NE} + 1)/36  \nonumber\\
    &&- p_{NW} (3 u_{NW}^2 - 9 u_{NW} v_{NW} + 3 u_{NW} + 3 v_{NW}^2 - 3 v_{NW} + 1)/36  \nonumber\\
    &&+ p_{SE} (3 u_{SE}^2 - 9 u_{SE} v_{SE} - 3 u_{SE} + 3 v_{SE}^2 + 3 v_{SE} + 1)/36  \nonumber\\
    &&+ p_{SW} (3 u_{SW}^2 + 9 u_{SW} v_{SW} + 3 u_{SW} + 3 v_{SW}^2 + 3 v_{SW}+ 1)/36\nonumber\\
    && 2/3(1/\omega - 1) \left[p_N v_N- 3 p_P v_P + p_S v_S\right.\nonumber\\
    &&+(p_{NE} u_{NE} - p_{NW} u_{NW} - p_{SE} u_{SE} + p_{SW} u_{SW})/4\nonumber\\
    &&\left.+ (p_{NE} v_{NE} + p_{NW} v_{NW}  + p_{SE} v_{SE} + p_{SW} v_{SW})/4\right].
\end{eqnarray}
\begin{eqnarray}\label{FDpressure}
 p_P^+ &=& - 2 p_P (3 u_P^2 + 3 v_P^2 - 2)/9 \nonumber\\
     &&- p_E (- 6 u_E^2 + 6 u_E + 3 v_E^2 - 2)/18+ p_W (6 u_W^2 + 6 u_W - 3 v_W^2 + 2)/18 \nonumber\\
     &&- p_N (3 u_N^2 - 6 v_N^2 + 6 v_N - 2)/18 + p_S (- 3 u_S^2 + 6 v_S^2 + 6 v_S + 2)/18 \nonumber\\
     &&+ p_{NE} (3 u_{NE}^2 + 9 u_{NE} v_{NE} - 3 u_{NE} + 3 v_{NE}^2 - 3 v_{NE} + 1)/36 \nonumber\\
     &&+ p_{NW} (3 u_{NW}^2 - 9 u_{NW} v_{NW} + 3 u_{NW} + 3 v_{NW}^2 - 3 v_{NW} + 1)/36 \nonumber\\
     &&+ p_{SE} (3 u_{SE}^2 - 9 u_{SE} v_{SE} - 3 u_{SE} + 3 v_{SE}^2 + 3 v_{SE} + 1)/36 \nonumber\\
     &&+ p_{SW} (3 u_{SW}^2 + 9 u_{SW} v_{SW} + 3 u_{SW} + 3 v_{SW}^2 + 3 v_{SW} + 1)/36 \nonumber\\
     &&+ 2/3(1/ \omega - 1)\left[ - p_E u_E + p_S v_S+ p_W u_W - p_N v_N\right.\nonumber\\
     &&+ (p_{SW} u_{SW}+p_{NW} u_{NW} - p_{NE} u_{NE} - p_{SE} u_{SE})/4 \nonumber\\
     &&\left.+ (p_{SW} v_{SW} - p_{NW} v_{NW}-p_{NE} v_{NE} + p_{SE} v_{SE})/4 \right],
\end{eqnarray}
It is important to stress that the above expressions (\ref{FDjx}), (\ref{FDjy}), (\ref{FDpressure}) for the considered lattice model are fully equivalent to (\ref{ACM2.0}) up to machine precision, since no asymptotic analysis is requested in their derivation.

From a computational perspective, optimal implementation requires that the number of floating point operations are reduced as much as possible by common subexpression elimination (CSE) {\cite{Hager2010}}. Moreover, for locating the macroscopic quantities $(p,u,v)$ contiguously in the memory, it is possible to collect them in a single array and to use the first index for addressing them, namely $\mathsf{M(1:3,1:N_x,1:N_y)}$ where $\mathsf{N_x} \times \mathsf{N_y}$ is the generic mesh. This leads to an optimized FD-style implementation. First of all, let us compute the following auxiliary quantities
%
%\red{
\begin{eqnarray}\label{CSE}
pu_P &=& \mathsf{M(1,i,j)}\,\mathsf{M(2,i,j)},\qquad \varphi u_P = 2\,pu_P,\nonumber\\
pv_P &=& \mathsf{M(1,i,j)}\,\mathsf{M(3,i,j)},\qquad \varphi v_P = 2\,pv_P,\nonumber\\
pv_N &=& \mathsf{M(1,i,j+1)}\,\mathsf{M(3,i,j+1)},\qquad pv_S = \mathsf{M(1,i,j-1)}\,\mathsf{M(3,i,j-1)},\nonumber\\
pu_E &=& \mathsf{M(1,i+1,j)}\,\mathsf{M(2,i+1,j)},\qquad pu_W = \mathsf{M(1,i-1,j)}\,\mathsf{M(2,i-1,j)},\nonumber\\
pu_{NE} &=& \mathsf{M(1,i+1,j+1)}\,\mathsf{M(2,i+1,j+1)},\nonumber\\
pu_{NW} &=& \mathsf{M(1,i-1,j+1)}\,\mathsf{M(2,i-1,j+1)},\nonumber\\
pu_{SE} &=& \mathsf{M(1,i+1,j-1)}\,\mathsf{M(2,i+1,j-1)},\nonumber\\
pu_{SW} &=& \mathsf{M(1,i-1,j-1)}\,\mathsf{M(2,i-1,j-1)},\nonumber\\
pv_{NE} &=& \mathsf{M(1,i+1,j+1)}\,\mathsf{M(3,i+1,j+1)},\nonumber\\
pv_{NW} &=& \mathsf{M(1,i-1,j+1)}\,\mathsf{M(3,i-1,j+1)},\nonumber\\
pv_{SE} &=& \mathsf{M(1,i+1,j-1)}\,\mathsf{M(3,i+1,j-1)},\nonumber\\
pv_{SW} &=& \mathsf{M(1,i-1,j-1)}\,\mathsf{M(3,i-1,j-1)},
\end{eqnarray}
%}
%
and
%
%\red{
\begin{eqnarray}\label{CSE2}
\phi_1 &=& pu_E\,(-\mathsf{M(2,i+1,j)} + 1) + \mathsf{M(1,i+1,j)}\,(\mathsf{M(3,i+1,j)}^2/2 - r_{13}),\nonumber\\
\phi_2 &=& pu_W\,( \mathsf{M(2,i-1,j)} + 1) - \mathsf{M(1,i-1,j)}\,(\mathsf{M(3,i-1,j)}^2/2 - r_{13}),\nonumber\\
\phi_3 &=& pv_N\,(-\mathsf{M(3,i,j+1)} + 1) + \mathsf{M(1,i,j+1)}\,(\mathsf{M(2,i,j+1)}^2/2 - r_{13}),\nonumber\\
\phi_4 &=& pv_S\,( \mathsf{M(3,i,j-1)} + 1) - \mathsf{M(1,i,j-1)}\,(\mathsf{M(2,i,j-1)}^2/2 - r_{13}),\nonumber\\
\phi_5 &=& pu_{NE}\,(\mathsf{M(2,i+1,j+1)}+3\,\mathsf{M(3,i+1,j+1)}-1)+\dots\nonumber\\
	&&\dots+pv_{NE}\,(\mathsf{M(3,i+1,j+1)}-1)+\mathsf{M(1,i+1,j+1)}\,r_{13},\nonumber\\
\phi_6 &=& pu_{NW}\,(\mathsf{M(2,i-1,j+1)} - 3\,\mathsf{M(3,i-1,j+1)} +1) +\dots\nonumber\\
  &&\dots+pv_{NW}\,(\mathsf{M(3,i-1,j+1)} - 1) + \mathsf{M(1,i-1,j+1)}\,r_{13},\nonumber\\
\phi_7 &=& pu_{SE}\,(\mathsf{M(2,i+1,j-1)} - 3\,\mathsf{M(3,i+1,j-1)} -1)+\dots\nonumber\\
  &&\dots+pv_{SE}\,(\mathsf{M(3,i+1,j-1)} + 1) + \mathsf{M(1,i+1,j-1)}\,r_{13},\nonumber\\
\phi_8 &=& pu_{SW}\,(\mathsf{M(2,i-1,j-1)} + 3\,\mathsf{M(3,i-1,j-1)} +1) +\dots\nonumber\\
  &&\dots+pv_{SW}\,(\mathsf{M(3,i-1,j-1)} + 1) + \mathsf{M(1,i-1,j-1)}\,r_{13},
\end{eqnarray}
%}
%
where $r_{13}=1/3$. By means of the quantities (\ref{CSE}) and (\ref{CSE2}), it is possible to compute the following additional quantities
\begin{eqnarray}\label{CSE3}
\phi_{Pp} &= pu_P + pv_P,\qquad \phi_{Pm} &= pu_P - pv_P,\nonumber\\
\phi_{NEp} &= pu_{NE} + pv_{NE},\qquad \phi_{SWp} &= pu_{SW} + pv_{SW},\nonumber\\
\phi_{NWm} &= pu_{NW} - pv_{NW},\qquad \phi_{SEm} &= pu_{SE} - pv_{SE},\nonumber\\
\Phi_{NE} &= \phi_{NEp} - \phi_{Pp},\qquad \Phi_{SW} &= \phi_{Pp} - \phi_{SWp},\nonumber\\
\Phi_{NW} &= \phi_{NWm} - \phi_{Pm},\qquad \Phi_{SE} &= \phi_{Pm} - \phi_{SEm},\nonumber\\
\Phi_{NWSE} &= \Phi_{NW}-\Phi_{SE},\qquad \Phi_{NESW} &= \Phi_{NE}-\Phi_{SW}.
\end{eqnarray}
Finally, auxiliary quantities are used in computing the updating formulas:
%
%\red{
\begin{eqnarray}\label{CSE4}
\rho_P^+ &=& \mathsf{M(1,i,j)}\,r_{43} - \varphi u_P\,\mathsf{M(2,i,j)} - \varphi v_P\,\mathsf{M(3,i,j)}+\phi_2-\phi_1+\phi_4-\phi_3+\dots\nonumber\\
  &&\dots+(\phi_5+\phi_6+\phi_7+\phi_8)/4-b_4\,(\Phi_{NW}+\Phi_{SE}-\Phi_{NE}-\Phi_{SW})+\dots\nonumber\\
  &&\dots+b\,(pu_E - pu_W + pv_N - pv_S),\nonumber\\
p_P^+&\equiv& \mathsf{M^+(1,i,j)} = \rho_P^+/3,\\
u_P^+&\equiv& \mathsf{M^+(2,i,j)} = ( \phi_1 + \phi_2 + (\phi_6 - \phi_5 - \phi_7 + \phi_8)/4+\dots\nonumber\\
&&\dots - a\,(pu_E - \varphi u_P + pu_W) - b_4\,(\Phi_{NESW}+\Phi_{NWSE}) )/\rho_P^+,\\
v_P^+&\equiv& \mathsf{M^+(3,i,j)} = ( \phi_3 + \phi_4 + (\phi_7 + \phi_8 - \phi_5 - \phi_6)/4+\dots\nonumber\\
&&\dots - a\,(pv_N - \varphi v_P + pv_S) - a_4\,(\Phi_{NESW}-\Phi_{NWSE}) )/\rho_P^+,
\end{eqnarray}
%}
%
where $r_{43}=4/3$, $b=2-2/\omega$, $a_4=a/4$, $b=2-2/\omega_p$ and $b_4=b/4$. The previous optimized formulas are consistent with Eqs. (\ref{FDjx}-\ref{FDpressure}) in case $\omega_p=\omega$, $b=a$ and $b_4=a_4$. Kinematic viscosity is controlled via the parameter $\omega$ according to (\ref{viscosity2}), whereas the parameter $\omega_p$ is responsible of the artificial compressibility. In particular, if $\omega_p\neq\omega$, the first term in Eq. (\ref{AArho4}) becomes proportional to $\epsilon^2/\nu_p$, where $\nu_p$ is the value obtained by using $\omega_p$ in Eq. (\ref{viscosity2}). Without additional computational costs, a proper choice of $\nu_p>\nu$ allows to reduce the compressibility error in case of under-resolved simulations.

\section{\label{appB}Appendix: Asymptotic analysis for energy equation}

Under the assumption of negligible viscous heating and conservation of internal energy, in the incompressible limit, the temperature field is governed by an advection-diffusion equation. Let us call $T$ the normalized temperature field such that $T\leq 1$ in all the domain of interest, with the quantities $f_i^{(e)}$ defined such that $p=\rho\,T/3$.

There were a number of suggestions aiming at enabling thermal fluid-dynamic simulations with the lattice Boltzmann method \cite{Lallemand2003}. Among the most interesting ones, we remind: (a) Increase of the number of velocities and inclusion of higher-order nonlinear terms (in flow velocity) in the
equilibrium distribution functions; (b) inclusion of finite difference corrections aiming at the fulfillment of energy conservation on standard lattices \cite{Prasianakis2007} and, (c) use of two sets of distribution functions for particle number density ($f_i$), and energy density ($g_i$), doubling the number of discrete velocities. Even though the first and the second approaches are preferable from the theoretical point of view, the last one is characterized by a much simpler implementation. When dealing with the incompressible limit, the pressure field is characterized by small variations. However, cases may be experienced where large temperature and density gradients compensate each other. If both pressure gradients and temperature gradients are small, a weak coupling between fluid dynamic and energy equations is realized and the simplified approach (c) discussed above can be adopted. This approach will be further extended below in the framework of the present LW-ACM. Extensions of the approach (b) in the framework of LW-ACM are currently under investigation as well.

The LW-ACM approach for (weak) thermal fluid dynamic simulations makes use of the following system of algebraic equations
\begin{equation}\label{ACM2.0t}
g_i(\hat{\mathbf{x}},\hat{t}+1)=
g_i^{(e)}(\hat{\mathbf{x}}-\hat{\mathbf{v}}_i,\hat{t})
+2\left(\frac{\omega_t-1}{\omega_t}\right)\left(
g_i^{(e,e)}(\hat{\mathbf{x}},\hat{t})-
g_i^{(e,e)}(\hat{\mathbf{x}}-\hat{\mathbf{v}}_i,\hat{t})\right),
\end{equation}
for $i=0,\dots,Q_t-1$ (the number of lattice velocities for solving the thermal field can be different from that used for solving the velocity field, i.e. $Q_t\neq Q$). The quantities $g_i^{(e)}$ are local functions of $T=\sum_i g_i$ and $T\mathbf{u}=\sum_i \hat{\mathbf{v}}_i g_i$ at the same point $\hat{\mathbf{x}}$ and time $\hat{t}$. The quantities $g_i^{(e)}$ are designed in order to recover the advection-diffusion equation, according to the constraints discussed below. In particular, recovering the advection-diffusion equation requires : $\sum_i g_i^{(e)}=T$ and $\sum_i\hat{\mathbf{v}}_i g_i^{(e)}=T\mathbf{u}$, i.e. the conservation of hydrodynamic moments, and $\sum_i\hat{\mathbf{v}}_i\hat{\mathbf{v}}_i g_i^{(e)}=\mathbf{\Pi}_t^{(e)}=\mathbf{u}\mathbf{u}+T/3\,\mathbf{I}$. On the other hand, the even parts of equilibria $g_i^{(e,e)}$ are defined as
\begin{equation}\label{feqe}
g_i^{(e,e)}(T,\mathbf{u})=\frac{1}{2}\left(g_i^{(e)}(T,\mathbf{u})+g_i^{(e)}(T,-\mathbf{u})\right).
\end{equation}

%Appendix \ref{appB} reports the asymptotic analysis of the previous formula in case of diffusive scaling, namely $\Delta x=\epsilon\ll 1$ and $\Delta t=\epsilon^2$ (see Appendix \ref{appA} for details about the diffusive scaling). If diffusive scaling is adopted, then the quantity $T$ satisfies an advection-diffusion equation with the diffusion coefficient $\alpha$ (proportional to the thermal diffusivity) given by
%
%\begin{equation}\label{diffusivity}
%\alpha=\frac{1}{3}\left(\frac{1}{\omega_t}-\frac{1}{2}\right).
%\end{equation}
%
%Accurately solving advection-diffusion equation requires that $\epsilon^2/\alpha\ll 1$, as usual in the Lattice Boltzmann method (LBM) as well (see Appendix \ref{appB} for details). Numerical evidences suggest that the previous numerical scheme is stable for $1\leq \omega_t<2$, which corresponds only to half of the stability range for the relaxation frequency of the LBM. However this is the most interesting half from the practical point of view, because it corresponds to the limit of vanishing diffusivity (i.e. high Peclet numbers). 

Eq. (\ref{ACM2.0t}) is formulated in terms of $\hat{\mathbf{x}}=\mathbf{x}/\Delta x$ and $\hat{t}=t/\Delta t$. In the following, let us consider the diffusive scaling, namely $\Delta x=\epsilon\ll 1$ and $\Delta t=\epsilon^2$ (see the Appendix \ref{appA} for further details on diffusive scaling). Taylor expansion of Eq. (\ref{ACM2.0t}) yields
\begin{eqnarray}\label{AAt}
g_i(t+\epsilon^2)&=&
g_i^{(e)}
-\epsilon\,\hat{\mathbf{v}}_i\cdot\nabla g_i^{(e)}
+\frac{\epsilon^2}{2}(\hat{\mathbf{v}}_i\cdot\nabla)^2 g_i^{(e)}\nonumber\\
&&+\left(2-\frac{2}{\omega_t}\right)\left(
\epsilon\,\hat{\mathbf{v}}_i\cdot\nabla g_i^{(e,e)}
-\frac{\epsilon^2}{2}(\hat{\mathbf{v}}_i\cdot\nabla)^2 g_i^{(e,e)}\right)+\dots,
\end{eqnarray}
where all the quantities are computed in the same point $\mathbf{x}$ and time $t$ and hence this is no more explicitly reported. Summation over $i$ of the equations (\ref{AAt}) yields
\begin{equation}\label{AAT}
\frac{\partial T}{\partial t}
+\frac{1}{\epsilon}\,\nabla\cdot (T\mathbf{u})
+\frac{\epsilon}{6}\,(\nabla\cdot)^3 \mathbf{Q}_t^{(e)}=
\left(\frac{1}{\omega_t}-\frac{1}{2}\right)(\nabla\cdot)^2\mathbf{\Pi}_t^{(e)}+O\left(\epsilon^2(\nabla\cdot)^2\mathbf{\Pi}^{(e)}_t\right),
\end{equation}
with $\mathbf{Q}_t^{(e)}=\sum_i\hat{\mathbf{v}}_i\hat{\mathbf{v}}_i\hat{\mathbf{v}}_i g_i^{(e)}$. In the rightmost term of the expression (\ref{AAT}), we rely upon the fact that spatial derivatives of the fourth-order moments have the same growth rate of spatial derivatives of the second-order moments. Moreover, in the same term, the diverge is raised to a power which is selected for consistency with the units of other terms. This is unessential, as far as the order of magnitude of the term is concerned.

Similarly to the fluid dynamic equations (see the Appendix \ref{appA}), we apply the following post-processing for scaling
all moments (arbitrary constants have been already omitted):
\begin{equation}\label{scaling2t}
\bar{\mathbf{u}}=\mathbf{u}/\epsilon,\qquad \bar{\mathbf{\Pi}}_t^{(e)}=\mathbf{\Pi}^{(e)}_t,\qquad \bar{\mathsf{Q}}_t^{(e)}=\mathsf{Q}^{(e)}_t/\epsilon. 
\end{equation}
One possibility for automatic implementation of the latter scaling is to assume
\begin{equation}\label{third_deft}
(\bar{\mathsf{Q}}^{(e)}_t)_{ijk}=\frac{1}{3}\left(\bar{u}_i\delta_{jk}+\bar{u}_j\delta_{ik}+\bar{u}_k\delta_{ij}\right),
\end{equation}
similarly to Eq. (\ref{third_def}), because thus all terms in $\mathsf{Q}^{(e)}_t$ become proportional to the velocity field (and they are automatically scaled by means of the first of the above scalings). 

Moreover, taking into account the second scaling and the definition of $\mathbf{\Pi}^{(e)}_t$ reported in the main text, we conclude that $\bar{\mathbf{\Pi}}_t^{(e)}=\epsilon^2\bar{\mathbf{u}}\bar{\mathbf{u}}+T/3\,\mathbf{I}$, where $T$ does not need to be scaled (or equivalently $\bar{T}=T$). Substituting the previous scalings into (\ref{AAt}), and taking into account (\ref{AArho2}), we obtain
\begin{equation}\label{AAT2}
\boxed{\frac{\partial \bar{T}}{\partial t}
+\bar{\mathbf{u}}\cdot\nabla \bar{T}=\alpha\nabla^2\bar{T}+O(\epsilon^2),}
\end{equation}
where
\begin{equation}\label{diffusivity2}
\alpha=\frac{1}{3}\left(\frac{1}{\omega_t}-\frac{1}{2}\right).
\end{equation}

%\section{\label{thermal}Appendix: Incompressible thermal fluid dynamics}

%\subsubsection{Thermal Couette flow}

%\begin{figure}[ht]
%\begin{center}
%\includegraphics[width=12cm]{thermalCouette}
%\end{center}
%\caption{Numerical results for the thermal Couette flow: results obtained by means of LW-ACM (blue line) and analytical solution (red dots).}
%\label{fig:thermal}
%\end{figure}

\vspace{0.5cm}
\begin{table}[ht]
\caption{Convergence analysis for thermal Couette flow in case of diffusive scaling, with Prandtl number $\mbox{Pr}=0.71$.}
\label{thermal-Couette-diffusive}
\vspace{0.5cm}
\begin{center}
\begin{tabular}{cccccc}
	\hline
\multicolumn{6}{c}{Link-wise ACM} \\
  \hline
$\epsilon\equiv\Delta x$ & $\mbox{Ma}\propto \Delta t/\Delta x$ & $\nu\propto\mbox{Re}^{-1}$ & $\alpha\propto\mbox{Pe}^{-1}$ 
& {Error $L^2[\bar{u}]$} & {Error $L^2[T]$} \\
  \hline
$1/5$ & $1.11\times 10^{-2}$ & $5.55\times 10^{-2}$ & $1.11\times 10^{-2}$ & $3.10\times10^{-3}$ & $1.61\times10^{-6}$\\
$1/10$ & $5.55\times 10^{-2}$ & $5.55\times 10^{-2}$ & $1.11\times 10^{-2}$ & $8.25\times10^{-4}$ & $4.28\times10^{-7}$\\
$1/20$ & $2.78\times 10^{-3}$ & $5.55\times 10^{-2}$ & $1.11\times 10^{-2}$ & $2.12\times10^{-4}$ & $1.01\times10^{-7}$%\\
%	\hline
%	\hline
%\multicolumn{6}{c}{\red{MRT-LBM}} \\
%$1/10$ & $5.55\times 10^{-2}$ & $5.55\times 10^{-2}$ & $1.11\times 10^{-2}$ & \red{$\times10^{-?}$} & \red{$\times10^{-?}$}\\
%  \hline
\end{tabular}
\end{center}
\end{table}

%\blue{
Here, a few numerical results are reported for the thermal Couette problem, which is realized by confining a viscous fluid in a gap between two parallel plates. Assuming that the one plate (hot wall located at $y=0$ and with temperature $\bar T_N$) moves in its own plane, whereas the other (cold wall located at $y=2L$ and with temperature $\bar T_S$) is at rest, the controlling parameters are the Prandtl number $\mbox{Pr}=\nu/\alpha$ (measuring the momentum diffusivity: $\nu$, to heat diffusivity: $\alpha$) and the Eckert number $\mbox{Ec} = \bar{u}^2/c_v \Delta \bar T$ (measuring the kinetic energy: $\rho \bar{u}^2/2$, to internal energy: $\rho c_v \Delta \bar{T}$). Thermal Couette flow admits the following analytical solution of the temperature field:
%%%
\begin{equation}\label{thermalcouette}
\bar T(y) = \bar T_S + \frac{y}{2L} \Delta \bar T + \frac{\mbox{Br} \Delta \bar T }{2} \frac{y}{2L} \left(1 - \frac{y}{2 L} \right),
\end{equation}
with $\Delta \bar T = \left( \bar T_N - \bar T_S \right)$, and the Brinkman number $\mbox{Br}= \mbox{Pr} \,\mbox{Ec}$.%}
%\end{equation}

%\blue{
Diffusive scaling was considered in our simulations, where the velocity field is scaled on meshes with different sizes, keeping fixed the relaxation frequency (see the Appendix \ref{appA} for details). Some preliminary numerical results are reported in Table \ref{thermal-Couette-diffusive} for $\mbox{Pr}=\nu/\alpha=0.71$.%}

\bibliographystyle{unsrt}

\end{document}